\documentclass[a4paper,11pt]{article}
\usepackage{jheppub} 
\usepackage{lineno}
\usepackage{cleveref}
\usepackage[normalem]{ulem}
\usepackage{amsmath, amssymb,braket}
\usepackage[dvipsnames]{xcolor}
\usepackage{float}
\usepackage{graphicx}
\usepackage{subfigure}
\usepackage{natbib}
\usepackage{adjustbox}
\usepackage{tikz}
\usepackage{soul}
\usetikzlibrary{shapes,arrows}
\usepackage[compat=1.1.0]{tikz-feynman}
\usepackage[colorlinks=true,citecolor=blue,urlcolor=blue]{hyperref}
\usepackage{slashed}
\usetikzlibrary{shapes.geometric, arrows}

\newcommand{\Po}{\Pi^{\mu\nu}}

\newcommand{\Trnl}{T^{\mu\nu}_{nl}}
\newcommand{\therm}{\mathrm{th}}
\newcommand{\sgn}{\mathrm{sgn}}

\newcommand{\herm}{\mathcal{H}}
\newcommand{\Aherm}{\bar{\mathcal{H}}}
\newcommand{\modk}{|\vec{k}|}

\DeclareRobustCommand{\Eq}[1]{Eq.~\eqref{#1}}

\DeclareRobustCommand{\Sec}[1]{Section~\ref{#1}}
\DeclareRobustCommand{\App}[1]{Appendix~\ref{#1}}
\DeclareRobustCommand{\Fig}[1]{Fig.~\ref{#1}}

\newcommand{\pimix}{\pi^{IJ}_{AA}}

\newcommand{\fourvec}[1]{#1}
\newcommand{\real}{\mathrm{Re}}
\newcommand{\imag}{\mathrm{Im}}

\newcommand{\Ap}{A^\prime}
\newcommand{\mAp}{m_{A^\prime}}

\newcommand{\eps}{\epsilon}
\newcommand{\epstil}{\tilde{\epsilon}}

\title{\boldmath Photon conversion to axions and dark photons in magnetized plasmas: a finite-temperature field theory approach}

\author{Nirmalya Brahma}
\author{and Katelin Schutz}
\affiliation{Department of Physics \& Trottier Space Institute, 
McGill University, Montr\'{e}al, Canada}

\emailAdd{nirmalya.brahma@mail.mcgill.ca}
\emailAdd{katelin.schutz@mcgill.ca}
\abstract{Some of the most stringent constraints on physics beyond the Standard Model (BSM) arise from considerations of particle emission from astrophysical plasmas. However, many studies assume that particle production occurs in an isotropic plasma environment. This condition is rarely (if ever) met in astrophysical settings, for instance due to the ubiquitous presence of magnetic fields. In anisotropic plasmas, the equations of motion are not diagonal in the usual polarization basis of transverse and longitudinal modes, causing a mixing of these modes and breaking the degeneracy in the dispersion relation of the two transverse modes. This behavior is captured by a $3\times3$ mixing matrix $\pi^{IJ}$, determined by projecting the response tensor of the plasma $\Po$ into mode space, whose eigenvectors and eigenvalues are related to the normal modes and their dispersion relations. In this work, we provide a general formalism for determining the normal modes of propagation that are coupled to axions and dark photons in an anisotropic plasma. As a key part of this formalism, we present detailed derivations of $\Po$ for magnetized plasmas in the long-wavelength limit using the real-time formalism of finite-temperature field theory. We provide analytic approximations for the normal modes and their dispersion relations assuming various plasma conditions that are relevant to astrophysical environments. These approximations will allow for a systematic exploration of the effects of plasma anisotropy on BSM particle production.
}

\begin{document}
\maketitle
\flushbottom
\section{Introduction}
\label{sec:intro}
Axions~\cite{peccei1977cp,weinberg1978new} and dark photons (DPs)~\cite{Holdom:1985ag} are among the most well-studied extensions of the Standard Model (SM) of particle physics. From the UV-physics perspective, string theories generally contain axions~\cite{witten1984some, svrcek2006axions, Arvanitaki:2009fg, acharya2010m, Cicoli:2012sz} and dark photons~\cite{Abel:2008ai,Goodsell:2009xc} as states in the spectrum. These particles can address outstanding issues in the SM, for instance the QCD axion can address the strong-$CP$ problem~\cite{peccei1977cp, weinberg1978new, wilczek1978problem, abbott1983cosmological, preskill1983cosmology, dine1983not}. From the IR perspective, both kinds of particles can serve as a portal to the dark sector via low-dimensional operators, most notably via interactions or kinetic mixing with the SM photon~\cite{Arkani-Hamed:2008hhe, Pospelov:2008zw,Nomura:2008ru,Hochberg:2018rjs,Chu:2011be,Arcadi:2017kky}. For the axion, $a$, these interactions are captured by the Lagrangian
\begin{equation}
\label{photon-axion-eom}
\mathcal{L} \supset \frac{1}{2}\partial_{\mu}a\partial^{\mu}a-\frac{1}{2}m_{a}^{2}a^{2}-\frac{g_{a\gamma}}{4} aF_{\mu\nu}\tilde{F}^{\mu\nu}
\end{equation}
where $\tilde{F}^{\mu\nu} = \frac{1}{2}\eps^{\mu\nu\alpha\beta} F_{\alpha\beta}$ is the dual of the electromagnetic field strength $F^{\mu\nu}$, and where the dimensionful coupling $g_{a\gamma}$ may not be fundamental but rather may arise from integrating out loops of other particles, depending on the axion model (e.g.~\cite{Kim:1979if,Shifman:1978bx,dine1983not,Zhitnitsky:1980tq}). Meanwhile, the behaviour of a kinetically mixed DP $A'$ can be described by the Lagrangian 
\begin{align}
    \mathcal{L} \supset 
    &-\frac{1}{4}\, F'_{\mu\nu} \, F'^{\mu\nu} +\frac{\chi}{2} \, F'_{\mu\nu} \, F^{\mu\nu} +
  \frac{1}{2} \, \mAp^{2} \, A'_{\mu} \, {A}'^{\mu} ~, 
\end{align}
where $\chi$ is the dimensionless mixing parameter and $\mAp$ is the DP mass generated through the St\"uckelberg mechanism~\cite{Stueckelberg:1938hvi}.\footnote{A dark Higgs mechanism can also generate the DP mass, along with additional dark-sector dynamics. For the sake of simplicity, we focus on the St\"uckelberg case in this work.} These interactions can manifest at a range of energies, including ones that are accessible in astrophysical environments and tabletop experiments. Notably, the interconversion of SM photons with axions (especially in the presence of an ambient magnetic field) and DPs provides a pathway for discovering or constraining BSM physics. 

Interconversion between photons and BSM states can be significantly enhanced when kinematic conditions are favorable, especially if there is a resonance or level crossing. Such resonances arise in the presence of a thermal background, where propagating photons can receive large corrections to their dispersion relations, leading to an effective photon mass~\cite{Das:1997gg, Bellac:2011kqa, Kapusta:2006pm}. The photon also acquires a third mode of propagation (beyond the two transverse vacuum modes), with its own distinct dispersion relations. Expressed in terms of the SM photon Lagrangian, $\mathcal{L} \supset -\frac{1}{4}\, F_{\mu\nu} \, F^{\mu\nu} + J^\mu A_\mu $, the current $J^\mu$ in a medium includes a contribution induced by the charges in the medium, $J^\mu = J^\mu_\text{ind} + J^\mu_\text{ext}$. Using linear response theory~\cite{Kapusta:2006pm, mallik2016hadrons, melrose2008quantum} we can express the induced current as
\begin{equation}\label{ind_current}
  J^\mu_\text{ind} = -\Pi^{\mu\nu}_R A_\nu \,.
\end{equation}
where $\Pi^{\mu\nu}_R = \Pi^{\mu\nu}_R(\omega, \vec{k})$ is the retarded photon self-energy which represents the linear response of the medium. This linear response can be computed in a number of different ways in different regimes, for instance using techniques from classical plasma physics, condensed matter physics, and thermal field theory. For example, in a classical isotropic plasma with $T/m_e\rightarrow0$, the transverse part of the self-energy $1/2 (\delta_{ij} - k_i k_j/k^2)\Pi^{ij}_R$ is equal to the squared plasma frequency $\omega_p^2 = {e^2 n_e/m_e}$, while the longitudinal part is $\Pi^{00}_R = \omega_p^2 k^2/\omega^2$ and the same results are obtained using classical, quantum, or field theoretic considerations~\cite{ashcroft1978solid,braaten1991calculation}. Moreover, the self-energy tensor (as defined in \Eq{ind_current}) can also be related to the dielectric 3-tensor $\eps^i_j$ (with spatial indices $i,j$) as 
\begin{equation}\label{dielec}
  \eps^i_j=\delta^i_j-\Pi^i_j/\omega^2, 
\end{equation}
or for instance the longitudinal dielectric function $\eps_L$ can be expressed as $\eps_L = 1-\Pi^{00}_R/k^2$ and can be obtained equivalently using Lindhard theory or finite-temperature field theory~\cite{dressel2002electrodynamics,Scherer:2024uui}. Imposing charge continuity ($\fourvec{K}^\mu J_\mu=0$ where $\fourvec{K}^\mu = (\omega, \vec{k})$ is the four-momentum) and gauge invariance induces a transversality condition on the self-energy 
\begin{equation}\label{transversality}
\fourvec{K}_\mu \Po=\fourvec{K}_\nu \Po =0\,,
\end{equation}
which is particularly useful for constructing the self-energy tensor. Ambient magnetic fields further enrich the physics of photons propagating in media, for instance giving rise to Faraday rotation and the Cotton-Mouton effect~\cite{rizzo1997cotton} and Landau diamagnetism~\cite{landau1930diamagnetismus} (which in turn gives rise to the Schubnikow-De Haas effect and De Haas-Van Alphen oscillations~\cite{de1930dependence, schubnikow1930new}, among other phenomena). Even in a vacuum, magnetic fields induce birefringence~\cite{klein1964birefringence,Raffelt:1987im} through non-linear terms in the Euler-Heisenberg Lagrangian~\cite{Heisenberg:1936nmg}. 

Many astrophysical and terrestrial searches for axions~\cite{CAST:2013bqn, CAST:2015qbl,Hook:2018iia,Leroy:2019ghm, Bondarenko:2022ngb,Millar:2021gzs,Noordhuis:2022ljw, Prabhu:2021zve, Prabhu:2020yif, Noordhuis:2023wid,Dessert:2019sgw, Wang:2021wae, Caputo:2020quz,OShea:2023gqn,An:2023wij,Todarello:2023ptf,Caputo:2021kcv,Lawson:2019brd} and DPs~\cite{Fischbach:1994ir,Redondo:2008aa,Vinyoles_2015,Redondo:2013lna,An:2013yfc,McDermott:2019lch,Mirizzi:2009iz,Caputo:2020bdy,Caputo:2020rnx,Li:2023vpv,Garcia:2020qrp,Gelmini:2020kcu} rely on the kinematic matching provided by plasmas when the effective photon mass is near the mass scale of the DP or axion. However, for simplicity, many studies assume an isotropic plasma. This assumption does not hold in the presence of external magnetic fields, which are ubiquitous in astrophysical systems. In recent years, various studies have quantified the effect of anisotropies in the context of axion conversion to photons, primarily in neutron star magnetospheres. These works obtain similar results using different methods, for instance using simulations of classical axion electrodynamics (including a simplified wave description using the WKB approximation) and quantum kinetic (Kadanoff-Baym) equations~\cite{Visinelli:2018zif,McDonald:2023ohd,Gines:2024ekm,McDonald:2024uuh}. However, these studies typically employ a dielectric tensor (response function) derived from classical kinetic theory, which is more similar to the hard thermal loop approximation (see Section 9.5 in Ref.~\cite{Kapusta:2006pm}), with relativistic generalizations discussed in Ref.~\cite{DGSwanson:2003}. 

Magnetic fields in thermal plasmas are known to heavily modify the properties of the background charged particles and consequently the photon self-energy, which has been extensively studied in the literature primarily using a quantum field theoretic framework~\cite{Ganguly:1999ts, DOlivo:2002omk, Shabad:2010hx, Hattori:2012je, Hattori:2022uzp} (see Ref.~\cite{Miransky:2015ava} for a detailed review). These modifications to the photon self-energy directly impact the independent normal modes of the equations of motion, which are no longer the transverse and longitudinal modes once the medium has anisotropy. The new normal modes, and their corresponding dispersion relations, can be directly determined from the anisotropic in-medium self-energy. Thus, the $B$-dependent self-energy tensor qualitatively alters the in-medium production of DPs and axions. 

In this work, we therefore provide a general description of the propagation of photons in anisotripic plasmas using the real-time formalism (RTF) of finite-temperature field theory (FTFT) . We work at order $\mathcal{O}(\alpha)$ in the long-wavelength limit where the photon has a small momentum $\vec{k}$, but we otherwise make no simplifying assumptions about the nature of the astrophysical plasma or magnetic field strength. In keeping the problem fairly general, we obtain a set of equations of motion that can be solved numerically as appropriate for the specific plasma under consideration. This is particularly useful since the magnetic fields can vary substantially between different systems used for BSM searches, both in terms of the magnetic field strength and the level of spatial coherence. We further consider various limiting cases of magnetized plasmas using our framework, for instance classical plasmas, degenerate plasmas, and plasmas with super-critical magnetic fields, making contact with previous work as appropriate. Given the behaviour of propagating photons described using this formalism, we can straightforwardly evaluate the impact on their mixing with DPs and axions in any magnetized plasma. 

The rest of this paper is organized as follows. In Section~\ref{sec:plasma_housekeeping}, we review the coupled equations of motion for the axion-photon and DP-photon systems in a unified framework and show how to obtain the normal modes of the system given the polarization tensor of the ambient medium. In Section~\ref{sec:emission}, we track how BSM particles are produced from these normal modes in a thermal medium, reviewing the familiar isotropic case before proposing a recipe (Section~\ref{sec:recipe}) for dealing with anisotropic plasmas. This requires knowledge of the plasma self-energy, which we calculate in a magnetized medium in Section~\ref{sec:selfenergy} using the real-time formalism in the long-wavelength limit. In Section~\ref{sec:limits}, we develop approximations that will lend themselves to different ambient environments, for instance with magnetic fields above the critical QED value or with highly degenerate electrons. Concluding remarks follow in Section~\ref{sec:conclusions}.

Readers who are most interested in applications of this formalism may wish to skip straight to the most relevant approximations as appropriate to their system of interest. For ease of utility, in Fig.~\ref{fig:flowchart} we provide a flow chart with hyperlinks to key equations pertaining to various environmental conditions. 

\tikzstyle{blob} = [rectangle, thick, rounded corners, minimum width=3.3cm, minimum height=1cm,text centered, text width=3.3cm, draw=teal, fill=teal!30]
\tikzstyle{arrow} = [ultra thick,->,draw=teal]
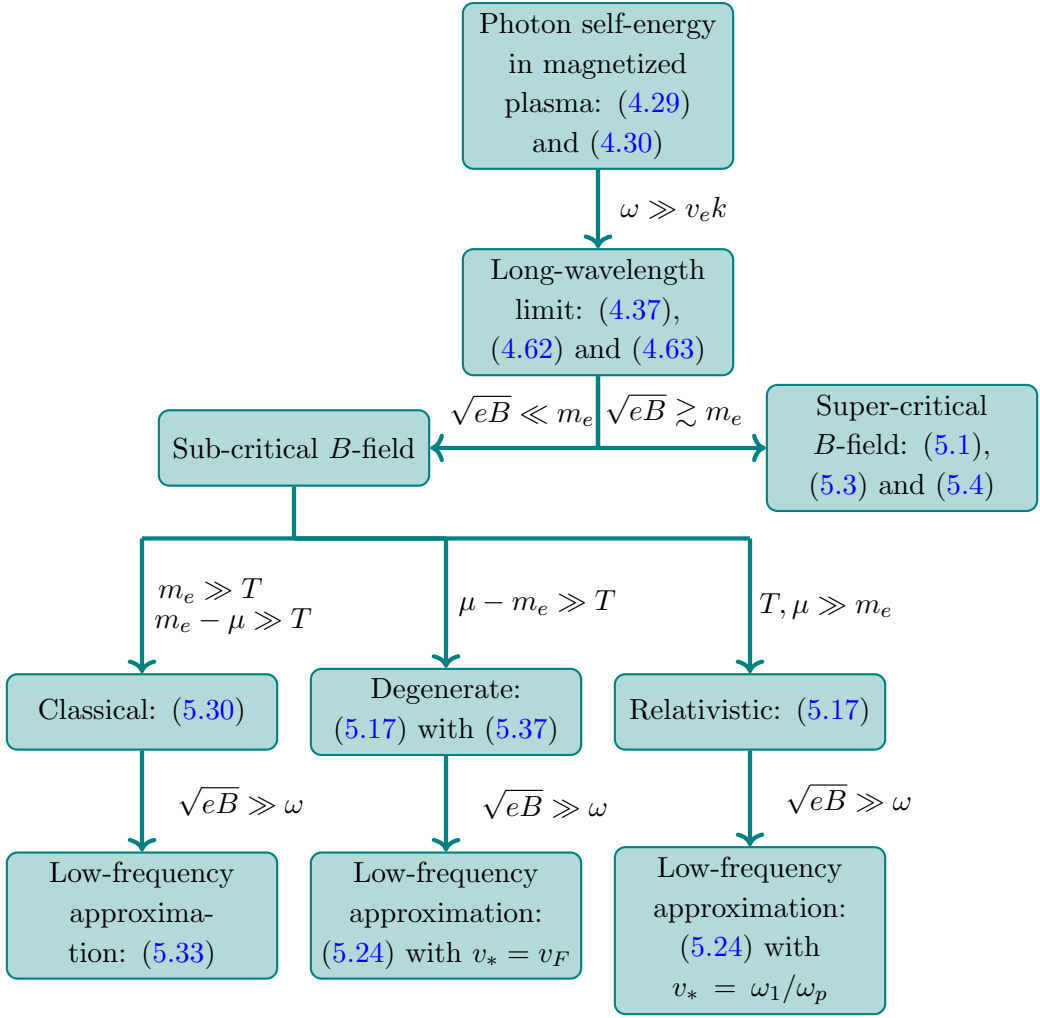
\begin{figure}\label{fig:flowchart}
\begin{center}
\begin{tikzpicture}[node distance=2cm]
\node (lwl) [blob, xshift=3cm, yshift=6cm] {Long-wavelength limit: \labelcref{RePiBLWL,Re-pi-i,piB-n}};
\node (selfenergy) [blob, above of=lwl, yshift=1cm] {Photon self-energy in magnetized plasma: \labelcref{RePiB_R,ImPiB_R}};
\node (supcrit) [blob, below of=lwl, xshift=4cm, yshift=0.2cm] {Super-critical $B$-field: \labelcref{RePiB_sup,pi_para_BV,pi_para_th}};
\node (subcrit) [blob, below of=lwl, xshift=-4cm, yshift=0.2cm] {Sub-critical $B$-field};
\node (classical) [blob, below of=subcrit, xshift=-2cm, yshift=-1.5cm] {Classical: \labelcref{pi-classical}};
\node (degn) [blob, below of=subcrit, xshift=2cm, yshift=-1.5cm] {Degenerate: \labelcref{pi_quasi_iso} with \labelcref{degenerate_approx}};
\node (rel) [blob, below of=subcrit, xshift=6cm, yshift=-1.5cm] {Relativistic: \labelcref{pi_quasi_iso}};
\node (classlowf) [blob, below of=classical, yshift=-0.7cm] {Low-frequency approximation: \labelcref{pi-classical-lowf}};
\node (degnlowf) [blob, below of=degn, yshift=-0.7cm] {Low-frequency approximation: \labelcref{pi_quasi_iso3} with $v_* = v_F$};
\node (rellowf) [blob, below of=rel, yshift=-0.9cm] {Low-frequency approximation: \labelcref{pi_quasi_iso3} with $v_* = \omega_1/\omega_p$};
\draw [arrow] (selfenergy) -- node[ xshift=1 cm] {$\omega \gg v_e k$} (lwl);
\draw [arrow] (lwl) |- node[ xshift=1cm, yshift=0.5cm] {$\sqrt{eB}\gtrsim m_e$} (subcrit);
\draw [arrow] (lwl) |- node[ xshift=-1cm, yshift=0.5cm] {$\sqrt{eB}\ll m_e$} (supcrit);
\draw[ultra thick, draw=teal] (subcrit)--(-1,3);
\draw[ultra thick,draw=teal] (-1,3)--(-3,3);
\draw[ultra thick,draw=teal] (-1,3)--(1,3);
\draw[ultra thick,draw=teal] (-1,3)--(5,3);
\draw [arrow] (-3,3) -- node[ xshift=0.9cm, yshift=0.2cm] {$m_e \gg T$} node[ xshift=1.2cm, yshift=-0.2cm] {$m_e-\mu \gg T$} (classical);
\draw [arrow] (1,3) -- node[ xshift=1.2cm] {$\mu - m_e \gg T$}(degn);
\draw [arrow] (5,3) -- node[ xshift=1cm] {$T,\mu\gg m_e$} (rel);
\draw [arrow] (classical) -- node[ xshift=1.3cm] {$\sqrt{eB}\gg \omega$} (classlowf);
\draw [arrow] (degn) -- node[ xshift=1.3cm] {$\sqrt{eB}\gg \omega$}(degnlowf);
\draw [arrow] (rel) -- node[ xshift=1.3cm] {$\sqrt{eB}\gg \omega$}(rellowf);
\end{tikzpicture}
\end{center}
\caption{A summary of key results as appropriate to different ambient environments.}
\vspace{-0.5cm}
\end{figure}

\section{Plasma Particle Production}
\label{sec:plasma_housekeeping}
\subsection{Coupled equations of motion}
\subsubsection{Photon-Axion System}
In the presence of a static, external magnetic field, the interaction Lagrangian for the photon-axion system can be written as a current term. Assuming that the external $A$-field is much larger than the axion field or the induced $A$-field, the interaction Lagrangian takes the form  $\mathcal{L}_\mathrm{int}=A_\mu J_a^\mu$ with $J_a^\mu = g_{a\gamma}\tilde{F}_{0}^{\mu\nu}\partial_\nu a$, where $\tilde{F}_{0}^{\mu\nu}$ is the dual field strength tensor of the external magnetic field $\vec{B}_0$. The EOM then obtained from the Lagrangian density in \Eq{photon-axion-eom} can be written in a manifestly covariant form as
\begin{equation}
\begin{aligned}
&\partial_{\mu}F^{\mu\nu}=J^{\nu}+g_{a\gamma}\tilde{F}_0^{\mu\nu}\partial_{\nu}a\\
&\left(\partial^{2}+m_{a}^{2}\right)a=-g_{a\gamma}\tilde{F}_0^{\mu\nu}\partial_\nu A_\mu\\
&\partial_{\mu}\tilde{F}^{\mu\nu}=0,
\end{aligned}
\end{equation}
where the inhomogeneous part of Maxwell's equations includes an axion-induced source term while the homogeneous part remains unmodified. The external magnetic field $\vec{B}_0$ can be promoted to a magnetic field four-vector $B^\mu_0 = (0,\vec{B}_0)$ and can be expressed in terms of the dual field tensor as $B^\mu_0=\tilde{F}_0^{\mu\nu}u_\nu$, where $u^\mu$ is the four-velocity of the medium. {If the properties of the background vary slowly enough such that the potential is essentially constant over the de Broglie wavelength -- essentially satisfying the WKB approximation (see Ref.~\cite{McDonald:2023ohd}) -- then we can treat the plasma as {locally homogeneous}.} Then in Fourier space, the modified axion-Maxwell's equations become
\begin{equation}\label{ax-EOM}
\begin{aligned}
&\left(\fourvec{K}^{2}g^{\mu\nu}-\fourvec{K}^{\mu}\fourvec{K}^{\nu}-\Pi^{\mu\nu}_R\right)A_{\nu}-i\fourvec{P} g_{a\gamma}\mathcal{B}^\mu_0(\fourvec{P}) a=0\\
&\left(\fourvec{P}^{2}-m_{a}^{2}\right)a+i \fourvec{K} g_{a\gamma}\mathcal{B}^\mu_0(\fourvec{K}) A_{\mu}=0
\end{aligned}
\end{equation}
where $\fourvec{K}^\mu=(\omega, \vec{k})$ and $\fourvec{P}^\mu=(\omega, \vec{p})$ are the photon and axion four-momentum respectively with $K^\mu = P^\mu + Q_B^\mu$, where $Q^\mu_B=(0,\vec{q}_B)$ is the momentum transfer from the magnetic field. Here we have defined the magnitude of a four-vector $\fourvec{V}^\mu$ as $\fourvec{V} \equiv \sqrt{\fourvec{V}^\mu \fourvec{V}_\mu}$ and where $\mathcal{B}^\mu_0(\fourvec{V}^\mu)\equiv \fourvec{V}^{-1}\tilde{F}_{0}^{\mu\nu}\fourvec{V}_\nu$ is the external magnetic field in the rest frame of the axion or photon. The photon and the axion have the same energy $\omega$ since magnetic fields do no work. Assuming the photon and axion propagation direction to be the same (see \cite{Millar:2021gzs, McDonald:2023ohd, McDonald:2024uuh} for generalizations to 3D) we have $q_B \sim \sqrt{\omega^2-m_\gamma^2}-\sqrt{\omega^2-m_a^2}$ where $m_\gamma$ denotes the in-medium effective mass of the photon. For on-shell resonant production, $m_\gamma \sim m_a $ implies that $K^\mu \sim P^\mu$ in \Eq{ax-EOM}.

\subsubsection{Photon-DP System}
For the photon-DP system we can perform a field redefinition to eliminate the kinetic mixing term, 
\begin{equation}
\boldsymbol{A}^\mu \equiv \begin{pmatrix}{\mathcal{A}}^\mu\\ \mathcal{S}^\mu\end{pmatrix}=
\begin{pmatrix}1 & 0\\ -\chi &1\end{pmatrix}\begin{pmatrix}A^\mu \\ A^{\prime \mu}\end{pmatrix}
+\mathcal{O}({\chi^2}).
\end{equation}
This yields the Lagrangian in the interaction basis for the active and sterile states up to $\mathcal{O}({\chi^2})$,
\begin{align}
 \label{dp_lagrangian}
\mathcal{L} \supset &-\frac{1}{4} \, F_{\mu\nu}^{\mathcal{A}} \, F^{\mu\nu}_{\mathcal{A}} - \frac{1}{4} \, F_{\mu\nu}^{\mathcal{S}} \, F^{\mu\nu}_{\mathcal{S}} + \frac{1}{2} (\boldsymbol{A}^\mu)^T \mathcal{M}^{2}_{\mu \nu} 
\boldsymbol{A}^\nu 
   ~.
\end{align}
where in the interaction basis, the mass-squared matrix is not diagonal,
\begin{equation}
  \mathcal{M}^{2}_{\mu \nu} = \begin{pmatrix} \Pi^{\mu\nu}_R & \chi \, \mAp^2 g^{\mu \nu}\\ \chi \, \mAp^2 g^{\mu \nu}& \mAp^2 g^{\mu \nu}\end{pmatrix} \,.
\end{equation}
In Fourier space (again assuming local homogeneity), in the absence of an external current, the Euler-Lagrange equations of motion of the coupled photon-DP system are \begin{equation}\label{dp-EOM}
\begin{aligned}
&\left(\fourvec{K}^{2}g^{\mu\nu}- \fourvec{K}^\mu \fourvec{K}^\nu -\Pi^{\mu\nu}_R\right)\mathcal{A}_{\nu}+\chi m_{A'}^{2}\mathcal{S}^{\mu}=0\\
&\left(\fourvec{K}^{2}-m_{A'}^{2}\right)\mathcal{S}^{\mu}+\chi m_{A'}^{2}\mathcal{A}^{\mu}=0.
\end{aligned}
\end{equation}

\subsubsection{Unified Treatment}
The coupled dynamics of axion-photon and DP-photon systems exhibit significant parallels. These similarities can be elucidated by expressing both sets of EOM in a unified form:
\begin{equation}
\label{EOM-general}
\left(\begin{array}{cc}
\fourvec{K}^{2}\tilde{g}^{\mu\nu}-\Pi_{R}^{\mu\nu} & \Pi_{AX}\\
\Pi_{XA} & \left(\fourvec{K}^{2}-m_{X}^{2}\right) {\delta}_{XX}-\Pi_{XX}
\end{array}\right)\left(\begin{array}{c}
{A}\\
{X}
\end{array}\right)=0.
\end{equation} 
where $\tilde{g}^{\mu\nu}=g^{\mu\nu}-\fourvec{K}^\mu \fourvec{K}^\nu/\fourvec{K}^2$ and the subscript $A$ represents (active) photons in the thermal plasma, while the subscript $X$ denotes BSM particles weakly interacting with photons. The notation for delineating axions and dark photons is provided in Table \ref{notation-ax-dp}. 
\begin{table}[h]
\centering
{\renewcommand{\arraystretch}{2}
\begin{tabular}{|c|c|c|c|}
\hline 
BSM State&
Degrees of Freedom $\left({A},{X}\right)$&
$\Pi_{AX}$&
${\delta}_{XX}$
\tabularnewline
\hline 
Axions & $\left(A^{\mu},a\right)$ & 
$\Pi_{Aa}^{\mu}= \left(\Pi_{aA}^{\mu}\right)^*=-ig_{a\gamma}\fourvec{K}\mathcal{B}_{0}^{\mu}$ & 
$1$\tabularnewline
\hline 
DPs & $\left(\mathcal{A}^{\mu},\mathcal{S}^{\mu}\right)$ &  
$\Pi_{\mathcal{A}\mathcal{S}}^{\mu\nu}=\left(\Pi_{\mathcal{S}\mathcal{A}}^{\mu\nu}\right)^* =\chi\mAp^{2}g^{\mu\nu}$ & 
$g^{\mu\nu}$\tabularnewline
\hline 
\end{tabular}}
\caption{Terms appearing in \Eq{EOM-general} for coupling axions and DPs to SM photons. }
\label{notation-ax-dp}
\end{table}
The in-medium self-energy $\Pi_{XX}$ contributes at higher order in the coupling, and hence does not appear in the leading-order EOM in \Eq{ax-EOM} and \Eq{dp-EOM}. Nevertheless, for the sake of completeness we still include it in \Eq{EOM-general}. 
The consolidated form of the EOM lends itself to a unified treatment of the independent normal modes in the plasma, which we describe below. 

\subsection{Normal Modes}\label{sec:2.2}
\subsubsection{Projection Tensors and Polarization Vectors}
In a vacuum, the photon self-energy can be represented in a covariant way in terms of $\fourvec{K}^\mu$ and $g^{\mu\nu}$. The four-velocity of an isotropic medium $u^\mu$ poses another independent four-vector. With these three quantities, we can formulate two independent projection tensors $P^{\mu \nu}$ that satisfy the gauge invariance condition of \Eq{transversality} in an \emph{isotropic} plasma. However, the selection of these two projection tensors is not unique. A common choice is to decompose the EOM in terms of modes that are transverse and longitudinal to the direction of propagation, using the projection tensors
\begin{equation}
P_{L}^{\mu\nu}=\epsilon_{L}^\mu \epsilon_{L}^\nu = \frac{\tilde{u}^{\mu}\tilde{u}^{\nu}}{\tilde{u}^{2}}; \quad \quad 
P_{T}^{\mu\nu}=\epsilon_{T1}^\mu \,\epsilon_{T1}^\nu + \epsilon_{T2}^\mu \,\epsilon_{T2}^\nu = g^{\mu\nu}-\frac{\fourvec{K}^{\mu}\fourvec{K}^{\nu}}{\fourvec{K}^{2}}-\frac{\tilde{u}^{\mu}\tilde{u}^{\nu}}{\tilde{u}^{2}}
\end{equation}
where $\epsilon_{I}^\mu$ is the polarization vector for the $I$th mode and $\tilde{u}^\mu = u^\mu - \frac{(\fourvec{K}\cdot u)}{\fourvec{K}^2}\fourvec{K}^\mu$. These projection tensors satisfy the orthogonality properties
\begin{equation}
P_{I}^{\mu\alpha}\left(P_{J}\right)_{\alpha}^{\nu}=\delta_{IJ}P^{\mu\nu}_I\text{ (no sum on $I$)},\quad \quad P_{I}^{\mu\nu}\left(P_{J}\right)_{\mu\nu}=\delta_{IJ}
\end{equation}
where $I$ and $J$ index over each of the modes. In an isotropic medium, the two transverse modes are degenerate and there is no unique transverse polarization vector decomposition. In the rest frame of the medium \textit{i.e} $u^\mu=(1,0,0,0)$, and in a coordinate system where the propagation direction is aligned towards the z-axis, \textit{i.e.} $\fourvec{K}^\mu=(\omega,0,0,\modk)$, one choice of orthogonal set of polarization vectors is:
\begin{equation}\label{TL_polvec}
\epsilon_{T1}^{\mu}=(0,1,0,0),\quad\epsilon_{T2}^{\mu}=(0,0,1,0),\quad
\epsilon_{L}^{\mu}=\frac{1}{\sqrt{\omega^{2}-\modk^{2}}}(\modk,0,0,\omega),
\end{equation} 
which satisfies the orthonormal properties
\begin{equation}\label{polvec-cond}
K^{\mu}\epsilon_{\mu}^{I}=0,\quad\left(\epsilon_{\mu}^{I}\right)^{*}\epsilon_{J}^{\mu}=-\delta_{J}^{I},\quad\sum_{I}\epsilon_{I}^{\mu}\left(\epsilon_{I}^{\nu}\right)^{*}=-\left(g^{\mu\nu}-\frac{K^{\mu}K^{\nu}}{K^{2}}\right).
\end{equation}
Using these polarization vectors, we can project quantities appearing in \Eq{EOM-general} onto the transverse and longitudinal directions. Lorentz scalars are unaffected by the decomposition, vectors are contracted with $-\left(\eps_\mu^{I}\right)^*$, and rank-two tensors are contracted with $-(\eps_\mu^{I})^* \eps_\nu^{J}$. From the rank-two tensors appearing in \Eq{EOM-general} and Table~\ref{notation-ax-dp}, we can see that we either contract the polarization vectors with the metric yielding something proportional to $\delta^{IJ}$ or we contract with $\Po_R$ which can yield a nontrivial mode-coupling structure. Explicitly, the projection of the photon self-energy becomes 
\begin{equation}
\label{mixmatrix}
  \pi^{IJ}_{AA}\equiv -(\eps_\mu^{I})^* \Pi^{\mu\nu}_R \epsilon^{J}_\nu
\end{equation}
and it couples the different modes $\{A^{I}\}$ present in the plasma. Henceforth, we will refer to $\pi^{IJ}_{AA}$ as the \textit{mixing matrix}. We emphasize the use of two distinct sets of indices in \Eq{mixmatrix}. The $\{\mu,\nu\}$ indices correspond to spacetime, whereas the $\{I,J\}$ indices belong to \textit{mode-space} and can assume three values representing the three modes of a propagating vector boson. 

To provide an explicit example of a mixing matrix, we note that in an \emph{isotropic} plasma, the self-energy can be expressed as
\begin{equation}
  \Po_R = \pi_T P^{\mu\nu}_T + \pi_L P^{\mu\nu}_L \label{eq:iso_decomp}
\end{equation}
where $\pi_T(\omega,\kappa)$ and $\pi_L(\omega,\kappa)$ are the transverse and longitudinal Lorentz-invariant form factors. Instead of the longitudinal polarization vector in \Eq{TL_polvec}, another suitable choice would be
\begin{equation}
\label{pol-vecL-2}
\epsilon_{L}^{\mu}(\fourvec{K})=\frac{k}{\omega\sqrt{\fourvec{K}^{2}}}\fourvec{K}^{\mu}+\frac{\sqrt{\fourvec{K}^{2}}}{\omega}\epsilon_{z}^{\mu}
\end{equation}
where the $\eps^\mu_z = (0,0,0,1)$ is the unit four-vector along the z-axis. The first term in \Eq{pol-vecL-2} drops out when contracting with $\Po$ owing to the gauge invariance condition \Eq{transversality}, making it clear that in an isotropic plasma, $\pi_{AA}$ can be shown to be diagonal with the diagonal entries given by the corresponding form factors, \textit{i.e.} $\pi_{AA}=\mathrm{diag}\{\pi_T,\pi_T,\pi_L\}$.

\subsubsection{Diagonalizing the Plasma Mixing Matrix $\pi_{AA}^{IJ}$}
In an \emph{anisotropic} plasma, the decomposition of \Eq{eq:iso_decomp} does not hold and the mixing matrix will not necessarily be diagonal as expressed in the basis of transverse and longitudinal modes. Therefore, it can be more convenient to work in the eigenbasis of the plasma mixing matrix $\pi_{AA}$, which we denote as $\pi^\mathcal{E
}_{AA}$. In this basis, the mixing matrix is diagonal with entries $(\pi^{\mathcal{E}}_{AA})^{I I}$ corresponding to the $I$th eigenvalue (no sum on $I$). Working in this basis allows us to decouple the EOM for the different plasma modes, enabling us to consider the evolution of each mode independently in a way that is completely analogous to finding the normal modes for a classical system. More precisely, our coupled EOM for the $A$-$X$ system in the eigenbasis becomes
\begin{equation}
\label{EOM-lambda}
\left(\begin{array}{cc}
\left(\fourvec{K}^{2} \mathbb{I}-\pi_{AA}^\mathcal{E}\right) & \pi_{AX}^\mathcal{E}\\
\pi_{XA}^\mathcal{E} & \left(\fourvec{K}^{2}-m_{X}^{2}\right){\delta}_{XX}^\mathcal{E}-\pi_{XX}^\mathcal{E}
\end{array}\right)\left(\begin{array}{c}
{A}^\mathcal{E}\\
{X}^\mathcal{E}
\end{array}\right)=0
\end{equation}
where all the quantities above with $\mathcal{E}$ in the superscript are projections in the plasma eigenbasis and where $\mathbb{I}$ is the $3\times 3$ identity matrix and $\pi^\mathcal{E}_{XA} = \left(\pi^\mathcal{E}_{AX}\right)^*$. Note that for axion-photon mixing, the matrix above is $4\times4$ while for DPs the matrix is $6\times6$. 

\subsubsection{Normal Mode Solutions}\label{NMS}
The matrix EOM in \Eq{EOM-lambda} have a propagating solution corresponding to a mix of in-medium SM plasma and BSM states. The solution exists if the determinant of the matrix vanishes, which gives the following two root equations for a given mode
\begin{subequations}
\begin{align}
\label{A-dispersion1}&\left([K^{0}]^I_A\right)^{2}=\modk^{2}+\Pi_A^I \\
\label{X-dispersion1}&\left([K^{0}]^I_X\right)^{2} =(\modk^{2}+m_{X}^{2})+\Pi_X^I ~,
\end{align}  
\end{subequations}
where $I$ indexes over the normal modes of the plasma and over the modes of the $X$-like state projected into the plasma eigenbasis. For the axion case, there is only one degree of freedom so $I=1$ and one can suppress the index in \Eq{X-dispersion1} if desired, while for DPs $I \in \{1, 2, 3\}$. In Eqs.~\eqref{A-dispersion1} and \eqref{X-dispersion1}, we take $[K^{0}]^I_{A,X}=\omega - i \Gamma^I_{A,X}$ to be complex valued with the real part given by the frequency of the $A-X$ system $\omega$ and with the imaginary part given in terms of the damping rate $\Gamma_{A,X}^I$.\footnote{Note that in the discussion in previous Sections, we were implicitly assuming undamped on-shell propagation with a purely real $K^{0}$.} Note that for the case of the DP, ${\delta}^\mathcal{E}_{XX}= \mathbb{I}$ because of the orthogonality properties of the polarization vectors. The self-energies appearing in \Eq{A-dispersion1} and \Eq{X-dispersion1} take the explicit form
\begin{equation}
\label{A-selfenergy1}
\Pi_{A}^{I}=\begin{cases}
\pi_{AA}^{I}, & X=a\\
\pi_{AA}^{I}+\frac{\pi_{A\mathcal{S}}^{I}\pi_{\mathcal{S}A}^{I}}{\pi_{AA}^{I}-m_X^{2}}+\mathcal{O}\left(\chi^{4}\right)\quad \text{(no sum on }I\text{)}, & X = \mathcal{S}
\end{cases}
\end{equation}
\begin{equation}
\label{X-selfenergy}
\Pi_{X}^{I}=\begin{cases}
\pi_{aa}-\sum\limits_{J}
\frac{\pi_{aA}^{J}\pi_{Aa}^{J}}{\pi_{AA}^{J}-m_X^{2}}, & X=a\\
\left(\pi_{\mathcal{S}\mathcal{S}}^I-\frac{\pi_{\mathcal{S}\mathcal{A}}^{I}\pi_{\mathcal{A}\mathcal{S}}^{I}}{\pi_{AA}^{I}-m_X^{2}}\right)
+\mathcal{O}\left(\chi^{4}\right) \quad \text{(no sum on }I\text{)}, & X = \mathcal{S}
\end{cases}
\end{equation} 
where the X self-energies have been decomposed into a one-particle irreducible (1PI) contribution and a \textit{mixing} contribution from the thermal plasma as described in Ref.~\cite{Hardy:2016kme} (see \App{App:thermal} for a derivation of these self-energies from thermal propagators). Note that for the sake of compactness, we have expressed the $I$th diagonal entry of $\pi^\mathcal{E}_{AA}$ as $\pi_{AA}^I$. Also note that in the plasma eigenbasis, $\pi^\mathcal{E}_{\mathcal{A}\mathcal{S}}$ and $\pi^\mathcal{E}_{\mathcal{S}\mathcal{A}}$ are also diagonalized because $\Pi^{\mu \nu}_{\mathcal{A}\mathcal{S}} = \Pi^{\mu \nu}_{\mathcal{S}\mathcal{A}} \sim g^{\mu \nu}$ and because of the orthogonality properties of the polarization vectors. Hence, for compactness, we also express the $I$th diagonal entry of $\pi^\mathcal{E}_{\mathcal{A}\mathcal{S}}$ as $\pi_{\mathcal{A}\mathcal{S}}^I$. %The presence of BSM particles does not affect the photon self-energy and dispersion relations at leading order since the mixing contributions in Eq.~\eqref{A-selfenergy1} are only $\mathcal{O}(\chi^2)$. Moreover, in \Eq{X-selfenergy} we have included the one-particle irreducible part of the DP self-energy, but it is $\mathcal{O}(\chi^4)$ in the absence of a direct coupling between kinetically mixed DPs and electrons. 
Therefore, at leading order in $\chi$, the matrix equation in Eq.~\eqref{X-dispersion1} as expressed in a basis of plasma normal modes also corresponds to a basis of DP normal modes whose dynamics are decoupled from each other. In other words, the same set of polarization vectors diagonalize both the photon and DP self-energies, implying that an $I$-mode DP can only be produced from a photon in the $I$th polarization. 
 
For weak damping $\Gamma \ll \omega$, the real parts of \Eq{A-dispersion1} and \Eq{X-dispersion1} give the dispersion relations for the respective particles
\begin{align}
\label{A-dispersion2}\text{SM plasma dispersion relation for polarization $I$: } \quad &\omega^{2}=\modk^{2}+\real(\Pi_A^I)\\
\label{X-dispersion2} \text{BSM dispersion relation for polarization $I$: } \quad &\omega^{2} =\modk^{2}+m_{X}^{2}+\real(\Pi_X^I)
\end{align}
such that $\real[\Pi^I_{A,X}]$ takes the form of an effective mass-squared, while the imaginary parts give the damping equations
\begin{align}
\label{damping1}
\Gamma_{A,X}^I=-\frac{1}{\omega}\imag\, (\Pi^I_{A,X}).
\end{align}
The real and imaginary parts of the eigenvalues correspond to the Hermitian and anti-Hermitian components of the plasma mixing matrix, respectively. These components are directly linked to the corresponding parts of the self-energy, as discussed in \App{App:dispersive}. 

Since this work primarily focuses on the effect of magnetized media on particle dispersion relations, we exclusively consider the Hermitian part of the self-energy, which can be decomposed as a sum of symmetric and anti-symmetric tensors associated with purely real and purely imaginary form factors, respectively (see \App{App:dispersive} for more details). 
In isotropic plasmas, the self-energy tensor is purely symmetric, rendering the distinction between the real and Hermitian parts of the self-energy unnecessary. However, in anisotropic environments, such as those with magnetic fields, time-reversal symmetry is broken, giving rise to anti-symmetric contributions with purely imaginary form factors. The dispersion relations are still governed by the Hermitian part of the self-energy, which now includes both symmetric and antisymmetric contributions (corresponding to real and imaginary form factors, see \Eq{disp&abs}). In fact, these imaginary form factors are responsible for the well-known Faraday rotation effect in magnetized plasmas \cite{Ganguly:1999ts,DOlivo:2002omk}. We keep the familiar notation used widely in the literature -- $\real\,\Po$ and $\imag\,\Po$, but it is important to note that these terms actually represent the Hermitian and anti-Hermitian parts of the self-energy tensor, respectively.

\section{Production in Plasmas}
\label{sec:emission}
\subsection{Rates from Detailed Balance}
Many searches for BSM particles involve their production in a thermal medium. For a bosonic particle $X$, we can use the principle of detailed balance to express the production rate in terms of the absorption rate, $\Gamma^P = e^{-\omega/T}\Gamma^A$, and the total damping rate in \Eq{damping1} is given by $\Gamma = \Gamma^A - \Gamma^P$. Thus, the production rate for particle $X$ from each $I$th plasma normal mode is
\begin{equation}\label{prodrate_I}
\begin{split}\frac{dn_{X}^{A^{I}\rightarrow X}}{dt}=\int\frac{d^{3}\vec{k}}{(2\pi)^{3}}\Gamma_{A^{I}\rightarrow X}^{P} & =\int\frac{d^{3}\vec{k}}{(2\pi)^{3}}\frac{1}{\left(e^{\omega/T}-1\right)}\times\begin{cases}
-\frac{1}{\omega}\imag\left[\frac{g_{a\gamma}^{2}\fourvec{K}^{2}\left|{\epsilon}_{I}^{\,\mathcal{E}}\cdot\mathcal{B}_{0}\right|^{2}}{\pi_{AA}^{I}-m_a^{2}}\right] & X=a\vspace{0.2cm}\\ 
 -\frac{1}{\omega}\imag\left[\frac{\chi^{2}\mAp^{4}}{\pi_{AA}^{I}-\mAp^{2}}\right] & X=\mathcal{S}^{I}
\end{cases}\vspace{0.5cm}\\
 & =\int\frac{d^{3}\vec{k}}{(2\pi)^{3}}\frac{1}{\left(e^{\omega/T}-1\right)}\times\begin{cases}
\frac{g_{a\gamma}^{2}\fourvec{K}^{2}\left|\epsilon_{I}^{\,\mathcal{E}}\cdot\mathcal{B}_{0}\right|^{2}\Gamma_{A}^{I}}{\left(m_a^{2}-\real[\pi_{AA}^{I}]\right)^{2}+\left(\omega\Gamma_{A}^{I}\right)^{2}} & X=a \vspace{0.2cm}\\
\frac{\chi^{2}\mAp^{4}\Gamma_{A}^{I}}{\left(\mAp^{2}-\real[\pi_{AA}^{I}]\right)^{2}+\left(\omega\Gamma_{A}^{I}\right)^{2}} & X=\mathcal{S}^{I}
\end{cases}
\end{split}
\end{equation}
where $n_X$ is the number density of X-particles in the plasma and the $X$-self-energies have been calculated from \Eq{X-selfenergy}. The total production rate can be written as a sum over the production rate from each plasma normal mode, $\Gamma^P_X = \sum_J \Gamma^P_{A^J\rightarrow X}$. From \Eq{prodrate_I} it is evident that the $X$ particles are resonantly produced from on-shell plasmons when there is a level crossing at $m_X \sim \sqrt{\real[\pi^I_{AA}]}$ for $\real[\pi^I_{AA}]>0$ which has been extensively used for constraining BSM physics. This resonance can be well-approximated as 
\begin{equation}\label{NWA}
  \frac{\Gamma_A^I}{\left(m_X^{2}-\real[\pi^I_{AA}]\right)^2 + \left(\omega \Gamma_A^I\right)^2} \longrightarrow \frac{\pi}{\omega m_X}\delta\left(m_X - \sqrt{\real[\pi^I_{AA}]}\right),
\end{equation}
making it straightforward to derive the resonant production rate from Eq.~\eqref{prodrate_I}. Away from the resonance, the off-shell production rate dominates and has been previously implemented to derive constraints on solar DPs \cite{Redondo:2008aa, Redondo:2013lna}. However, the expression for the photon self-energy used in earlier studies only applies to on-shell excitations, and has to be modified according to Ref.~\cite{Scherer:2024uui} for the off-shell case. In this work, we make no such assumptions on the kinetmatics, so the photon self-energy in magnetized plasma that we compute is also valid off shell.

\subsection{Isotropic Case}
The case of an isotropic plasma has been extensively studied to establish most of the current astrophysical bounds on axions and DPs (a comprehensive list is presented in Ref.~\cite{ciaran}). The axion production rate from longitudinal and transverse (summing the two degenerate $T$-modes) plasmons in an isotropic plasma can be written from Eq.~\eqref{prodrate_I} as 
\begin{align}
&\frac{dn_a^{A^L\rightarrow a}}{dt}=\int \frac{k^2dk}{2\pi^2}\frac{g_{a\gamma}^{2}\fourvec{K}^{2}\left(B_{0}^{L}\right)^{2}}{ \left(e^{\omega/T}-1\right)}\frac{\Gamma_A^L}{\left(m_a^{2}-\real[\pi^{L}]\right)^{2}+\left(\omega\Gamma_A^L\right)^{2}} \\
&\frac{dn_a^{A^T\rightarrow a}}{dt}=\int \frac{k^2dk}{2\pi^2}\frac{g_{a\gamma}^{2}\omega^{2}\left(B_{0}^{T}\right)^{2}}{\left(e^{\omega/T}-1\right)}\frac{\Gamma_A^T}{\left(m_a^{2}-\real[\pi^{T}]\right)^{2}+\left(\omega \Gamma_A^T\right)^{2}}
\end{align}
where $\Gamma_A^{T,L}$ are the damping rates for $T,L$ photons and $B^L_0 = \hat{\fourvec{K}}\cdot \vec{B}_0$ is the longitudinal component (with respect to the propagation direction) of the magnetic field and $B^T_0 = \sqrt{B_0^2 - (B^L_0)^2}$ is the transverse component. 

Similarly, for DPs, the production rate from $L$-plasmons and each of the $T$-plasmons are
\begin{equation}
\frac{dn_\mathcal{S}^{A^{T,L}\rightarrow \mathcal{S}^{T,L}}}{dt}=\int \frac{k^2dk}{2\pi^2}\frac{\chi^{2}\mAp^{4}}{\left(e^{\omega/T}-1\right)}\frac{\Gamma_{T,L}}{\left(\mAp^{2}-\real[\pi^{T,L}]\right)^{2}+\left(\omega\Gamma_{T,L}\right)^{2}} 
\end{equation}
where $\mathcal{S}^{T,L}$ are the corresponding transverse and longitudinal modes of the DPs.
Note that for an isotropic plasma, the form factors $\pi_T$ and $\pi_L$ are Lorentz-invariant and do not depend on the orientation of $\vec{k}$ and hence the integral measure can be written in spherical coordinates as $d^3\vec{k}\rightarrow 4\pi k^2 dk$. 

After being produced, the vast majority of $I$-mode DPs (where $I=T,L$ only if the production environment is isotropic) propagate out of the plasma, since the probability of converting back to a photon is $\mathcal{O}(\chi^4)$. Generally, terrestrial experiments search for these rare conversions of astrophysically produced DPs back into SM photons in the detector. If the detector is made of an isotropic material, then transverse (longitudinal) DPs would be converted exlusively to transverse (longitudinal) photon signals. 

\subsection{Recipe for BSM Emission from Anisotropic Plasmas}
\label{sec:recipe} 
As an extension to the procedure for the case of isotropic plasmas described above, the strategy we propose for computing the production rate of BSM particles in an \emph{anisotropic} plasma can be summarized as follows:
\begin{itemize}
\item compute the mixing matrix $\pi_{AA}^{IJ}$ in the basis of transverse and longitudinal modes, which can be a spatially varying quantity in realistic systems;
\item diagonalize $\pi_{AA}^{IJ}$ to determine the normal modes of the plasma that evolve independently and their in-medium polarization vectors $(\epsilon_I^\mathcal{E})^\mu$, which will also depend on spatial location (this step can be skipped for isotropic plasmas);
\item compute the production rate from each plasma normal mode from \Eq{prodrate_I} (for the DP case, the plasma normal modes only produce the corresponding DP polarization);
\item determine how the BSM particle $X$ propagates through the anisotropic medium, which can be qualitatively different from the isotropic case: as $X$ moves spatially, the local plasma normal modes may evolve (both in their dispersion relations and in the direction of their polarization vectors), which may lead to additional resonances with nontrivial interference~\cite{Brahma:2023zcw};
\item determine how detection is impacted by the production of $X$ from a set of plasma normal modes that may be different from those in the experiment (e.g. DP normal modes produced in an anisotropic plasma will mix with a superposition of $L$ and $T$ modes in an isotropic detector material). The probability for a $I$ mode dark photon to convert to a $T$ or $L$ photon in an isotropic detector material is given by
\begin{equation}
P_{\mathcal{S}^{I}\rightarrow A^{T,L}}^{D}=\left|\Lambda_{T,L}^{I}\right|^{2}P_{\mathcal{S}^{T,L}\rightarrow A^{T,L}}^{D}=\left|\Lambda_{T,L}^{I}\right|^{2}\frac{\chi^{2}\mAp^{4}}{\left(\fourvec{K}^{2}-\real\left[\pi_{D}^{T,L}\right]\right)^{2}+\left(\omega\Gamma_{D}^{T,L}\right)^{2}}
\end{equation}
where $\Lambda$ is the diagonalizing matrix used to transform from the basis of transverse and longitudinal modes to the eigenbasis, $\eps^I_\mathcal{E}= \eps^i \Lambda^I_i$ with $i$ running over the $T,L$ modes, and $\real\left[\pi_D^{T,L}\right]$ and $\Gamma^{T,L}_D$ are the effective photon mass and damping rate respectively of the detector material.

\end{itemize}
This recipe requires detailed knowledge of the photon self-energy $\Pi_R^{\mu \nu}$ in an anisotropic plasma. In the next Section, we present a derivation of the self-energy tensor at $\mathcal{O}(\alpha)$ under the assumption that the plasma anisotropy is generated by a magnetic field. This self-energy can then be computed numerically if desired. We additionally show that further analytic simplifications can be made in the long-wavelength limit. We finally discuss the limiting behaviour in different plasma environments in Section~\ref{sec:limits} (readers who are more interested in applications of this formalism can go directly to this Section).

\section{Self-energy in the Real-Time Formalism}
\label{sec:selfenergy}
\subsection{Brief Review of the Real-Time Formalism}
The real-time formalism (RTF) provides a powerful framework for studying the dynamics of quantum fields at finite temperatures and densities. One of the built-in advantages of the RTF is its ability to describe the time evolution of thermal systems in a natural way, without the need for analytical continuation from imaginary time as in the imaginary time formalism (ITF). This makes the RTF particularly well-suited for studying non-equilibrium processes such as thermalization, phase transitions, dissipative processes, and more \cite{Das:1997gg, Bellac:2011kqa, Kapusta:2006pm, mallik2016hadrons}. Furthermore, in the ITF, summing over frequencies can become challenging, especially for multi-loop calculations. In contrast, the momentum integrals in the RTF are more akin to the familiar vacuum case, where established techniques are available.
For computing quantities like the response functions of a system in thermal equilibrium, either formalism is applicable. However, to maintain generality and address the possibility of incorporating non-thermal processes in the future, we utilize the RTF throughout our calculations.

\begin{figure}[t]
\centering
\includegraphics[width=0.5\textwidth]{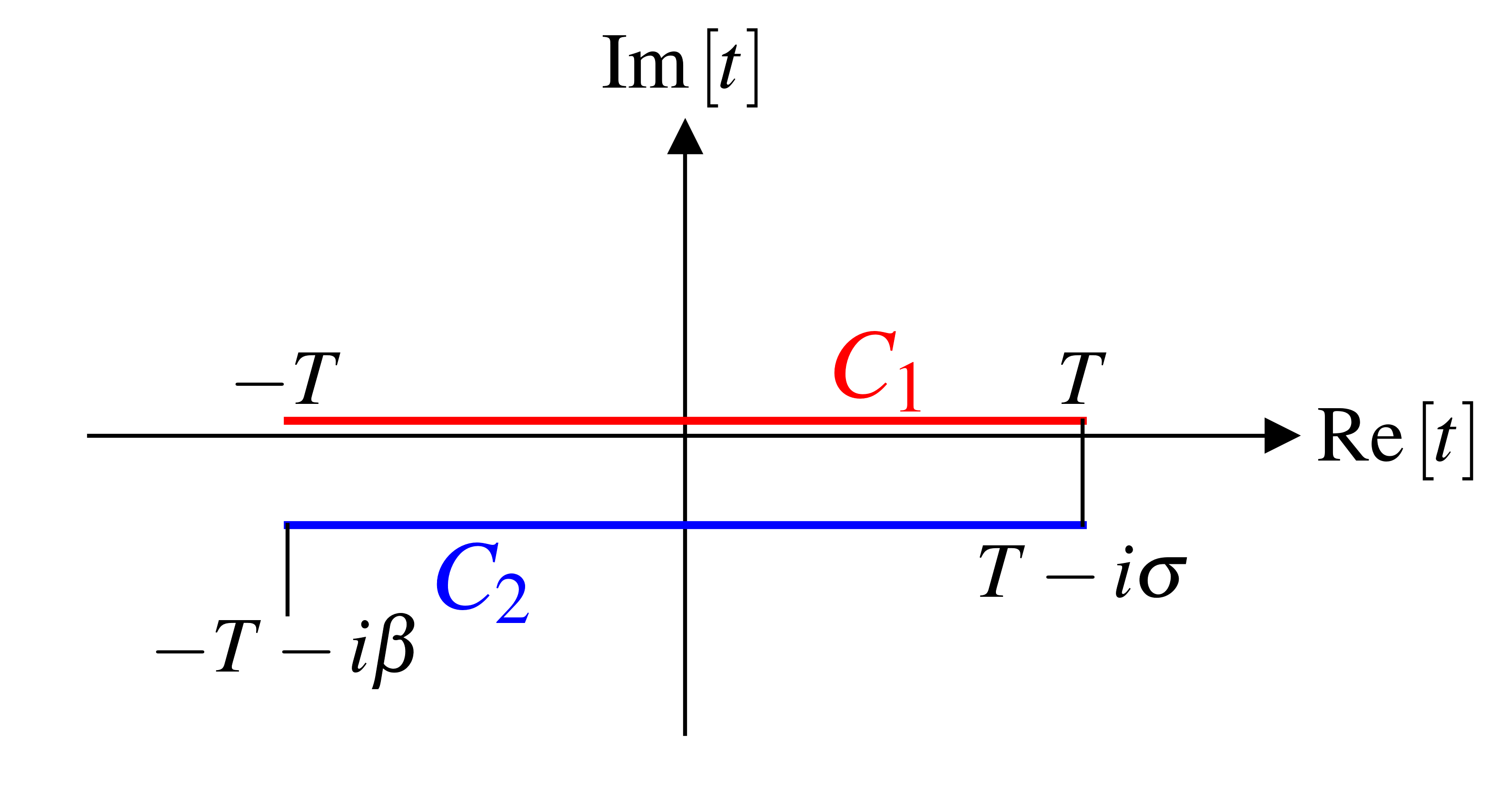}
\caption{Real-time contour. \label{fig:1}}
\end{figure}

A widely used formulation of RTF is the closed-time path (CTP) formalism, where the thermal propagator for a Dirac spinor field $\psi^\alpha(x)$ is given by 
\begin{equation}
iG^{\alpha\beta}_{(ab)}(x^\mu,y^\nu)=\left\langle T_{C}\,\psi^\alpha(x^\mu_a)\bar{\psi}^\beta(y^\nu_b)\right\rangle 
\end{equation}
with the time-ordering $T_C$ taken along the complex time path shown in \Fig{fig:1} and where $\bar{\psi}^\alpha(x)$ is the adjoint field. Here, $\alpha$ and $\beta$ are the spinor indices of the field, while the subscripts $a,b=1,2$ (often called the \textit{thermal indices}) denote the time branch of the complex contour $C$ on which the temporal component of $x^\mu_a$ or $y^\nu_b$ lies. For example, for $G_{(12)}(x,y)$, $x^0$ lies on $C_1$ and $y^0$ lies on $C_2$. Four non-trivial CTP propagator structures are possible. In momentum space, the free-field propagators with four-momentum transfer $Q^\mu$ in a thermal medium with four-velocity $u^\mu$ can be represented as~\cite{Das:1997gg, Bellac:2011kqa}
\begin{subequations}\label{free-med-prop-noB}
\begin{align}
\label{G11}i\left(G_{(11)}^{0}\right)^{\alpha\beta}&=\left[i\Delta_F(\fourvec{Q})+\eta N(\fourvec{Q}\cdot u)2\pi\delta\left(\fourvec{Q}^{2}-m^{2}\right)\right]d^{\alpha\beta}~~~~\\
i\left(G_{(22)}^{0}\right)^{\alpha\beta}&=\left[i\left(G_{(11)}^{0}\right)^{\alpha\beta}\right]^{*}\\
i\left(G_{(12)}^{0}\right)^{\alpha\beta}&=\bigg[\eta e^{\sigma \fourvec{K}^{0}}\Big(N(\fourvec{Q}\cdot u) +\eta\Theta(-\fourvec{Q} \cdot u)\Big) 2\pi\delta\left(\fourvec{Q}^{2}-m^{2}\right)\bigg]d^{\alpha\beta}\\
i\left(G_{(21)}^{0}\right)^{\alpha\beta}&=\left[\eta e^{-\beta\mu}e^{\left(\beta-2\sigma\right)\fourvec{Q}^{0}}\right]i\left(G_{(12)}^{0}\right)^{\alpha\beta}
\end{align}
\end{subequations}
where the superscript $0$ denotes the free-field case, $\eta = \pm 1$ for bosons and fermions, and $\Delta_F(\fourvec{Q})$ is the scalar vacuum Feynman propagator 
\begin{equation}
  \Delta_F(\fourvec{Q})=\frac{1}{\fourvec{Q}^2 - m^2 +i\eps}
\end{equation}
with the covariant structure contained in the Klein-Gordon divisor $d^{\alpha\beta}$~\cite{Landsman:1986wp, Landsman:1986uw}, which is $(\fourvec{Q}^\mu \gamma_\mu^{\alpha\beta}+m \delta^{\alpha\beta})$ for Dirac fermions. Finally, the matter and anti-matter distribution functions are combined into the thermal function $N$ given by
\begin{equation}
  N(x) = \Theta(x) f_+(x) + \Theta(-x) f_-(-x),
\end{equation}
with the matter and antimatter distribution functions given by 
\begin{equation}
\label{dist-func}
  f_\pm(x) = \frac{1}{e^{\beta(x\mp \mu)}-\eta}
\end{equation}
with inverse temperature $\beta$ and chemical potential $\mu$. 
As evident from \Eq{G11}, the propagators in the RTF manifestly separate into a zero-temperature vacuum part $\Delta_F$, which represents the exchange of virtual particles, and a temperature-dependent part, which represents an on-shell contribution giving rise to dissipative effects in a plasma. 

The four propagators can be assembled to a single $2\times 2$ matrix,\footnote{The two dimensional matrix structure of the RTF propagators is related to the doubling of degrees of freedom at finite temperature, where the entire system is actually composed of two parts: our quantum system of interest and the associated heat bath~\cite{Das:1997gg}.} which can be expressed in a diagonal form as
\begin{equation}
\begin{split}
i\mathcal{G}_{0}^{\alpha\beta}&\equiv i\left(\begin{array}{cc}
\left(G_{(11)}^{0}\right)^{\alpha\beta} & \left(G_{(12)}^{0}\right)^{\alpha\beta}\\
\left(G_{(21)}^{0}\right)^{\alpha\beta} & \left(G_{(22)}^{0}\right)^{\alpha\beta}
\end{array}\right) =\mathcal{U}_{\therm}\left(\begin{array}{cc}
d^{\alpha\beta}i\Delta_F & 0\\
0 & d^{\alpha\beta}\left(i\Delta_F\right)^{*}
\end{array}\right)\mathcal{U}_{\therm}
\end{split}
\end{equation}
where $\mathcal{U}_\therm$ is the thermal Bogoliubov matrix (in the $(1,2)$-basis) that transforms the vacuum propagator to a thermal propagator and is given by~\cite{Landsman:1986uw,Lundberg:2020mwu}
\begin{equation}
\mathcal{U}_{\mathrm{th}}(\fourvec{Q})=\left(\begin{array}{cc}
\cos(h)\theta_{\mathrm{th}}^\fourvec{Q} & \eta e^{\beta\mu/2}e^{-(\beta-2\sigma)\fourvec{Q}^{0}/2}\sin(h)\theta_{\mathrm{th}}^\fourvec{Q}\\
e^{-\beta\mu/2}e^{(\beta-2\sigma)\fourvec{Q}^{0}/2}\sin(h)\theta_{\mathrm{th}}^\fourvec{Q} & \cos(h)\theta_{\mathrm{th}}^\fourvec{Q}
\end{array}\right)
\end{equation}
where the trigonometric (hyperbolic) functions are for fermions (bosons) with the thermal angle $\theta_\therm^\fourvec{Q}$ for four momentum $\fourvec{Q}$ defined as
\begin{equation}
\begin{array}{cc}
\sin(h)\theta^{\fourvec{Q}}_{\therm}=\sqrt{N(\fourvec{Q}\cdot u)}\\\\
\cos(h)\theta^{\fourvec{Q}}_{\therm}=\left[\Theta(\fourvec{Q}\cdot u)+\eta\Theta(-\fourvec{Q}\cdot u)\right]\sqrt{1+\eta N(\fourvec{Q}\cdot u)}.
\end{array}
\end{equation}
The full-propagator $\mathcal{G}^{\alpha\beta}_{(ab)}$ in an interacting theory is given by the Schwinger-Dyson equation
\begin{equation}
i\left(\mathcal{G}^{-1}\right)_{(ab)}^{\alpha\beta}=i\left(\mathcal{G}_{0}^{-1}\right)_{(ab)}^{\alpha\beta}-i\Pi_{(ab)}^{\alpha\beta},
\end{equation}
where the inverse of $\mathcal{G}$ is in the covariant indices $(\alpha \beta)$ (as opposed to the thermal indices $(a, b)$, \textit{i.e.} $\mathcal{G}^{\alpha\rho}\left(\mathcal{G}^{-1}\right)_{\rho\beta}=g_{\beta}^{\alpha}$), and the real-time self-energy matrix components $\Pi^{\alpha\beta}_{(ab)}$ arise from the sum of all the 1PI two-point diagrams with one external vertex of type $\alpha,a$ and the other of type $\beta,b$. The same thermal matrix $\mathcal{U}_\therm$ can also be used to diagonalize the self-energy matrix (see Refs.~\cite{Bellac:2011kqa, Landsman:1986uw, Lundberg:2020mwu} for a detailed discussion), 
\begin{equation}
\label{realtime-self}
i\Pi^{\alpha\beta}=\left(\begin{array}{cc}
i{\Pi}^{\alpha\beta}_{(11)} & i{\Pi}^{\alpha\beta}_{(12)}\\
i{\Pi}^{\alpha\beta}_{(21)} & i{\Pi}^{\alpha\beta}_{(22)}
\end{array}\right)=\mathcal{U}_{\therm}^{-1}\left(\begin{array}{cc}
i\bar{\Pi}^{\alpha\beta} & 0\\
0 & \left(i\bar{\Pi}^{\alpha\beta}\right)^{*}
\end{array}\right)\mathcal{U}_{\therm}^{-1},
\end{equation}
and thus the self energy matrix has only one independent component, $\bar{\Pi}^{\alpha\beta}$. The real (Hermitian) and imaginary (anti-Hermitian) parts of this real-time self energy are related to the retarded self-energy by~\cite{Bellac:2011kqa}
\begin{equation}
\label{diagtoretard}
\mathrm{Re}\bar{\Pi}^{\alpha\beta}=\mathrm{Re}\Pi_{R}^{\alpha\beta},\quad\mathrm{Im}\bar{\Pi}^{\alpha\beta}=\sgn(\fourvec{Q}\cdot u)\mathrm{Im}\Pi_{R}^{\alpha\beta}
\end{equation}
where $\sgn(\ldots)$ is the sign function. Using \Eq{realtime-self} and \Eq{diagtoretard}, we can thus write the retarded self-energy in terms of the $(11)$-part of the real-time self-energy 
\begin{align}
&\mathrm{Re}\Pi_{R}^{\alpha\beta}\left(\fourvec{Q}\right)=\mathrm{Re}\Pi_{(11)}^{\alpha\beta}\left(\fourvec{Q}\right), \\ &\mathrm{Im}\Pi_{R}^{\alpha\beta}\left(\fourvec{Q}\right)=\tanh\left(\frac{\beta (\fourvec{Q}\cdot u)}{2}\right)\mathrm{Im}\Pi_{(11)}^{\alpha\beta}\left(\fourvec{Q}\right).
\end{align}
Operationally, in the calculations presented in \Sec{unmagnetized} and \Sec{magnetized}, we first evaluate the $(11)$-part of the self-energy in RTF and then use the above relations to evaluate the retarded self energy. 

\subsection{RTF Photon Self-Energy in an Unmagnetized Plasma}\label{unmagnetized}
Equipped with the RTF, in this Section we explicitly calculate the photon self-energy in an isotropic \textit{electron-positron} ($e^-e^+$) plasma\footnote{Since ions are significantly heavier than electrons, their response to propagating electromagnetic fields is reduced and hence their contribution to the self-energy is negligible. Consequently, the electron-ion plasma can be viewed as a special case of the $e^-e^+$ plasma where the positron contributions are disregarded.} with no external magnetic fields. We present the simpler unmagnetized case before progressing to the analogous (but more complicated) magnetized case. The results in this Section are perhaps more easily obtained using the ITF~\cite{Das:1997gg,Bellac:2011kqa}. However, as discussed above, we use the RTF due to its wider applicability and similarity with vacuum calculations including renormalization procedures (see \App{App:vac}). Note that while the calculations presented below are specifically tailored for photons within an electron background, they can be straightforwardly adapted to the self-energy of any gauge boson with an arbitrary coupling to either a scalar or fermion.
\begin{figure}[t]
\centering
\includegraphics[width=.5\textwidth]{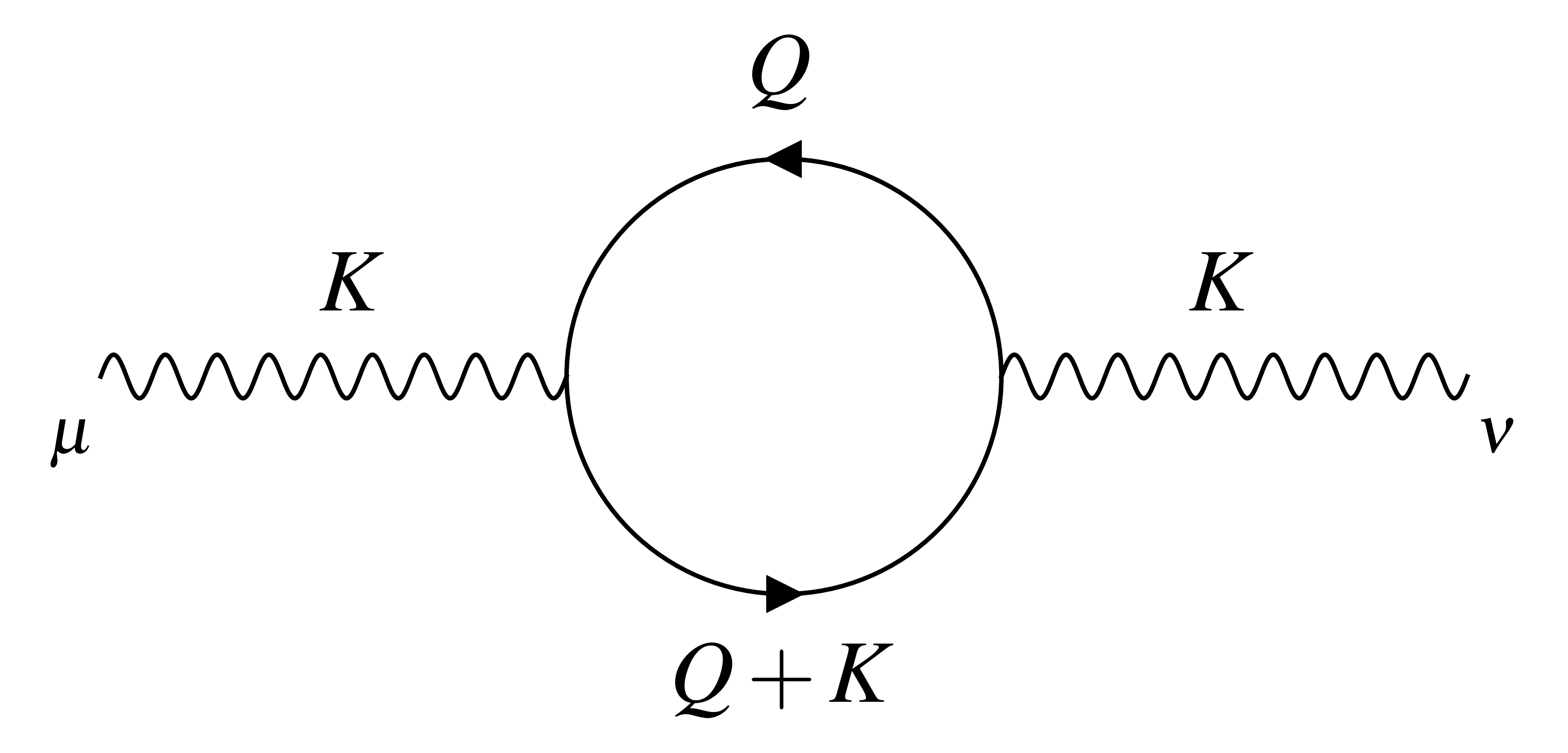}
\caption{Photon self-energy diagram.
\label{fig:loop}}
\end{figure}

The $(11)$-part of the one-loop photon self-energy in the presence of an unmagnetized $e^-e^+$ plasma, depicted in \Fig{fig:loop}, can be expressed as \begin{equation}\label{self1-noB}
\Pi^{\mu\nu}_{(11)}(\fourvec{K})=-ie^{2}\int\frac{d^{4}\fourvec{Q}}{(2\pi)^{4}}\mathrm{Tr}\left[\gamma^{\mu}iS^0_{(11)}(\fourvec{K}+\fourvec{Q})\gamma^{\nu}iS^0_{(11)}(\fourvec{Q})\right]
\end{equation}
where $S^0_{(11)}(\fourvec{Q})$ is the $(11)$-part of the free electron propagator. Using \Eq{free-med-prop-noB}, $S^0_{(11)}(\fourvec{Q})$ can be written as
\begin{equation}
  iS^0_{(11)}(\fourvec{Q}) = iS_{V}^0(\fourvec{Q}) + iS_{\therm}^0(\fourvec{Q})
\end{equation}
where $iS^0_V(\fourvec{Q})$ is the vacuum electron propagator 
\begin{equation}
\label{S_V}
iS^0_V(\fourvec{Q})=i\frac{\slashed{\fourvec{Q}}+m}{\fourvec{Q}^2 - m^2 +i\eps}
\end{equation}
and the medium contribution is
\begin{equation}
\label{thermal-e-prop}
  i S_{\therm}^0(\fourvec{Q}) = -N(\fourvec{Q}\cdot u)[iS_{V}^0(\fourvec{Q}) - i\bar{S}_{V}^0(\fourvec{Q})]
\end{equation}
where $\bar{S}_{V}^{0}\equiv\gamma^{0}S_{V}^{0\dagger}\gamma^{0}$. The integrand in \Eq{self1-noB} has three qualitatively distinct terms: (i) the vacuum term $\propto S^V S^V$, which ultimately yields a pure-vacuum self-energy $\big(\Pi^{\mu\nu}_{(11)}\big)^V$ that is worked out in \App{App:vac}, (ii) the {cross terms} $\propto S^V S^\therm$, which we evaluate below, and (iii) the purely thermal term $\propto S^\therm S^\therm$, which we also evaluate below.

It is evident from \Eq{thermal-e-prop} that the thermal part of the propagator ($S^\therm$) is real. Meanwhile, the vacuum part ($S^V$) can be decomposed into a real and imaginary parts\footnote{We emphasize once again that the real and imaginary parts of the self-energy tensors correspond to the Hermitian and anti-Hermitian components, respectively.} using the Plemelj formula~\cite{Bellac:2011kqa, melrose2008quantum} 
\begin{equation}\label{vac_prop}
iS^0_{V}(\fourvec{Q})=\left[i\mathcal{P}\left(\frac{1}{\fourvec{Q}^{2}-m^{2}}\right)+\pi\delta\left(\fourvec{Q}^{2}-m^{2}\right)\right]\left(\slashed{\fourvec{Q}}+m\right), 
\end{equation}
where $\mathcal{P}(\ldots)$ is the principal-value function and $m$ is the electron mass. The above decomposition can be used to separate the real and imaginary parts of the trace in the integrand in \Eq{self1-noB}. In the rest-frame of the medium, \textit{i.e.} $u^\mu=(1,0,0,0)$, the different terms give
\begin{align}\label{RePi-11}
\mathrm{Re}\Pi_{(11)}^{\mu\nu}=\mathrm{Re}\left(\hat{\Pi}_{(11)}^{\mu\nu}\right)^{V}-e^{2}\int\frac{d^{3}\vec{q}}{(2\pi)^{3}}\int dq^{0}&\bigg\{ N(q^{0})\mathcal{P}\left[\frac{T^{\mu\nu}(\fourvec{Q},\fourvec{K})}{\fourvec{P}^{2}-m^{2}}\right]\delta\left(\fourvec{Q}^{2}-m^{2}\right)\\&+N(p^{0})\mathcal{P}\left[\frac{T^{\mu\nu}(\fourvec{Q},\fourvec{K})}{\fourvec{Q}^{2}-m^{2}}\right]\delta\left(\fourvec{P}^{2}-m^{2}\right)\bigg\} \nonumber
\end{align}
\begin{align}\label{ImPi-11}
\mathrm{Im}\Pi_{(11)}^{\mu\nu}=\mathrm{Im}\left(\hat{\Pi}_{(11)}^{\mu\nu}\right)^{V}-e^{2}\pi\int\frac{d^{3}\vec{q}}{(2\pi)^{3}}\int dq^{0}&\left\{ 2N(q^{0})N(p^{0})-N(q^{0})-N(p^{0})\right\}\\ &\times T^{\mu\nu}\delta\left(\fourvec{Q}^{2}-m^{2}\right)\delta\left(\fourvec{P}^{2}-m^{2}\right),\nonumber
\end{align}
where $P^\mu=Q^\mu+K^\mu$ and for any four-vector we have $q^0=Q\cdot u$ with the components of the four-vector given by $Q^\mu=(q^0,\vec{q})$. The $\hat\Pi$ denotes the renormalized self-energy, and the fermionic trace $T^{\mu\nu}$ is defined by
\begin{equation}\label{Tmunu}
  T^{\mu\nu}(\fourvec{Q},\fourvec{K})=\mathrm{Tr}\left[\gamma^\mu (\slashed{\fourvec{P}}+m)\gamma^\nu (\slashed{\fourvec{Q}}+m)\right].
\end{equation}
By exploiting the properties of the $\delta$-function, each $\delta$-function with a quadratic argument in the expressions above can be split into two $\delta$-functions with linear arguments. The integration over $\fourvec{q}^0$ can then be carried out using these $\delta$-functions. The selection of the appropriate phase factors ($f_e$ or $f_{\bar{e}}$) is determined by the $\Theta$-functions contained in factors involving $N(\fourvec{q}^0)$ and $N(\fourvec{p}^0)$. Using \Eq{RePi-11}, the real part of the retarded self-energy can be expressed as
\begin{align}
\mathrm{Re}\,\Pi_{R}^{\mu\nu} &= \mathrm{Re}\left(\hat{\Pi}_{(11)}^{V}\right)^{\mu\nu} \nonumber \\
&\quad - e^{2} \int \frac{d^{3}q}{(2\pi)^{3}}\, \mathcal{P} \left\{
\frac{1}{2E_{q}} \left[ f_{e}\left(E_{q}\right) 
\left(\frac{T^{\mu\nu}}{P^{2} - m^{2}}\right)_{q^{0} = E_{q}} + 
f_{\bar{e}}\left(E_{q}\right) 
\left(\frac{T^{\mu\nu}}{P^{2} - m^{2}}\right)_{q^{0} = -E_{q}} \right] \right. \nonumber \\
&\quad \left. + \frac{1}{2E_{p}} \left[
f_{e}\left(E_{p}\right) 
\left(\frac{T^{\mu\nu}}{Q^{2} - m^{2}}\right)_{q^{0} + k^{0} = E_{p}} + 
f_{\bar{e}}\left(E_{p}\right) 
\left(\frac{T^{\mu\nu}}{Q^{2} - m^{2}}\right)_{q^{0} + k^{0} = -E_{p}} \right] 
\right\}
\end{align}
where $E_q=\sqrt{|\vec{q}|^2 + m^2}$ is the electron energy. Meanwhile, using the following identities for a thermal distribution,
\begin{align}
&\left.\left(1-f_{\bar{e}}\left(E_{\fourvec{q}}\right)-f_{e}\left(E_{\fourvec{p}}\right)\right)\frac{1}{2}\left(\coth\left(\frac{\beta \fourvec{k}^{0}}{2}\right)-1\right)=f_{\bar{e}}\left(E_{\fourvec{q}}\right)f_{e}\left(E_{\fourvec{p}}\right)\right|_{\fourvec{k}^{0}=E_{\fourvec{q}}+E_{\fourvec{p}}}\\
&\left.f_{\bar{e}}\left(E_{\fourvec{q}}\right)f_{e}\left(-E_{\fourvec{p}}\right)=e^{-\beta \fourvec{k}^{0}}f_{\bar{e}}\left(E_{\fourvec{p}}\right)f_{e}\left(-E_{\fourvec{q}}\right)=2\frac{\left(f_{\bar{e}}\left(E_{\fourvec{q}}\right)-f_{\bar{e}}\left(E_{\fourvec{p}}\right)\right)}{1-\coth\left(\frac{\beta \fourvec{k}^{0}}{2}\right)}\right|_{\fourvec{k}^{0}=E_{\fourvec{q}}-E_{\fourvec{p}}},
\end{align}
we can express the imaginary part of the retarded self-energy as
\begin{align}
\label{ImPi2}
\mathrm{Im}\,\Pi_{R}^{\mu\nu} &= -\pi e^{2} \tanh\left(\frac{\beta k^{0}}{2}\right) 
\int \frac{d^{3}q}{(2\pi)^{3}} \frac{1}{4E_{p} E_{q}} \nonumber \\
&\quad \times \Bigg\{ 
\left[ 1 - f_{\bar{e}}\left(E_{q}\right) - f_{e}\left(E_{p}\right) \right] 
T^{\mu\nu} \bigg|_{q^{0} = -E_{q}} \delta\left(k^{0} - E_{q} - E_{p}\right) \nonumber \\
&\quad + \left[ 1 - f_{e}\left(E_{q}\right) - f_{\bar{e}}\left(E_{p}\right) \right] 
T^{\mu\nu} \bigg|_{q^{0} = E_{q}} \delta\left(k^{0} + E_{q} + E_{p}\right) \nonumber \\
&\quad + \left[ f_{\bar{e}}\left(E_{q}\right) - f_{\bar{e}}\left(E_{p}\right) \right] 
T^{\mu\nu} \bigg|_{q^{0} = -E_{q}} \delta\left(k^{0} - E_{q} + E_{p}\right) \nonumber \\
&\quad + \left[ f_{e}\left(E_{q}\right) - f_{e}\left(E_{p}\right) \right] 
T^{\mu\nu} \bigg|_{q^{0} = E_{q}} \delta\left(k^{0} + E_{q} - E_{p}\right) 
\Bigg\}.
\end{align}
Here, the first two terms in square brackets represent pair annihilation and production processes while the last two terms represent Landau damping~\cite{landau1946vibrations, Das:1997gg, Bellac:2011kqa}. Moreover, the terms that lack any phase space factors of $f_e$ or $f_{\bar{e}}$ arise from the vacuum term (see Sec 5.3 in \cite{mallik2016hadrons}). The expression in \Eq{ImPi2} matches exactly with what is obtained from the {imaginary-time formalism} (ITF). 

\subsection{Photon self-energy in a magnetized plasma}
\label{magnetized}
Extending the calculation presented above to account for an external magnetic field is relatively straightforward in the RTF.\footnote{The ITF calculation of the retarded photon self-energy in a magnetized plasma has been performed explicitly for the absorptive part in Refs.~\cite{Wang:2021eud,Wang:2021ebh}.} We assume that the magnetic field is constant and uniform, which greatly simplifies the formalism and is a useful approximation in astrophysical systems where the magnetic fields vary on length scales much larger than the photon wavelength and on time scales that are much longer than the photon frequency. Beyond the scope of previous works dealing with photons propagating in magnetized plasmas~\cite{Ganguly:1999ts, DOlivo:2002omk, Hattori:2022uzp}, the inclusion of such magnetic fields has also been extensively studied in the context of quark-gluon plasmas (QGP)~\cite{Bandyopadhyay:2016fyd, Hattori:2017xoo, Karmakar:2018aig, Wang:2021eud} and $\rho$-mesons (see \cite{Ghosh:2019fet} and references therein) at heavy-ion colliders. 

Given the similarity with the earlier \Sec{unmagnetized}, here we briefly outline only the key features of the calculation for the case of a magnetized plasma. The main modification enters into the free vacuum propagator $S^0_V$, which can be adapted to account for the effects of magnetic fields. {One key effect is that when magnetic fields exceed their critical value ($B_c\equiv m_e^2/e \sim 4.4 \times 10^{13} \mathrm{G}$), the usual perturbative expansions of QED break down. Each insertion of external magnetic field into the fermionic propagator enhances the diagram by a factor of $\mathcal{O}(B/B_c)$~\cite{Hattori:2012je}. Hence, in the \textit{super-critical} field limit ($B \gtrsim B_c$), one must sum up diagrams at all orders, resulting in non-linear dependencies of the observables to the external field (the regime of non-linear QED~\cite{Heisenberg:1936nmg, Adler:1971wn}). Another qualitatively new effect is that in the presence of external magnetic fields, the energy levels of charged particles are quantized into degenerate Landau Levels (LLs)~\cite{landau1930diamagnetismus}. Even in a vacuum, magnetic fields induce Landau quantization in quantum fluctuations, leading to a variety of physical phenomena (e.g. vacuum birefringence~\cite{Heisenberg:1936nmg}, the quantum Hall effect~\cite{girvin1999quantum}, de Haas van Alphen oscillations~\cite{de1930dependence}, and more). All of these effects involving magnetic fields can be accounted for by using a modified free vacuum propagator, $S^0_{BV}$, which can be expressed using the \textit{Landau representation}\footnote{The well-known Schwinger proper time representation~\cite{PhysRev.82.664} of the electron propagator in an external uniform magnetic field can be transformed to the Landau representation as detailed in Ref.~\cite{kuznetsov2013electroweak} and the references therein.} as~\cite{Chodos:1990vv,Chyi:1999fc,Gusynin:1995nb, Miransky:2015ava}
\begin{equation}\label{S_BV}
iS^0_{BV}\left(\fourvec{Q}\right)=i\sum_{n=0}^{\infty}\frac{\left(-1\right)^{n}e^{-\alpha_{q}}\mathcal{D}_{n}\left(\fourvec{Q}\right)}{q_{\parallel}^{2}-M_{n}^{2}+i\epsilon}
\end{equation}
where $n$ indexes the Landau levels with magnetic effective mass $M_n=\sqrt{m^2+2neB}$. The four-vector $Q^\mu$ has been decomposed into a longitudinal component $q^{\mu}_\parallel = g^{\mu\nu}_\parallel Q_\nu$ and a transverse component $q^{\mu}_\perp = g^{\mu\nu}_\perp Q_\nu$, \textit{with respect to the magnetic field}. Here we have used $g_\parallel^{\mu\nu}\equiv u^\mu u^\nu - b^\mu b^\nu$ (where $b^\mu$ is the unit four-vector along the direction of external magnetic field four-vector, \textit{i.e.} $b^\mu\equiv B^\mu_0/\sqrt{B_0^2}$) and $g_\perp^{\mu\nu}\equiv g^{\mu\nu}-g^{\mu\nu}_\parallel$ to project along the longitudinal and transverse direction respectively. More explicitly, the longitudinal component can be written as $q_\parallel^\mu = q^0 u^\mu + q_{\parallel,b} b^\mu$ where $q_{\parallel,b}\equiv -Q^\mu b_\mu$ is the momentum component parallel to the magnetic field four-vector. So, for instance, if the the magnetic field is oriented along the $z$-axis such that $b^\mu = (0,0,0,1)$, then $q_{\parallel,b}=q_z$ which has been the convention used in much of the past literature. The function $\mathcal{D}_n$ is given by
\begin{align}
\mathcal{D}_{n}(\fourvec{Q})&=-4\slashed{q}_{\perp}L_{n-1}^{1}\left(2\alpha_{q}\right) +\left(\slashed{q}_{\parallel}+m\right)\left[\mathcal{S}^{+}L_{n}\left(2\alpha_{q}\right)-\mathcal{S}^{-}L_{n-1}\left(2\alpha_{q}\right)\right]
\end{align}
where $L^\alpha_n$ denotes the generalized Laguerre polynomials (with $L_n \equiv L^0_n$ and $L_{-1}=0$), which are evaluated with the argument $\alpha_q \equiv -q_\perp^2/eB$. The spin-projection operators are given by $\mathcal{S}^\pm = (1\pm i\gamma^1 \gamma^2)$. The full $(11)$-propagator in the presence of a magnetized medium then becomes
\begin{align}
iS_{(11)}^{B}(\fourvec{Q})&=\sum_{n=0}^{\infty}\left(-1\right)^{n}e^{-\alpha_{q}}\mathcal{D}_{n}\left(\fourvec{Q}\right) \left[\frac{i}{q_{\parallel}^{2}-M_{n}^{2}+i\epsilon}-2\pi N\left(q_\parallel\cdot u\right)\delta\left(q_{\parallel}^{2}-M_{n}^{2}\right)\right] .
\end{align}
By utilizing this modified propagator and conducting the RTF calculations in the same manner as the previous subsection, we obtain the following expressions for the real and imaginary components of the self-energy in a magnetized plasma,
\begin{align}\label{RePiB_R}
\mathrm{Re}\left(\Pi_{R}^B\right)^{\mu\nu} &= \mathrm{Re}\left(\hat{\Pi}_{(11)}^{BV}\right)^{\mu\nu} - e^{2}\sum_{l,n=0}^{\infty} \int \frac{d^{3}\vec{q}}{(2\pi)^{3}}\times\nonumber\\
& \quad\mathcal{P} \Bigg\{ 
\frac{1}{2E_{q}^{l}} \left[
  f_{e}\left(E_{q}^{l}\right) 
  \left(\frac{T_{nl}^{\mu\nu}}{p_{\parallel}^{2} - M_{n}^{2}}\right)_{q^{0} = E_{q}^{l}} + 
  f_{\bar{e}}\left(E_{q}^{l}\right) 
  \left(\frac{T_{nl}^{\mu\nu}}{p_{\parallel}^{2} - M_{n}^{2}}\right)_{q^{0} = -E_{q}^{l}} 
\right] \nonumber \\
&\quad + \frac{1}{2E_{p}^{n}} \left[
  f_{e}\left(E_{p}^{n}\right) 
  \left(\frac{T_{nl}^{\mu\nu}}{q_{\parallel}^{2} - M_{l}^{2}}\right)_{q^{0} + k^{0} = E_{p}^{n}} + 
  f_{\bar{e}}\left(E_{p}^{n}\right) 
  \left(\frac{T_{nl}^{\mu\nu}}{q_{\parallel}^{2} - M_{l}^{2}}\right)_{q^{0} + k^{0} = -E_{p}^{n}} 
\right] 
\Bigg\}
\end{align}
\begin{align}
\label{ImPiB_R}
\mathrm{Im}\left(\Pi_{R}^{B}\right)^{\mu\nu} &= -\pi e^{2}\sum_{l,n=0}^\infty 
\int \frac{d^{3}q}{(2\pi)^{3}} \frac{1}{4E_{p}^{n}E_{q}^{l}} \times \nonumber \\
&\quad\Bigg\{ 
\left[ 1 - f_{\bar{e}}\left(E_{q}^{l}\right) - f_{e}\left(E_{p}^{n}\right) \right] 
T^{\mu\nu}_{nl} \bigg|_{q^{0} = -E_{q}^{l}} \delta\left(k^{0} - E_{q}^{l} - E_{p}^{n}\right) \nonumber \\
&\quad + \left[ 1 - f_{e}\left(E_{q}^{l}\right) - f_{\bar{e}}\left(E_{p}^{n}\right) \right] 
T^{\mu\nu}_{nl} \bigg|_{q^{0} = E_{q}^{l}} \delta\left(k^{0} + E_{q}^{l} + E_{p}^{n}\right) \nonumber \\
&\quad + \left[ f_{\bar{e}}\left(E_{q}^{l}\right) - f_{\bar{e}}\left(E_{p}^{n}\right) \right] 
T^{\mu\nu}_{nl} \bigg|_{q^{0} = -E_{q}^{l}} \delta\left(k^{0} - E_{q}^{l} + E_{p}^{n}\right) \nonumber \\
&\quad + \left[ f_{e}\left(E_{q}^{l}\right) - f_{e}\left(E_{p}^{n}\right) \right] 
T^{\mu\nu}_{nl} \bigg|_{q^{0} = E_{q}^{l}} \delta\left(k^{0} + E_{q}^{l} - E_{p}^{n}\right) 
\Bigg\}
\end{align}
where the energy of the $n$th Landau level is 
\begin{equation}\label{landau_energy}
  E^n_q\equiv \sqrt{q_{\parallel,b}^2+m_e^2+2neB}
\end{equation}
and the modified trace term $T^{\mu\nu}_{nl}$ is defined as
\begin{equation}
\label{Tmunu-nl1}
\Trnl\left(\fourvec{Q},\fourvec{K}\right)\equiv\left(-1\right)^{n+l}e^{-\left(\alpha_{p}+\alpha_{q}\right)}\mathrm{Tr}\left[\gamma^{\mu}\mathcal{D}_{n}\left(\fourvec{P}\right)\gamma^{\nu}\mathcal{D}_{l}\left(\fourvec{Q}\right)\right]. \quad \quad
\end{equation}
In analogy to the unmagnetized case presented in the subsection above, the magnetized self-energy also divides into two parts: a vacuum component in the presence of magnetic fields (denoted by $\Pi^{BV}$) and an in-medium component. {The expressions in \Eq{RePiB_R} and \Eq{ImPiB_R} are very general in the sense that any one-loop self-energy of a gauge boson coupled to a charged fermion (for instance, gluons and quarks) in the presence of a magnetic field can be written in this form. The only difference would show up in the trace term $\Trnl$ which encodes the information about the couplings. }

\subsection{Long-Wavelength Limit}
A particularly relevant regime is the long-wavelength limit (LWL), $\Vec{k}\rightarrow 0$, which is valid when $\omega \gg v_e k$ where $v_e$ is the typical velocity of an electron in the plasma. This specific kinematic regime ensures that the wavelength of electromagnetic excitations in the plasma is longer than the distance travelled by thermal electrons during one oscillation, and thus the effects of backreactions and diffusion can be neglected in determining how these excitations propagate. Note that even though we are taking the LWL, we still assume a separations of scales such that the properties of the background vary on even longer length scales. Also note that this kinematic regime is independent of both the nature of the plasma and the strength of the magnetic field. In Section~\ref{sec:limits}, we examine our LWL results in various limits corresponding to magnetic fields above and below the critical value for plasmas that are classical, relativistic, and degenerate. 

In this Section, we instead derive the most general expressions for the photon self-energy in the LWL for a magnetized plasma, expressed in terms of a minimal set of form factors. Since $\vec{k}\rightarrow 0$ (and hence $\fourvec{K}^\mu \rightarrow \omega u^\mu$), we are left with only three independent quantities in this regime: $u^\mu$, $b^\mu$ (the unit external magnetic field vector in the rest frame of the medium), and the metric tensor $g^{\mu\nu}$. We can construct projection tensors $P_i^{\mu\nu}$ transverse to $K^\mu$ (so that $K_\mu P_i^{\mu\nu}=u_\mu P_i^{\mu\nu}=0$ preserving gauge invariance) to serve as a basis~\cite{DOlivo:2002omk,Wang:2021ebh},
\begin{align}
&P_{\parallel}^{\mu\nu}\equiv g_{\parallel}^{\mu\nu}-u^\mu u^\nu = -b^{\mu}b^{\nu},\\
&P_{\perp}^{\mu\nu}\equiv g^{\mu\nu}-u^{\mu}u^{\nu}-P_{\parallel}^{\mu\nu}=g_{\perp}^{\mu\nu}, \\ &P_{\times}^{\mu\nu}\equiv i\epsilon^{\mu\nu\alpha\beta}u_{\alpha}b_{\beta}
\end{align}
where the subscripts $\parallel$ and $\perp$ denote quantities that are projected longitudinal and transverse to the external magnetic field, respectively, for instance $g_\parallel = u^\mu u^\nu - b^\mu b^\nu$. These projection tensors satisfy the following orthonormal and multiplicative properties:
\begin{equation}
P_{\parallel}P_{\perp}=P_{\parallel}P_{\times}=0;~~ P_{\perp}P_{\times}=P_{\times};~~
P_{\parallel}^{2}=P_{\parallel};~~ P_{\perp}^{2}=P_{\times}^{2}=P_{\perp};~~ 
P_{\parallel\mu}^{\mu}=1;~~P_{\perp\mu}^{\mu}=2.
\end{equation}
The projection tensors fully encapsulate the covariant structure of the self-energy tensor, which can thus be expressed as 
\begin{equation}\label{RePiBLWL}
\left(\Pi_{R}^B\right)^{\mu\nu}=\left(\hat{\Pi}^{BV}\right)^{\mu\nu}+\pi_{\parallel}P_{\parallel}^{\mu\nu}+\pi_{\perp}P_{\perp}^{\mu\nu}+\pi_{\times}P_{\times}^{\mu\nu} 
\end{equation}
where $\pi_i$'s are the associated Lorentz-invariant form factors for $i=\parallel, \perp, \times$ and the magnetized vacuum term $\hat{\Pi}^{BV}$ is derived in \App{App:vac}.
 
In the discussion below for the remainder of this Section, we provide details of the simplifications used to derive \Eq{RePiBLWL} from \Eq{RePiB_R}. Readers primarily interested in the final result and its phenomenological applications may choose to bypass the rest of this Section.

\subsubsection{Trace Technology}
In order to simplify the trace term $\Trnl$ in the LWL, we first evaluate the $k_\perp \rightarrow 0$ limit which is relevant for the lowest Landau level (LLL) approximation used in approximating the self-energy in the limit of super-critical magnetic fields (see \Sec{sec:4.1} for more details). We then take the $k_{\parallel,b} = -K^\mu b_\mu \rightarrow 0$ limit to complete the LWL approximation. 

We use the following properties of the spin-projection operators 
\begin{align}
\left(\mathcal{S}^{\pm}\right)^{2}=2\mathcal{S}^{\pm},\quad\mathcal{S^{\pm}\mathcal{S^{\mp}}}=0,\quad\left[\mathcal{S}^{\pm},\gamma_{\parallel}^{\mu}\right]=0\\
\mathcal{S}^{\pm}\gamma^{\mu}\mathcal{S}^{\pm}=2\mathcal{S}^{\pm}\gamma_{\parallel}^{\mu},\quad\mathcal{S}^{\pm}\gamma^{\mu}\mathcal{S}^{\mp}=2\mathcal{S}^{\pm}\gamma_{\perp}^{\mu}
\end{align}
to write the following identities
\begin{equation}\label{traceiden1}
\begin{array}{ccc}
\frac{1}{2}\mathrm{Tr}\left[\gamma^{\mu}\left(\slashed{\fourvec{p}}_{\parallel}+m\right)\mathcal{S}^{\pm}\gamma^{\nu}\left(\slashed{\fourvec{q}}_{\parallel}+m\right)\mathcal{S}^{\pm}\right]=\left(\tilde{T}_{S}^{a}\right)^{\mu\nu}\pm i\left(\tilde{T}_{A}^{a}\right)^{\mu\nu}\\
\frac{1}{2}\mathrm{Tr}\left[\gamma^{\mu}\left(\slashed{\fourvec{p}}_{\parallel}+m\right)\mathcal{S}^{\pm}\gamma^{\nu}\left(\slashed{\fourvec{q}}_{\parallel}+m\right)\mathcal{S}^{\mp}\right]=\left(\tilde{T}_{S}^{b}\right)^{\mu\nu}\mp i\left(\tilde{T}_{A}^{b}\right)^{\mu\nu},
\end{array}
\end{equation}
where the terms with a tilde, $\tilde{T}_{S,A}^{a,b}$, are defined as
\begin{subequations}
\begin{align}
\left(\tilde{T}_{S}^{a}\right)^{\mu\nu}&\equiv \mathrm{Tr}\left[\gamma_{\parallel}^{\mu}\left(\slashed{\fourvec{p}}_{\parallel}+m\right)\gamma_{\parallel}^{\nu}\left(\slashed{\fourvec{q}}_{\parallel}+m\right)\right] \\
\left(\tilde{T}_{A}^{a}\right)^{\mu\nu}&\equiv \mathrm{Tr}\left[\gamma^{1}\gamma^{2}\gamma_{\parallel}^{\mu}\left(\slashed{\fourvec{p}}_{\parallel}+m\right)\gamma_{\parallel}^{\nu}\left(\slashed{\fourvec{q}}_{\parallel}+m\right)\right]\\
\left(\tilde{T}_{S}^{b}\right)^{\mu\nu}&\equiv \mathrm{Tr}\left[\gamma_{\perp}^{\mu}\left(\slashed{\fourvec{p}}_{\parallel}+m\right)\gamma_{\perp}^{\nu}\left(\slashed{\fourvec{q}}_{\parallel}+m\right)\right] \\
\left(\tilde{T}_{A}^{b}\right)^{\mu\nu}&\equiv \mathrm{Tr}\left[\gamma^{1}\gamma^{2}\gamma_{\perp}^{\mu}\left(\slashed{\fourvec{p}}_{\parallel}+m\right)\gamma_{\perp}^{\nu}\left(\slashed{\fourvec{q}}_{\parallel}+m\right)\right].
\end{align}
\end{subequations}
Here $S$ and $A$ signify the symmetric and anti-symmetric parts respectively. Using the identities of \Eq{traceiden1} in \Eq{Tmunu-nl1}, we can write $\Trnl$ as
\begin{align}
\label{Tmunu-nl2}
\Trnl\left(k_{\perp}\rightarrow 0\right)\supset&(-1)^{n+l}\,2e^{-2\alpha_{\fourvec{q}}}\bigg\{ \left(\tilde{T}_{S}^{a}\right)^{\mu\nu}\left[L_{n}L_{l}+L_{n-1}L_{l-1}\right]\nonumber\\
&-\left(\tilde{T}_{S}^{b}\right)^{\mu\nu}\left[L_{n}L_{l-1}+L_{n-1}L_{l}\right]
+8\mathrm{Tr}\left[\gamma^{\mu}\gamma^{\rho}\gamma^{\nu}\gamma^{\sigma}\right]\left[q_{\rho}^{\perp}q_{\sigma}^{\perp}L_{n-1}^{1}L_{l-1}^{1}\right]\nonumber\\
&+i\left(\tilde{T}_{A}^{a}\right)^{\mu\nu}\left[L_{n}L_{l}-L_{n-1}L_{l-1}\right]+i\left(\tilde{T}_{A}^{b}\right)^{\mu\nu}\left[L_{n}L_{l-1}-L_{n-1}L_{l}\right]\bigg\} 
\end{align}
where the $\supset$ signifies that we have ignored terms like $\mathrm{Tr}\left[\gamma^{\mu}\left(\slashed{\fourvec{p}}_{\parallel}+m\right)\mathcal{S}^{\pm}\gamma^{\nu}\slashed{\fourvec{q}}_{\perp}\right]$ since they are linear in $\vec{\fourvec{q}}_\perp$ and will therefore vanish upon integration over $\int d^2 \vec{q}_\perp $. Evidently, the transverse and longitudinal momenta are separated -- the exponentials and the Laguerre polynomials in the square brackets carry the transverse parts while the $\tilde{T}$ factors are composed of purely longitudinal momenta. We can thus decompose the integral measure into $\int d^3 \vec{\fourvec{q}}\rightarrow \int d^2 \vec{\fourvec{q}}_\perp \int dq_{\parallel,b}$ and the transverse and longitudinal integrals can be carried out independently. Employing the following orthogonality properties of the generalized Laguerre polynomials
\begin{subequations}
\begin{align}
&\int\frac{d^{2}q_{\perp}}{(2\pi)^{2}}e^{-2\alpha_{\fourvec{q}}}L_{n}(2\alpha_{\fourvec{q}})L_{l}(2\alpha_{\fourvec{q}})=\frac{eB}{8\pi}\delta_{l}^{n}\\
&\int\frac{d^{2}q_{\perp}}{(2\pi)^{2}}e^{-2\alpha_{\fourvec{q}}}L_{n}^{\alpha}(2\alpha_{\fourvec{q}})L_{l}^{\beta}(2\alpha_{\fourvec{q}})q_{\perp}^{\mu}=0\\
&\int\frac{d^{2}q_{\perp}}{(2\pi)^{2}}e^{-2\alpha_{\fourvec{q}}}L_{n-1}^{1}(2\alpha_{\fourvec{q}})L_{l-1}^{1}(2\alpha_{\fourvec{q}})q_{\perp}^{\mu}q_{\perp}^{\nu}=-g_{\perp}^{\mu\nu}\frac{e^{2}B^{2}}{32\pi}n\delta_{l-1}^{n-1},
\end{align}
\end{subequations}
we can perform the transverse integrals in \Eq{RePiB_R} to obtain
\begin{equation}\label{int_tmununl}
\int\frac{d^{3}\vec{q}}{(2\pi)^{3}}\Trnl(k_\perp \rightarrow 0)=\frac{eB}{4\pi}\int\frac{dq_{\parallel,b}}{2\pi}\tilde{T}^{\mu\nu}_{n,l}
\end{equation}
with
\begin{multline}\label{tilde_Tmunu}
\tilde{T}^{\mu\nu}_{n,l}\equiv (-1)^{n+l}\left\{ \left(\tilde{T}_{S}^{a}\right)^{\mu\nu}\left[\delta^{n}_{l}+\delta^{n-1}_{l-1}\right]-\left(\tilde{T}_{S}^{b}\right)^{\mu\nu}\left[\delta^{n}_{l-1}+\delta^{n-1}_{l}\right]
+\left(\tilde{T}_{S}^{c}\right)^{\mu\nu}\delta_{l-1}^{n-1}\right.\\
\left.+i\left(\tilde{T}_{A}^{a}\right)^{\mu\nu}\left[\delta^{n}_{l}-\delta^{n-1}_{l-1}\right]+i\left(\tilde{T}_{A}^{b}\right)^{\mu\nu}\left[\delta^{n}_{l-1}-\delta^{n-1}_{l}\right]\right\}. 
\end{multline}
Here, we have defined
\begin{equation}
\left(\tilde{T}_{S}^{c}\right)^{\mu\nu}\equiv2neB\,\mathrm{Tr}\left[\gamma^{\mu}\gamma^{\rho}\gamma^{\nu}\gamma^{\sigma}\right]\left(-g_{\rho\sigma}^{\perp}\right)=16neB\,g_{\parallel}^{\mu\nu},\quad \quad
\end{equation}
and the rest of the traces in Eq.~\eqref{int_tmununl} have been evaluated using \texttt{FeynCalc}~\cite{Shtabovenko:2016sxi} to finally obtain
\begin{subequations}
\begin{align}
&\left(\tilde{T}_{S}^{a}\right)^{\mu\nu}=4\left\{ g_{\parallel}^{\mu\nu}\left(m^{2}-q_{\parallel}\cdot\left(q_{\parallel}+k_{\parallel}\right)\right)+q_{\parallel}^{\mu}\left(q_{\parallel}+k_{\parallel}\right)^{\nu}+q_{\parallel}^{\nu}\left(q_{\parallel}+k_{\parallel}\right)^{\mu}\right\}; \\
&\left(\tilde{T}_{A}^{a}\right)^{\mu\nu}=0;\\
&\left(\tilde{T}_{S}^{b}\right)^{\mu\nu}=4g_{\perp}^{\mu\nu}\left(m^{2}-q_{\parallel}\cdot\left(q_{\parallel}+k_{\parallel}\right)\right);\\
&\left(\tilde{T}_{A}^{b}\right)^{\mu\nu}=-4\epsilon^{\mu\nu\alpha\beta}u_{\alpha}b_{\beta}\left(m^{2}-q_{\parallel}\cdot\left(q_{\parallel}+k_{\parallel}\right)\right).
\end{align}
\end{subequations}
Finally combining the decomposition in \Eq{int_tmununl} with the self-energy expression in \Eq{RePiB_R} we obtain
\begin{equation}
\label{RePiB2}
\left(\Pi_{R}^{B}\right)^{\mu\nu}=\mathrm{Re}\left(\hat{\Pi}^{BV}\right)^{\mu\nu}+\left(\Pi_{S}^{a}\right)^{\mu\nu}+\left(\Pi_{S}^{b}\right)^{\mu\nu}
+\left(\Pi_{S}^{c}\right)^{\mu\nu}+\left(\Pi_{A}^{b}\right)^{\mu\nu}
\end{equation}
where $\Pi^{a,b,c}_{S,A}$ are the contributions to the self energy from the corresponding traces.

\subsubsection{Form factors and covariant structure of $\Pi^{\mu\nu}$ }
Below, we evaluate and further simplify the Hermitian part of each term in \Eq{RePiB2} (note that this is only the same as the real part as expressed in the plasma eigenbasis). {In the remainder of this paper, we concentrate exclusively on the Hermitian component of the self-energy which contributes to the dispersion relation of the photon-BSM system (see \Sec{NMS}) and leave the discussion for the anti-Hermitian part for future work (see e.g. Refs.~\cite{Hattori:2022uzp, Wang:2021ebh} for some recent progress).}

It is possible to evaluate $\Pi^a_S$ from \Eq{RePiB_R} and the form of the corresponding trace by shifting the integration variables for some of the terms ($q_\parallel \rightarrow q_\parallel -k_\parallel$ and $q_{\parallel,b} \rightarrow -q_{\parallel,b}$) and using the Kronecker deltas from the integral over $\vec{q}_{\perp}$ to carry out the sum over Landau levels. We obtain 
\begin{align}
\left(\Pi_{S}^{a}\right)^{\mu\nu}=-\frac{e^{3}B}{4\pi}\sum_{n=0}^{\infty}\left(2-\delta_{0}^{n}\right)\int\frac{dq_{\parallel,b}}{2\pi}&\frac{f_{e}\left(E_{\fourvec{q}}^{n}\right)+f_{\bar{e}}\left(E_{\fourvec{q}}^{n}\right)}{2E_{\fourvec{q}}^{n}}\\ &\times \left\{ \left[\frac{\left(\tilde{T}_{S}^{a}\right)_{\fourvec{q}^{0}=E_{\fourvec{q}}^{n}}^{\mu\nu}}{\left(q_{+}^{n}+k_{\parallel}\right)^{2}-M_{n}^{2}}\right]+\left[\frac{\left(\tilde{T}_{S}^{a}\right)_{\fourvec{q}^{0}=-E_{\fourvec{q}}^{n}}^{\mu\nu}}{\left(q_{-}^{n}+k_{\parallel}\right)^{2}-M_{n}^{2}}\right]\right\} \nonumber
\end{align}
where we have defined $q_\parallel |_{\fourvec{q}^0=\pm E_q^l}\equiv \fourvec{q}^l_\pm = \left(\pm E^l_q,0,0,q_{\parallel,b}\right)$ for compactness. Following Appendix A.2 of Ref.~\cite{Hattori:2022uzp}, we can further express $\Pi^a_S$ as $\left(\Pi_{S}^{a}\right)^{\mu\nu}=\pi_{S}^{a}\tilde{P}_{\parallel}^{\mu\nu}-\pi^c_S k^{\mu}_\parallel k^{\nu}_\parallel k^{2}_\parallel$ with $\tilde{P}^{\mu\nu}_\parallel = g^{\mu\nu}_\parallel - \fourvec{k}^{\mu}_\parallel \fourvec{k}^{\nu}_\parallel/\fourvec{k}^{2}_\parallel$ as the projection tensor parallel to the external magnetic field. Recalling the definition $k_{\parallel,b} = -K^\mu b_\mu$ (as opposed to $k_\parallel^\mu = g^{\mu \nu}_\parallel K_\nu$), we find that
\begin{equation}
\pi_{S}^{a}=\frac{e^{3}B}{4\pi}\sum_{n=0}^{\infty}\left(2-\delta_{0}^{n}\right)\int
\frac{dq_{\parallel,b}}{2\pi}
\frac{\left[f_{e}\left(E_{\fourvec{q}}^{n}\right)+f_{\bar{e}}\left(E_{\fourvec{q}}^{n}\right)\right]4m_{e}^{2}\left(k_{\parallel}^{2}+2q_{\parallel,b}k_{\parallel,b}\right)}{2E_{\fourvec{q}}^{n}\,\left[ k_{\parallel}^{2}\left(q_{\parallel,b}-\frac{1}{2}k_{\parallel,b}\right)^{2}-\frac{\omega^{2}}{4}\left(k_{\parallel}^{2}-4M_{n}^{2}\right)\right]},
\end{equation}
\begin{equation}
\pi_{S}^{c}=\frac{e^{3}B}{4\pi}\sum_{n=0}^{\infty}\left(1-\delta_{0}^{n}\right)\int
\frac{dq_{\parallel,b}}{2\pi}
\frac{\left[f_{e}\left(E_{\fourvec{q}}^{n}\right)+f_{\bar{e}}\left(E_{\fourvec{q}}^{n}\right)\right]16neB}{2E_{\fourvec{q}}^{n}\,\left[ k_{\parallel}^{2}\left(q_{\parallel,b}-\frac{1}{2}k_{\parallel,b}\right)^{2}-\frac{\omega^{2}}{4}\left(k_{\parallel}^{2}-4M_{n}^{2}\right)\right]}.
\end{equation}
The term $\Pi^c_S$ can be similarly evaluated to be $\left(\Pi^c_S\right)^{\mu\nu} = \pi^c_S g_\parallel^{\mu\nu}$ so that 
\begin{equation}
  \left(\Pi^a_S\right)+ \left(\Pi^c_S\right) = (\pi^a_S+\pi^c_S)\tilde{P}^{\mu\nu}_\parallel
\end{equation}
Eventually, we perform the final step of the LWL approximation and set $k_{\parallel,b} \rightarrow 0$, which reduces $\tilde{P}^{\mu\nu}_\parallel\rightarrow P^{\mu\nu}_\parallel$ and the form factors simplify to
\begin{equation}\label{pi_a_S}
\pi_{S}^{a}=\frac{e^{3}B}{4\pi}\sum_{n=0}^{\infty}\left(2-\delta_{0}^{n}\right)\int\frac{dq_{\parallel,b}}{2\pi}
\frac{\left[f_{e}\left(E_{\fourvec{q}}^{n}\right)+f_{\bar{e}}\left(E_{\fourvec{q}}^{n}\right)\right] }{2E_{\fourvec{q}}^{n}}\frac{4m_{e}^{2}}{\left[\left(E_{\fourvec{q}}^{n}\right)^{2}-\frac{\omega^{2}}{4}\right]}~,
\end{equation}\label{pi_c_S}
\begin{equation}
\pi_{S}^{c}=\frac{e^{3}B}{4\pi}\sum_{n=0}^{\infty}\left(1-\delta_{0}^{n}\right)\int\frac{dq_{\parallel,b}}{2\pi}\frac{f_{e}\left(E_{\fourvec{q}}^{n}\right)+f_{\bar{e}}\left(E_{\fourvec{q}}^{n}\right)}{2E_{\fourvec{q}}^{n}}\frac{16neB}{\left(E_{\fourvec{q}}^{n}\right)^{2}-\frac{\omega^{2}}{4}}.
\end{equation}
Now, focusing on the term $\Pi^b_S$ in \Eq{RePiB_R} and again performing similar shifts in the variables of integration, we can express the contribution of this term in the photon self-energy as 
\begin{align}
\left(\Pi_{S}^{b}\right)^{\mu\nu} & =\frac{e^{3}B}{4\pi}\int\frac{dq_{\parallel,b}}{2\pi}\sum_{s=+,-} \, \sum_{l,n=0}^{\infty}\left(-1\right)^{n+l}\left(\delta_{l-1}^{n}+\delta_{l}^{n-1}\right) \nonumber\\ &\times \left\{ \frac{f_{e}^{(s)}\left(E_{\fourvec{q}}^{l}\right)}{2E_{\fourvec{q}}^{l}}\left[\frac{\left(\tilde{T}_{S}^{b}\right)_{\fourvec{q}^{0}=sE_{\fourvec{q}}^{n}}^{\mu\nu}}{\left(q_{s}^{l}+k_{\parallel}\right)^{2}-M_{n}^{2}}\right]+\frac{f_{e}^{(-s)}\left(E_{\fourvec{q}}^{n}\right)}{2E_{\fourvec{q}}^{n}}\left[\frac{\left(\tilde{T}_{S}^{b}\right)_{\fourvec{q}^{0}=sE_{\fourvec{q}}^{n}}^{\mu\nu}}{\left(q_{s}^{n}+k_{\parallel}\right)^{2}-M_{l}^{2}}\right]\right\} \nonumber\\
 & =-\frac{e^{3}B}{4\pi}\int\frac{dq_{\parallel,b}}{2\pi}\sum_{s=+,-} \,\sum_{n=0}^{\infty}\frac{f_{e}^{(s)}\left(E_{\fourvec{q}}^{n}\right)+f_{e}^{(-s)}\left(E_{\fourvec{q}}^{n}\right)}{2E_{\fourvec{q}}^{n}}\left(\tilde{T}_{S}^{b}\right)_{\fourvec{q}^{0}=sE_{\fourvec{q}}^{n}}^{\mu\nu}\nonumber\\ 
 & \qquad \qquad\qquad \qquad \times \left\{ \frac{1-\delta_{0}^{n}}{\left(q_{s}^{n}+k_{\parallel}\right)^{2}-M_{n-1}^{2}} +\frac{1}{\left(q_{s}^{n}+k_{\parallel}\right)^{2}-M_{n+1}^{2}}\right\} 
\end{align}
where $s=+,-$ and we have defined $f^{(+)}_e=f_e$ and $f^{(-)}_e=f_{\bar{e}}$. Setting $k_{\parallel,b}\rightarrow 0$, the trace can be evaluated to be
\begin{equation}
\left(\tilde{T}_{S}^{b}\right)_{\fourvec{q}^{0}=sE_{\fourvec{q}}^{n}}^{\mu\nu}=-4\left(2neB+s\omega E_{\fourvec{q}}^{n}\right)P_{\perp}^{\mu\nu}
\end{equation}
where $P_\perp^{\mu\nu}=g_\perp^{\mu\nu}$ is the projection tensor transverse to the external magnetic field and the denominators become
\begin{equation}
\left(q_{s}^{n}+k_{\parallel}\right)^{2}-M_{n\mp1}^{2}=\omega^{2}+2s\omega E_{\fourvec{q}}^{n}\pm2eB.
\end{equation}
Putting everything together, we can express $\left(\Pi^b_S\right)^{\mu\nu} = \pi^b_S P_\perp^{\mu\nu}
$ with
\begin{align}\label{pi_b_S}
\pi_{S}^{b}&=\frac{e^{3}B}{4\pi}\sum_{n=0}^{\infty}\int\frac{dq_{\parallel,b}}{2\pi}\frac{f_{e}\left(E_{\fourvec{q}}^{n}\right)+f_{\bar{e}}\left(E_{\fourvec{q}}^{n}\right)}{2E_{\fourvec{q}}^{n}} \nonumber\\
&\times \sum_{s=+,-}\left[4\left(2neB+s\omega E_{\fourvec{q}}^{n}\right)\left\{ \frac{1}{\omega^{2}+s2\omega E_{\fourvec{q}}^{n}-2eB}+\frac{1-\delta_{0}^{n}}{\omega^{2}+s2\omega E_{\fourvec{q}}^{n}+2eB}\right\} \right].
\end{align}
The term $\Pi^b_A$ in \eqref{RePiB2} can be evaluated in a similar way as $\Pi^b_S$. Again, in the $k_{\parallel,b} \rightarrow 0$ limit $\tilde{T}_{A}^{b}$ becomes
\begin{equation}
i\left(\tilde{T}_{A}^{b}\right)_{\fourvec{q}^{0}=sE_{\fourvec{q}}^{n}}^{\mu\nu}=4\left(2neB+s\omega E_{\fourvec{q}}^{n}\right)P_{\times}^{\mu\nu}
\end{equation}
where $P_\times^{\mu\nu} = i\eps^{\mu\nu\alpha\beta}u_{\alpha}b_\beta$ and the total contribution can be expressed as $\left(\Pi^b_A\right)^{\mu\nu} = \pi^b_A P_\times^{\mu\nu}$ with
\begin{align}\label{pi_b_A}
\pi_{A}^{b}&=\frac{e^{3}B}{4\pi}\sum_{n=0}^{\infty}\int\frac{dq_{\parallel,b}}{2\pi}\frac{f_{e}\left(E_{\fourvec{q}}^{n}\right)-f_{\bar{e}}\left(E_{\fourvec{q}}^{n}\right)}{2E_{\fourvec{q}}^{n}} \nonumber\\
&\times\sum_{s=+,-}s\left[4\left(2neB+s\omega E_{\fourvec{q}}^{n}\right)\times\left\{ \frac{1}{\omega^{2}+s2\omega E_{\fourvec{q}}^{n}-2eB}-\frac{1-\delta_{0}^{n}}{\omega^{2}+s2\omega E_{\fourvec{q}}^{n}+2eB}\right\} \right]
\end{align}

Ultimately, adding all the contributions 
and defining 
\begin{equation}
\label{def-formfac}
  \real[\pi_\parallel] \equiv \pi^a_S + \pi^c_S, \quad \real[\pi_\perp]\equiv \pi^b_S,\quad \real[\pi_\times] \equiv \pi^b_A,
\end{equation}
the Hermitian part of the retarded self energy expression in \Eq{RePiB_R} (which is the real part as expressed in the eigenbasis) reduces to the form in \Eq{RePiBLWL} for a magnetic plasma in the LWL limit. For the sake of clarity, these form factors are compactly expressed as 
\begin{equation}\label{Re-pi-i}
  \real[\pi_i(\omega,B)]=\frac{e^3 B}{4\pi}\sum_{n=0}^\infty \pi_i^{(n)}(\omega,B), \quad i=\perp,\parallel,\times
\end{equation}
where the contribution from the $n$th Landau level (with energy $E^n_q$ defined in \Eq{landau_energy}) is given by
\begin{subequations}\label{piB-n}
\begin{equation}
\label{pipara-n-1}
\pi_\parallel^{(n)}=\int\frac{dq_{\parallel,b}}{2\pi}\frac{f_{e}\left(E_{q}^{n}\right)+f_{\bar{e}}\left(E_{q}^{n}\right)}{2E_{q}^{n}}\frac{\left(2-\delta_{0}^{n}\right)4m_{e}^{2} + \left(1-\delta_{0}^{n} \right)16neB}{\left(E_{q}^{n}\right)^{2}-\frac{\omega^{2}}{4}}
\end{equation}
\begin{multline}
\label{piperp-n-1}
\pi_{\perp}^{(n)}=\int\frac{dq_{\parallel,b}}{2\pi}\frac{f_{e}\left(E_{q}^{n}\right)+f_{\bar{e}}\left(E_{q}^{n}\right)}{2E_{q}^{n}}\sum_{s=+,-}4\left(2neB+s\omega E_{q}^{n}\right) \\
\times\left\{ \frac{1}{\omega^{2}+2s\omega E_{q}^{n}-2eB}+\frac{1-\delta_{0}^{n}}{\omega^{2}+2s\omega E_{q}^{n}+2eB}\right\} 
\end{multline}
\begin{multline}
\label{picross-n-1}
\pi_{\times}^{(n)}=\int\frac{dq_{\parallel,b}}{2\pi}\frac{f_{e}\left(E_{q}^{n}\right)-f_{\bar{e}}\left(E_{q}^{n}\right)}{2E_{q}^{n}}\sum_{s=+,-}4s\left(2neB+s\omega E_{q}^{n}\right)\\
\times \left\{ \frac{1}{\omega^{2}+2s\omega E_{q}^{n}-2eB}-\frac{1-\delta_{0}^{n}}{\omega^{2}+2s\omega E_{q}^{n}+2eB}\right\} .
\end{multline} 
\end{subequations}

\subsection{Plasma mixing matrix in a magnetized plasma}
Referring back to the discussion of plasma normal modes from \Sec{sec:2.2}, and now equipped with analytic expressions for the self-energy in a magnetic plasma, we can readily compute the corresponding mixing matrix $\pi^{IJ}_{AA}$. In a coordinate system where the propagation direction aligns with the $z$-axis and the magnetic field is inclined at an angle $\theta_B$ to the propagation direction (i.e., $\fourvec{K}^\mu =(\omega,0,0,\modk)$ and $b^\mu = (0,0,-\sin \theta_B, \cos \theta_B$)), the Hermitian component of the mixing matrix can be expressed as \cite{DOlivo:2002omk}
\begin{align}
\label{pimixB}
\real & \left[\pi_{AA}^{IJ}\right]=\real\left[-(\eps^I)^*\cdot\hat{\Pi}^{BV}\cdot\eps^J\right] \nonumber \\
&\,\, +\left[\begin{array}{ccc}
\real[\pi_{\perp}] & -i\real[\pi_{\times}]c_\theta & -i\frac{\sqrt{K^{2}}}{\omega}\real[\pi_{\times}]s_\theta\\
i\real[\pi_{\times}]c_\theta & \real[\pi_{\perp}]c^{2}_\theta+\real[\pi_{\parallel}]s^{2}_\theta & \frac{\sqrt{K^{2}}}{\omega}\left(\real[\pi_{\perp}]-\real[\pi_{\parallel}]\right)c_\theta s_\theta\\
i\frac{\sqrt{K^{2}}}{\omega}\real[\pi_{\times}]s_\theta & \frac{\sqrt{K^{2}}}{\omega}\left(\real[\pi_{\perp}]-\real[\pi_{\parallel}]\right)c_\theta s_\theta & \frac{K^{2}}{\omega^{2}}\left(\real[\pi_{\parallel}]c^{2}_\theta+\real[\pi_{\perp}]s^{2}_\theta\right)
\end{array}\right] \qquad \qquad
\end{align}
where for compactness we have defined $c_\theta \equiv \cos \theta_B$ and $s_\theta \equiv \sin \theta_B$ and the factors of $\sqrt{\fourvec{K}^2/\omega^2}$ in front of the form factors originate from the polarization vectors ($\eps_{T,L}$) while the form factors themselves are evaluated at $\vec{k}\rightarrow 0$. To obtain the plasma normal modes we need to go one step further and diagonalize the mixing matrix, as discussed in \Sec{sec:2.2}. {For the simplest configuration where the propagation is parallel to the magnetic field ($\theta_B =0$), the plasma mixing matrix (ignoring the vacuum contributions, see the discussion around \Fig{fig:ratio_piBV_wp} in \App{App:vac}) can be readily diagonalized to obtain the following dispersion relations
\begin{equation}\label{disp_theta0}
  \omega^2 = k^2 +\real[\pi_\perp(\omega)+\pi_\times(\omega)],\quad
  \omega^2 = k^2 + \real[\pi_\perp(\omega)-\pi_\times(\omega)],\quad
  \omega^2 = \real[\pi_\parallel(\omega)],
\end{equation}
with the eigenvectors given by
\begin{equation}\label{eigenvec_theta0}
  \eps_{\perp 1} = \frac{1}{\sqrt{2}}\{0,1,i,0\},\quad
  \eps_{\perp 2} = \frac{1}{\sqrt{2}}\{0,1,-i,0\},\quad
  \eps_{\parallel} = \frac{1}{\sqrt{\omega^2-k^2}}\{k,0,0,\omega\}.
\end{equation}
Evidently, in the presence of a magnetic field, the transverse modes split and become circularly polarized due to Faraday rotation. For parallel propagation, the longitudinal dispersion in a magnetized medium remains analogous to that in an unmagnetized one, with the substitution $\pi_L \rightarrow \frac{K^2}{\omega^2}\pi_\parallel$.} For any arbitrary configuration the diagonalization can be accomplished numerically in principle (see \cite{Pal:2020bwo} for some analytical progress). However, the most significant obstacle to this lies in evaluating the infinite sums within the form factor expressions, as they present formidable numerical challenges and are frequently plagued by convergence issues \cite{Kohri:2001wx, Hattori:2012ny}. In the following Section, we introduce various approximations regarding the ambient plasma properties, which will pave the way for analytical advancements in calculating the form factors.

\section{Limiting Behaviour of the Self-Energy in Plasma Environments} 
\label{sec:limits}
\subsection{Super-critical magnetic fields $B\gg B_c$}\label{sec:4.1}
When the external magnetic field is very strong ($\sqrt{eB}\gg m_e, T, \mu$), the energies of all the higher Landau levels (HLL), as shown in \Eq{landau_energy}, are considerably larger compared to the energy of the lowest Landau level (LLL) with $n=0$, which is independent of the external $B$-field~\cite{melrose1976vacuum,Miransky:2015ava,Bandyopadhyay:2016fyd, Hattori:2012je,Hattori:2022uzp}. Hence, the occupation of the HLLs are very suppressed compared to the LLL, which dominates the self-energy sum. This can also be inferred from \Eq{S_BV} where only the $n=0$ terms survive in the $eB \rightarrow \infty$ limit~\cite{Bandyopadhyay:2016fyd, Hattori:2022uzp}. Particles in the LLL have vanishing (classical) cyclotron orbits and are constrained to move along the magnetic field lines effectively reducing the dimensions of the system from ($3+1$) to ($1+1$)~\cite{Gusynin:1995nb}.

As can be seen from the expression of self-energies in the previous section (for instance \Eq{pi_a_S}, \Eq{pi_b_S} or \Eq{pi_b_A}), the photon self-energy can be written as a sum over the Landau levels $n$ and $l$ of the two electrons running in the self-energy loop. In the LLL approximation, we just take the contribution from the $n=l =0$ term. Then only the parallel component proportional to ($\delta^n_l + \delta^{n-1}_{l-1}$) contributes, while the $\pi_\perp, \pi_\times \propto \left(\delta^n_{l-1} + \delta^{n-1}_{l}\right)$ terms vanish when $n=l=0$. The magnetized vacuum contribution to the LLL self-energy can be calculated from Eq.~\eqref{Pi_BV2},
\begin{align}\label{pi_para_BV}
\left(\Pi_{(11)}^{BV}\right)^{\mu\nu}_{LLL}=\left(\pi_\parallel^{BV}\right)_{LLL} \tilde{P}^{\mu\nu}_\parallel,
\end{align}
where the form factor $\left(\pi_\parallel^{BV}\right)_{LLL}$ is finite and gauge-invariant and is given by\footnote{For a calculation in the $R/A$-basis where the self-energy is written in terms of the retarded and advanced propagator see Ref.~\cite{Hattori:2022uzp}.}
\begin{equation}\label{pi_BV_form}
  \left(\pi_\parallel^{BV}\right)_{LLL} =- \frac{e^{3}B}{2\pi^{2}}e^{-\frac{\left|\vec{k}_{\perp}\right|^{2}}{2eB}} k_\parallel^2 \int_{0}^{1}dx\,\frac{x(1-x)}{\Delta_-}
\end{equation}
while from \labelcref{Re-pi-i,pipara-n-1} the thermal part (which captures the medium effects) becomes $\real[\pi_\parallel^{LLL}] \tilde{P}_\parallel^{\mu\nu}$ with
\begin{equation}\label{pi_para_th}
\real[\pi_\parallel^{LLL}]=\frac{e^{3}B}{4\pi}e^{-\frac{\left|\vec{k}_{\perp}\right|^{2}}{2eB}}\mathcal{P}\left[\int\frac{dq_{\parallel,b}}{2\pi}\frac{f_{e}\left(E_{q}^{0}\right)+f_{\bar{e}}\left(E_{q}^{0}\right)}{2E_{q}^{0}}\frac{4m_{e}^{2}\left(k_{\parallel}^{2}+2q_{\parallel,b}k_{\parallel,b}\right)}{k_{\parallel}^{2}\left(q_{\parallel,b}-\frac{1}{2}k_{\parallel,b}\right)^{2}-\frac{\omega^{2}}{4}\left(k_{\parallel}^{2}-4m_{e}^{2}\right)}\right].
\end{equation}
The LLL self-energy then can be expressed as 
\begin{equation}\label{RePiB_sup}
\real(\Pi_{R}^B)^{\mu\nu}_{LLL}=\real[\tilde{\pi}^{LLL}_\parallel ]\tilde{P}_\parallel^{\mu\nu}
\end{equation}
where the total longitudinal form factor is given by $\real[\tilde{\pi}_\parallel] = \real[\pi^{BV}_\parallel]+\real[\pi_\parallel]$. 

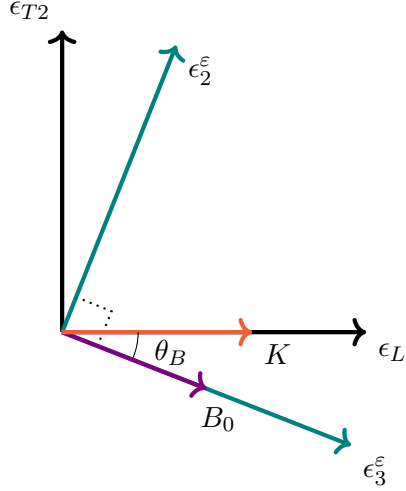
\begin{figure}
\centering
\begin{tikzpicture}[node distance=2cm]
\draw[thick, dotted, rotate around={-22:(0,0)}] (0,0) rectangle (0.5,0.5);

\draw[ultra thick,->] (0,0) -- (4,0) node[anchor=north west] {$\epsilon_L$};
\draw[ultra thick,->] (0,0) -- (0,4) node[anchor=south east] {$\epsilon_{T2}$};
\draw[ultra thick,->, draw=teal] (0,0) -- (1.5,3.8) node[anchor=north west] {$\epsilon^\varepsilon_2$};
\draw[ultra thick,->, draw=teal] (0,0) -- (3.8,-1.5) node[anchor=north west] {$\epsilon^\varepsilon_3$};

\draw[ultra thick,->, draw=violet] (0,0) -- (1.9, -0.75) node[xshift=0.15cm, yshift=-0.4cm] {$B_0$};

\draw[ultra thick,->, draw=RedOrange] (0,0) -- (2.5, 0) node[anchor=north west] {$K$};

\draw (1,0) arc (0:-22:0.9cm) node[xshift=0.5cm, yshift=0.07cm] {$\theta_B$};

\end{tikzpicture}

\caption{Eigenmodes of propagation in a plasma with a super-critical magnetic field oriented at angle $\theta_B$ relative to the direction of propagation. Note that the transverse mode of propagation that is perpendicular to the $K-B$ plane is not affected by the magnetic field. }
\label{fig:eigenmodes_orientation}
\end{figure}

In the coordinate system where $\fourvec{K}^\mu =(\omega,0,0,\modk)$ and $b^\mu = (0,0,-\sin \theta_B, \cos \theta_B$), and in the basis of T/L, the Hermitian part of the mixing matrix for super-critical magnetic fields becomes
\begin{equation}
\real\left[\pi^{IJ}_{AA}\right]=\frac{\real[\tilde{\pi}_{\parallel}^{LLL}]}{\omega^{2}-\modk^{2}\cos^{2}\theta_{B}}\left[\begin{array}{ccc}
0 & 0 & 0\\
0 & \omega^{2}\sin^{2}\theta_{B} & -\omega\sqrt{\fourvec{K}^2}\sin\theta_{B}\cos\theta_{B}\\
0 & -\omega\sqrt{\fourvec{K}^2}\sin\theta_{B}\cos\theta_{B} & \fourvec{K}^2\cos^{2}\theta_{B}
\end{array}\right].
\end{equation}
This mixing matrix can be analytically diagonalized to obtain the normal modes of dispersion. The diagonal mixing matrix becomes $\hat{\pi}^{\mathcal{E}}_{AA} = \mathrm{diag}\{0,0,\tilde{\pi}_\parallel^{LLL}\}$ with the corresponding new polarization vectors given by
\begin{subequations}
\begin{align}
&\epstil_{1}^{\mu}=(0,1,0,0),\\
&\epstil_{2}^{\mu}=\frac{\omega^{2}\sin\theta_{B}}{\sqrt{\omega^{2}-\modk^{2}}\sqrt{\omega^{2}-\modk^{2}\cos^{2}\theta_{B}}} \left(\frac{\modk}{\omega},0,(1-\frac{\modk^{2}}{\omega^{2}})\cot\theta_{B},1\right),\\
&\epstil_{3}^{\mu}=\frac{\omega\cos\theta_{B}}{\sqrt{\omega^{2}-\modk^{2}\cos^{2}\theta_{B}}}\left(\frac{\modk}{\omega},0,-\tan\theta_{B},1\right).
\end{align}
\end{subequations}
One can check that the these new polarization vectors, which are depicted in Fig.~\ref{fig:eigenmodes_orientation}, satisfy all the conditions in \Eq{polvec-cond}. In fact, the dielectric tensor for such a plasma can be evaluated in the $\big|{\vec{k}}\big|\rightarrow0$ limit from \Eq{dielec} to be 
\begin{equation}\label{dielec_supB}
\eps^i_j=\left[\begin{array}{ccc}
1 & 0 & 0\\
0 & 1-\frac{\real[\tilde{\pi}_{\parallel}]}{\omega^{2}}\sin^{2}\theta_{B} & \frac{\real[\tilde{\pi}_{\parallel}]}{\omega^{2}}\sin\theta_{B}\cos\theta_{B}\\
0 & \frac{\real[\tilde{\pi}_{\parallel}]}{\omega^{2}}\sin\theta_{B}\cos\theta_{B} & 1-\frac{\real[\tilde{\pi}_{\parallel}]}{\omega^{2}}\cos^{2}\theta_{B}
\end{array}\right],
\end{equation}
which has a similar form to what has been used in previous literature dealing with axion conversion in NS magnetospheres (see e.g. Refs.~\cite{Hook:2018iia,Noordhuis:2022ljw,Millar:2021gzs}). One important difference with earlier works is that in LWL, the Hermitian part of the mixing matrix is proportional to $\real[\tilde{\pi}_{\parallel}^{LLL}]$ rather than the result $\omega_p^2/\langle \gamma_\parallel \rangle$ obtained from classical kinetic theory \cite{Gurevich_Beskin_Istomin_1993, Millar:2021gzs}, where the plasma frequency $\omega_p$ in a relativistic, magnetized plasma is defined in terms of the Landau levels as
\begin{equation}
\omega_{p}^{2}=\frac{e^{3}B}{4\pi^{2}}\sum_{n}\left(2-\delta_{0}^{n}\right)\int_{-\infty}^{\infty}dq_{\parallel,b}\frac{f_{e}\left(E_{q}^{n}\right)+f_{\bar{e}}\left(E_{q}^{n}\right)}{E_{q}^{n}}\left(1-\frac{q_{\parallel,b}^{2}+2neB}{3\left(E_{q}^{n}\right)^{2}}\right)
\end{equation}
and $\langle \gamma_\parallel\rangle$ is the thermal average of the Lorentz factor $\gamma_\parallel = 1/\sqrt{1-v_\parallel^2}$. In the $b\gg1$ regime, the vacuum contribution $\real[{\pi}_{\parallel}^{BV}]$ dominates for relativistic particles (see \Fig{fig:ratio_piBV_wp} in \App{App:vac}) and overall $\real[\tilde{\pi}_{\parallel}]$ is negative, indicating that resonant mixing with BSM particles is not possible~\cite{Raffelt:1987im}. For the thermal medium contribution, our numerical results show that $\real[{\pi}_{\parallel}^{LLL}]$ decreases more gradually compared to $\omega_p^2/\langle \gamma_\parallel \rangle$ for $T \gg m_e$ in the presence of a super-critical magnetic field, as illustrated in \Fig{fig:piLLL}.

\begin{figure}[t]
\centering
\includegraphics[width=.55\textwidth]{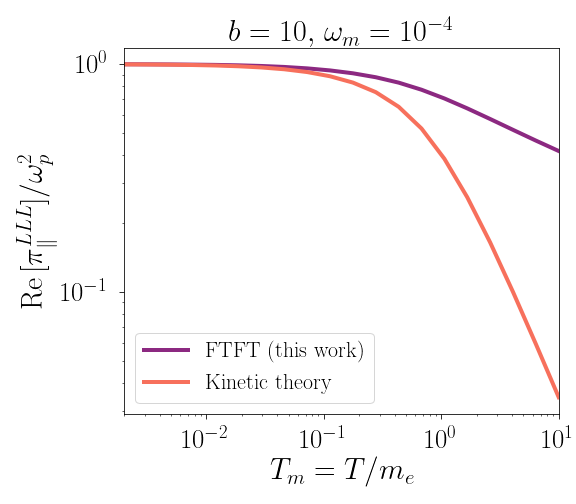}
\qquad
\caption{Form factors in a super-critical magnetized plasma are shown as a function of $T_m= T/m_e$ for $b=10$ and $\omega_m = 10^{-4}$. At high temperatures, the results obtained using FTFT methods deviates substantially from those obtained using classical kinetic theory~\cite{Gurevich_Beskin_Istomin_1993}.} 
\label{fig:piLLL}
\end{figure}

\subsection{Sub-critical magnetic fields}
Below the critical magnetic field threshold ($b=eB/m_e^2\ll 1$), we must keep the sum over the HLL. In a magnetized plasma, the electron (or positron) number density can be written as a sum over Landau levels as \cite{Kernan:1995bz, 1991ApJ...383..745L}
\begin{equation}\label{neB}
  n_{e/\bar{e}}(T,\mu,B)=\frac{eB}{4\pi^2}\sum_{n=0}^\infty (2-\delta^n_0)\int_{-\infty}^{\infty} dq_{\parallel,b} f_{e/\bar{e}}(E_q^n).
\end{equation}

In general, summing over the Landau levels in Eqs.~\eqref{Re-pi-i} and \eqref{neB} is analytically challenging and poses issues with numerical convergence. However, in the sub-critical field limit, the relative change in energy between neighbouring Landau levels is small,
\begin{equation}
  \frac{1}{E_n}\frac{\Delta E_n}{\Delta n}\sim \frac{eB}{E_n^2} < b \ll 1,
\end{equation}
where the second relation holds since $E_n\geq m_e$. We can therefore approximate the sum over $n$ as an integral,
\begin{equation}
  \sum_n \rightarrow \int \frac{d^2 \vec{q}_\perp}{2\pi eB}.
\end{equation}
Here $\vec{q}_{\perp}^{\,2} \equiv 2neB$ can be thought of as the momentum transverse to the magnetic fields (so that $\vec{q}_{\perp}$ is a 2D vector), and the Landau energies become $E(q_{\parallel,b}, \vec{q}_{\perp})=\sqrt{m_e^2+q_{\parallel,b}^2+\vec{q}_{\perp}^2}$. Thus, all relevant quantities can be expressed as 3D-integrals: one over the longitudinal momentum and the other two over the two components of the transverse momentum. However, evaluating these integrals numerically to obtain the form factors and subsequently diagonalizing the mixing matrix is computationally demanding. To facilitate further analytical progress, in the discussion below we systematically explore various limiting cases involving different plasma properties.

Since the dispersion of the LLL is qualitatively different from the higher Landau levels, we decompose the electron number density as $n_e=n_e^{(n=0)}+n_e^{(n>0)}$ with 
\begin{subequations}\label{neB2}
\begin{align}
&\label{neB_0} n_{e}^{(n=0)}=\frac{eB}{2\pi}\int_{-\infty}^{\infty}\frac{dq_{\parallel,b}}{2\pi}f_{e}(E_{0}),\\
&\label{neB_neq0}n_{e}^{(n>0)}=2\int_{\sqrt{2eB}}^{\infty}\frac{d^{2}\vec{q}_{\perp}}{(2\pi)^{2}}\int_{-\infty}^{\infty}\frac{dq_{\parallel,b}}{2\pi}f_{e}(E)
\end{align} 
\end{subequations}
where the lower limit of the $q_\perp$-integration in \Eq{neB_neq0} starts at a value corresponding to $n=1$. Note the use of cylindrical coordinates because of the asymmetry between the quantized transverse and continuous longitudinal momenta. The integrands in the expressions of the form factors can be further simplified. For example, the integrands in \Eq{piperp-n-1} and \Eq{picross-n-1} can be expanded (for $n \geq 1$) as
\begin{multline}\label{simplify}
\text{$\perp$ form factor integrand:}\\
\sum_{s=+,-}\left(2neB+s\omega E_{q}^{n}\right)\left\{ \frac{1}{\omega^{2}+s2\omega E_{q}^{n}-2eB}+\frac{1}{\omega^{2}+s2\omega E_{q}^{n}+2eB}\right\}\\
=\left(\omega^{2}\left(\left(E_{q}^n\right)^{2}-neB\right)+2ne^{2}B^{2}\right)C_- +\left(\omega^{2}\left(\left(E_{q}^n\right)^{2}-neB\right)-2ne^{2}B^{2}\right)C_+
\end{multline}
\begin{multline}
\text{$\times$ form factor integrand:}\\
\sum_{s=+,-}s\left(2neB+s\omega E_{q}^{n}\right)\times\left\{ \frac{1}{\omega^{2}+s2\omega E_{q}^{n}-2eB}-\frac{1}{\omega^{2}+s2\omega E_{q}^{n}+2eB}\right\} \\
=\omega E_{q}^n\left(-\left(\frac{\omega^{2}}{2}-eB\right)+2neB\right)C_-+\omega E_{q}^n\left(\left(\frac{\omega^{2}}{2}+eB\right)-2neB\right)C_+
\end{multline}
where
\begin{equation}
C_{\mp}(E_{q}^n)=\frac{1}{\left(\omega E_q^n\right)^{2}-\left(\frac{\omega^{2}}{2}\mp eB\right)^{2}}.
\end{equation}

\subsubsection{Quasi-isotropic and sharp peak approximations}\label{sec:quasi_isotropic}
The LLL is the state of maximum anisotropy since the motion of charged particles is constrained to follow specific cyclotron orbits along magnetic field lines with negligible transverse motion ($\vec{q}_{\perp}^{\,2} \rightarrow 0$) \cite{Gusynin:1995nb,Hattori:2022uzp}. Hence the ratio $\zeta \equiv n_e^{(n= 0)}/n_e^{(n>0)}$ serves as a rough indicator of anisotropy in the magnetic plasma. As shown in \Fig{fig:zeta}, the LLL occupancy is negligible for $T_m\equiv T/m_e\gtrsim b$, which can be intuitively attributed to the isotropization of the plasma for temperatures higher than the cyclotron frequency, $\omega_B (=eB/m_e)$, since higher temperatures enhance the random motion of background particles. Thus, for $T_m\lesssim b$, the LLL number density is significant in the low-temperature and non-degenerate regime and hence has to be taken into account appropriately. Conversely, in high-$T$ or degenerate plasmas, the contribution from the LLL can be safely ignored and henceforth we will refer to such plasmas as \textit{quasi-isotropic}. 

The electron number density in such restored, quasi-isotropic plasmas can be expressed as 
\begin{equation}\label{neBrel}
  n_{e/\Bar{e}}(T,\mu,B)=2\int \frac{d^3q}{(2\pi)^3} f_{e/\Bar{e}}(E)
\end{equation}
where now the momentum integral measure can be converted to spherical coordinates with $\vec{q}_{\perp,b}^2 = q^2\sin^2 \vartheta$ and $q_{\parallel,b}^2 = q^2\cos^2 \vartheta$ where $\vartheta$ is the angle between the magnetic field and the momentum of the background particles. 

\begin{figure}[t]
\centering
\includegraphics[width=.495\textwidth]{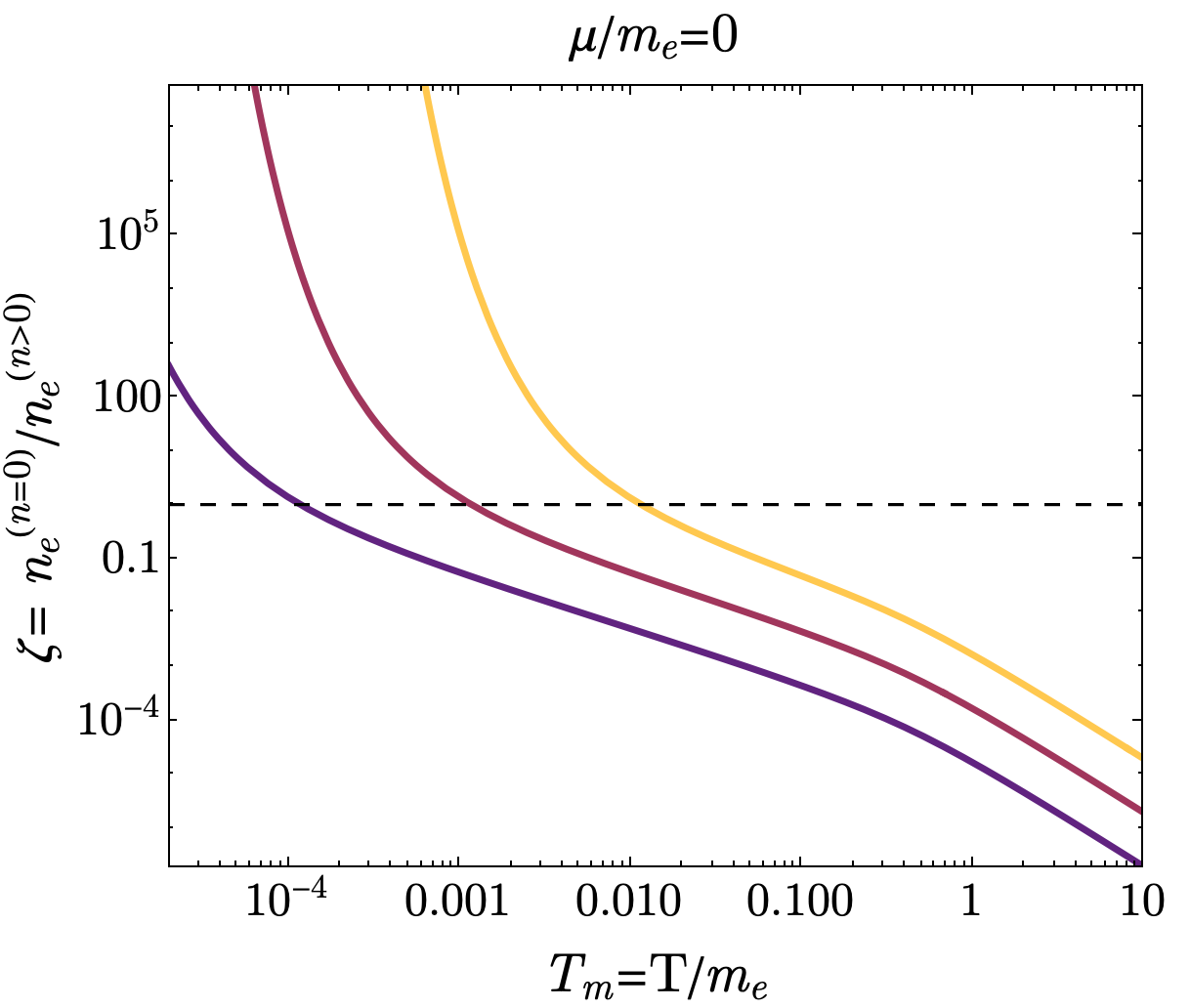}
\includegraphics[width=.495\textwidth]{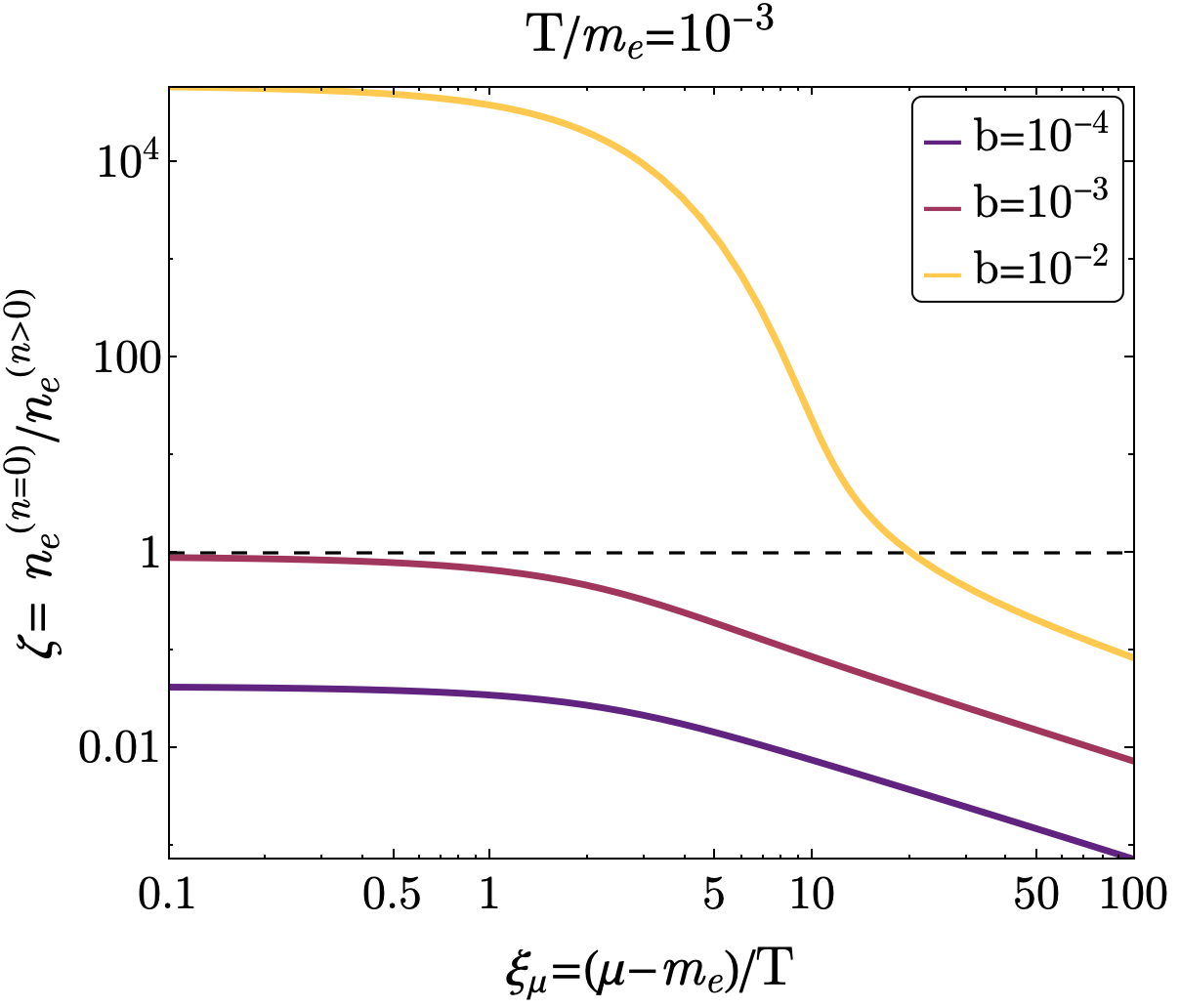}
\caption{Ratio of $\zeta = n_e^{(n> 0)}/ n_e^{(n=0)}$ in the sub-critical field limit as as function of $T_m=T/m_e$ in a non-degenerate plasma ($\mu/m_e = 0$; left) and as a function of $\xi_\mu=(\mu-m_e)/T$ in a low-temperature plasma ($T/m_e = 10^{-3}$; right). This $\zeta$ parameter tracks how the anisotropy depends on the relativistic and degenerate nature of the plasma.
\label{fig:zeta}}
\end{figure}

In this quasi-isotropic regime, the form factors in a high-$T$ or high-$\mu$ plasma (ignoring the LLL contribution) can be expressed in spherical coordinates by replacing $2neB \rightarrow q^2 \sin^2\vartheta$ and performing the angular integrals. We obtain
\begin{subequations}\label{pi_quasi_iso}
\begin{equation}\label{piperprel}
\real[\pi_{\perp}]\approx 2 e^{2}\int\frac{dq}{(2\pi)^{2}}\left[f_{e}+f_{\bar{e}}\right]\frac{q^2}{E}\left[\omega^{2}\left(E^{2}-\frac{q^{2}}{3}\right)\left(C_-+C_+\right)
+\, eB \frac{2 q^{2}}{3}(C_- - C_+)\right]
\end{equation}
\begin{equation}\label{picrossrel}
\real[\pi_{\times}]\approx 2 e^{2}\int\frac{dq}{(2\pi)^{2}}\left[f_{e}-f_{\bar{e}}\right]\frac{q^2}{E}\left[(\omega E)(eB)\left(C_-+C_+\right) -\omega E\left(\frac{\omega^{2}}{2}-\frac{2 q^{2}}{3}\right)\left(C_--C_+\right)\right]
\end{equation}
\begin{equation}\label{pipararel}
\real[\pi_{\parallel}]\approx 4 e^{2}\int\frac{dq}{(2\pi)^{2}}\left[f_{e}+f_{\bar{e}}\right]\frac{q^2}{E}\frac{(\frac{2q^{2}}{3} + m_e^2)}{\left(E^{2}-\omega^{2}/4\right)}.
\end{equation}
\end{subequations}
Notably, in this regime, the $\pi_\parallel$ form factor becomes independent of the magnetic field strength. 

Further simplifications in the quasi-isotropic regime can be done if we consider low ($\omega^2 \ll eB$) or high ($\omega^2 \gg eB$) frequency limits compared to the magnetic field: 
\begin{subequations}
\begin{equation}
C_{-}+C_{+}\approx\begin{cases}
\frac{2}{\left(\omega^{2}E^{2}-e^{2}B^{2}\right)} & \omega^{2}\ll eB\\
\frac{2}{\omega^{2}\left(E^{2}-\omega^{2}/4\right)} & \omega^{2}\gg eB
\end{cases}
\end{equation}
\begin{equation}
C_{-}-C_{+}\approx\begin{cases}
\frac{-2\omega^{2}eB}{\left(\omega^{2}E^{2}-e^{2}B^{2}\right)^{2}} & \omega^{2}\ll eB\\
\frac{-2eB}{\omega^{2}\left(E^{2}-\omega^{2}/4\right)^{2}} & \omega^{2}\gg eB,
\end{cases}
\end{equation}
\end{subequations}
which when implemented reduces the form factors to
\begin{equation}\label{pi_perp_quasi_iso}
\pi_{\perp}\left(\omega,B\right)\approx\begin{cases}
\frac{e^{2}}{\pi^{2}}\int dq\left[f_{e}+f_{\bar{e}}\right]\frac{q^{2}}{E}\left[\frac{E^{2}-\frac{q^{2}}{3}}{E^{2}-\left(\frac{eB}{\omega}\right)^{2}}-\frac{2}{3}\left(\frac{eB}{\omega}\right)^{2}\frac{q^{2}}{\left(E^{2}-\left(\frac{eB}{\omega}\right)^{2}\right)^{2}}\right] & \omega^{2}\ll eB\\
\frac{e^{2}}{\pi^{2}}\int dq\left[f_{e}+f_{\bar{e}}\right]\frac{q^{2}}{E}\left[\frac{E^{2}-\frac{q^{2}}{3}}{E^{2}-\frac{\omega^{2}}{4}}-\frac{2}{3}\left(\frac{eB}{\omega}\right)^{2}\frac{q^{2}}{\left(E^{2}-\frac{\omega^{2}}{4}\right)^{2}}\right] & \omega^{2}\gg eB
\end{cases}
\end{equation}

\begin{equation}\label{pi_cross_quasi_iso}
\pi_{\times}\left(\omega,B\right)\approx\begin{cases}
\frac{e^{2}}{\pi^{2}}\frac{eB}{\omega}\int dq\left[f_{e}-f_{\bar{e}}\right]q^{2}\left[\frac{1}{E^{2}-\left(\frac{eB}{\omega}\right)^{2}}+\frac{\frac{\omega^{2}}{2}-\frac{2q^{2}}{3}}{\left(E^{2}-\left(\frac{eB}{\omega}\right)^{2}\right)^{2}}\right] & \omega^{2}\ll eB\\
\frac{e^{2}}{\pi^{2}}\frac{eB}{\omega}\int dq\left[f_{e}-f_{\bar{e}}\right]q^{2}\left[\frac{1}{E^{2}-\frac{\omega^{2}}{4}}+\frac{\frac{\omega^{2}}{2}-\frac{2q^{2}}{3}}{\left(E^{2}-\frac{\omega^{2}}{4}\right)^{2}}\right] & \omega^{2}\gg eB.
\end{cases}
\end{equation}
Evidently, the high-frequency limit can be obtained by replacing $eB \rightarrow \omega^2/2$ in the denominators of the low-frequency expressions. Moreover, we can also take the $B\rightarrow 0$ limit in the quasi-isotropic expression in \Eq{pi_quasi_iso} which further reduces (for $\omega \ll m_e$) to
\begin{equation}\label{pi_bequal0}
\pi_{\perp}\approx \pi_\parallel \rightarrow \omega_p^2 = 2 e^{2}\int_{0}^{\infty}\frac{dq}{2\pi^{2}}\left[f_{e}(E)+f_{\bar{e}}(E)\right]\frac{q^{2}}{E}\left[1-\frac{q^{2}}{3 E^2}\right], \quad
\pi_\times \approx \mathcal{O}\left(\frac{eB}{\omega m_e}\right)
\end{equation}
and which matches well with the well-known form factors for transverse and longitudinal modes in an ordinary isotropic plasma without any magnetic fields (see Ref.~\cite{Braaten:1993jw}). The gyrotropic term goes to zero linearly with the magnetic field strength.

To analytically estimate the integrals in \Eq{pi_quasi_iso} for the quasi-isotropic limit, building on the approximations developed in Ref.~\cite{Braaten:1993jw}, we can perform integration by parts to obtain
\begin{subequations}\label{pi_quasi_iso2}
\begin{align}
&\real\left[\pi_{a}\right]=-m_{e}^{2}\frac{e^{2}}{3\pi^{2}}\int_{0}^{1}dv\,\frac{v^{3}}{1-v^{2}}\frac{d}{dv}\left(f_{e}+f_{\bar{e}}\right)J_{a}(v),\qquad a=\perp, \parallel \\
&\real\left[\pi_{\times}\right]=-m_{e}^{2}\frac{e^{2}}{3\pi^{2}}\int_{0}^{1}dv\,\frac{v^{3}}{1-v^{2}}\frac{d}{dv}\left(f_{e}-f_{\bar{e}}\right)J_{\times}(v),
\end{align}
\end{subequations}
where we have changed the integration variable to $v=q/E$ and in the low-frequency regime ($\omega_m^2 \ll b$) the functions $J_a(v)$ are given by
\begin{subequations}\label{J_a}
\begin{align}
&J_{\perp}(v)=\frac{1}{1-\left(1-v^{2}\right)\rho_{m}^{2}},\\
&J_{\times}(v)=\rho_{m}\frac{3\left(1-v^{2}\right)}{v^{3}}\left[\sqrt{\frac{v^{2}}{1-v^{2}}}\frac{\frac{v^{2}}{3}-\frac{\omega_{m}^{2}}{4}\left(1-v^{2}\right)}{1-\left(1-v^{2}\right)\rho_{m}^{2}}-\frac{\omega_{m}^{2}}{4\sqrt{\rho_{m}^{2}-1}}\tanh^{-1}\left(\frac{\sqrt{\frac{1}{1-v^{2}}-1}}{\sqrt{\rho_{m}^{2}-1}}\right)\right],\\
&J_{\parallel}(v)=\frac{1-v^{2}}{v^{3}}\left[\frac{v}{1-v^{2}}+\frac{\omega_{m}^{2}}{2}\tanh^{-1}v-\left(1+\frac{\omega_{m}^{2}}{2}\right)\frac{\sqrt{4-\omega_{m}^{2}}}{\omega_{m}}\tan^{-1}\left(\frac{\omega_{m}v}{\sqrt{4-\omega_{m}^{2}}}\right)\right],
\end{align}
\end{subequations}
where we have defined $\rho_m \equiv b/\omega_m$. Utilizing the fact that $\frac{d}{dv}(f_e \pm f_{\bar{e}})$ is peaked at some velocity $v_*$, the integrals in \Eq{pi_quasi_iso2} can be evaluated by approximating the functions $J_a(v)$ at $v=v_*$ and subsequently pulling them out of the integral. This is known as the \textit{sharp peak approximation} \cite{Braaten:1993jw} and can be employed to simplify the quasi-isotropic form factors to
\begin{subequations}\label{pi_quasi_iso3}
\begin{align}
&\real\left[\pi_{a}\right]\simeq \omega_p^2 J_{a}(v_*),\qquad a=\perp, \parallel \\
&\real\left[\pi_{\times}\right]= \omega_{p\times}^2 J_{\times}(v_*),
\end{align}
\end{subequations}
where the plasma frequency $\omega_p$ is written in an alternate form using integration by parts in \Eq{pi_bequal0}:
\begin{equation}
  \omega_{p/p\times}^2 = -m_{e}^{2}\frac{e^{2}}{3\pi^{2}}\int_{0}^{1}dv\,\frac{v^{3}}{1-v^{2}}\frac{d}{dv}\left(f_{e} \pm f_{\bar{e}}\right).
\end{equation}
where the $+$ ($-$) sign in the phase space factors corresponds to the definition of $\omega_p$ ($\omega_{p\times}$). Following the prescription of Ref.~\cite{Braaten:1993jw}, we take $v_* = \omega_1/\omega_p$ in the LWL or small $k$ limit, with 
\begin{equation}
   \omega_{1}^2 = m_{e}^{2}\frac{e^{2}}{\pi^{2}}\int_{0}^{1}dv\,\frac{v^{2}}{(1-v^{2})^2}\left(\frac{5}{3}v^2 - v^4\right)\left(f_{e} + f_{\bar{e}}\right).
\end{equation}

\subsubsection{Classical regime}
For a classical plasma, the temperature and chemical potential are negligible compared to the electron mass. More precisely, the plasma is non-relativistic ($m_e \gg T$) and non-degenerate ($m_e-\mu \gg T$). Also, a classical plasma is generally an electron-ion plasma and hence we can ignore the positron contributions in our expressions for the form factors.
In this limit, the Fermi distribution for electrons reduces to a Boltzmann distribution 
\begin{equation}
  f_e(E_n,\mu) \propto e^{-(E_n-\mu)/T}.
\end{equation} 
In the classical, sub-critical ($b\ll 1$) limit, the Landau energy levels can be expressed as
\begin{equation}
  E_n \rightarrow E \approx m_e+\frac{q_{\parallel,b}^2}{2m_e}+n b m_e.
\end{equation}
To leading order, we can thus approximate $E_n \rightarrow m_e$ throughout the integrands, except within the momentum-dependent distribution functions. This simplifies the 
integrals over $q_{\parallel, b}$ in Eqs.~\labelcref{piperp-n-1,picross-n-1,pipara-n-1,neB} to Gaussian integrals. Also, the sum over $n$ \textit{can be done analytically} in this case (no need to approximate as an integral), since it converges rapidly due to the $e^{-nb/T_m}$ factor coming from the phase space distribution. Performing the integrals and the summation in \Eq{neB}, we obtain
\begin{subequations}\label{neB_classical}
\begin{align}
& n_e^{(n=0)}=\left(\frac{m_{e}T}{2\pi}\right)^{3/2}b\frac{m_e}{T}e^{-(m_{e}-\mu)/T} \\
& n_e^{(n> 0)}=2\left(\frac{m_{e}T}{2\pi}\right)^{3/2} \frac{b\frac{m_e}{T}}{e^{b m_e/T}-1}e^{-(m_{e}-\mu)/T}.
\end{align}
\end{subequations}
Summing these two contributions yields the usual nonrelativistic number density for $b\rightarrow 0$, as expected. From \Eq{neB_classical}, $\zeta \sim \left(e^{b/T_m}-1\right)$ becomes greater than 1 only when $b \gtrsim T/m_e$ (\textit{i.e.} the LLL number density becomes significant in more highly magnetized plasmas), as illustrated in \Fig{fig:zeta}.

\begin{figure}[t]
\centering
\includegraphics[width=.55\textwidth]{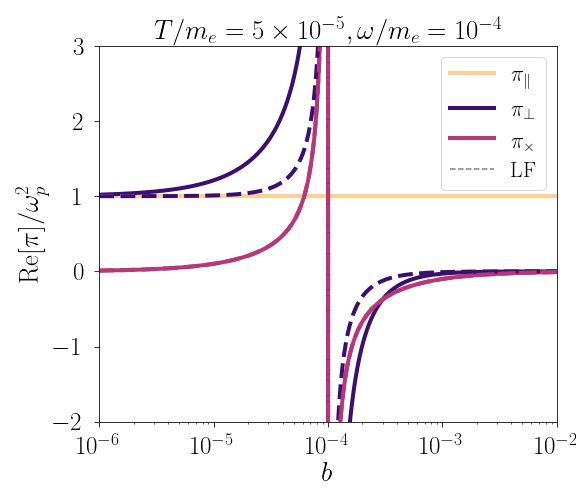}
\qquad
\caption{form factors in a classical magnetized plasma (in the limit $m_e - \mu \gg T$) as a function of the magnetic field value. The solid lines are the general expressions from \Eq{pi-classical} in the sub-critical field limit while the dashed lines are the low-frequency counterparts (note that for $\pi_\parallel$ and $\pi_\times$ the solid and dashed lines are not distinguishable). 
\label{fig:classical_modes_vs_b}}
\end{figure}

In the perpendicular form factor in \Eq{simplify}, the $\left(E_n^2-neB\right)$ term can be approximated to $m_e^2+neB$ by using the Landau energy expression in \Eq{landau_energy} and ignoring the negligible $q_\parallel^2/m_e^2$ term. Performing the Gaussian integrals, the form factors for the classical case reduce to
\begin{subequations}\label{pi-classical}
\begin{equation}
\real[\pi_{\perp}]\approx\frac{e^{2}n_{e}^{(n=0)}}{m_{e}}\omega_{m}^{2}c_{-}+\frac{e^{2}n_{e}^{(n>0)}}{2m_{e}}e^{b/T_{m}}\left\{ \omega_{m}^{2}\left(c_{-}+c_{+}\right)+\frac{2b^{2}}{e^{b/T_{m}}-1}(c_{-}-c_{+})\right\} 
\end{equation}
\begin{multline}
\real[\pi_{\times}]\approx\frac{e^{2}n_{e}^{(n=0)}}{m_{e}}\omega_{m}\left(b-\frac{\omega_{m}^{2}}{2}\right)c_{-}+\frac{e^{2}n_{e}^{(n>0)}}{2m_{e}}\left\{ \omega_{m}b\left(c_{-}+c_{+}\right)\right.\\
\left.+\omega_{m}\frac{\left(2b-\frac{\omega_{m}^{2}}{2}\right)e^{b/T_{m}}+\frac{\omega_{m}^{2}}{2}}{e^{b/T_{m}}-1}(c_{-}-c_{+})\right\} 
\end{multline}
\begin{equation}
\real[\pi_{\parallel}]\approx\frac{e^{2}/m_{e}}{1-\frac{\omega_{m}^{2}}{4}}\left(n_{e}^{(n=0)}+n_{e}^{(n>0)}\frac{(1+2b)e^{b/T_{m}}-1}{e^{b/T_{m}}-1}\right)
\end{equation}
\end{subequations}
where $\omega_m = \omega/m_e$ and 
\begin{equation}
c_{\mp}=\frac{1}{\omega_{m}^{2}-\left(\frac{\omega_{m}^{2}}{2}\mp b\right)^{2}}
\end{equation}
are the simplified form of the factors $C_{\mp}$. These form factors are depicted in Fig.~\ref{fig:classical_modes_vs_b} as a function of the rescaled magnetic field $b=eB/m_e^2$. For small enough $b$, the values of $\pi_\perp$ and $ \pi_\parallel$ reduce to their isotropic values corresponding to the plasma frequency $\omega_p^2=e^2n_e/m_e$~\cite{Braaten:1993jw} for a classical plasma, while the Faraday rotation term $\pi_\times$ vanishes as expected. For large $b$, only the longitudinal form factor, $\pi_\parallel$ survives due to the dimensional reduction in the presence of strong magnetic fields~\cite{Gusynin:1995nb}. 

{For a more detailed comparison between the magnetized plasma eigenmodes and the corresponding isotropic case, we plot the ratio $\pi^I_{AA}/\omega_p^2$ in the \(\omega_m\)-\(b\) and \(T_m\)-\(b\) planes in \Fig{fig:classical_modes_contour}. We observe that at relatively low frequencies (\(\omega_m \lesssim b\)), the form factors become non-positive, excluding the possibility of level-crossing and resonant conversion. However, for \(\omega_m \gtrsim b\), the form factors become positive, and the mostly transverse form factors \(\pi_1\) and \(\pi_2\) deviate significantly from the plasma frequency in regions with sufficiently strong magnetic fields (\(b \gtrsim T_m\)). This behavior is expected, as magnetic fields enhance anisotropy where as high temperatures tend to restore isotropy. The mostly longitudinal form factor remains roughly congruent with the isotropic result.}

\begin{figure}[t!]
\includegraphics[width=\textwidth]{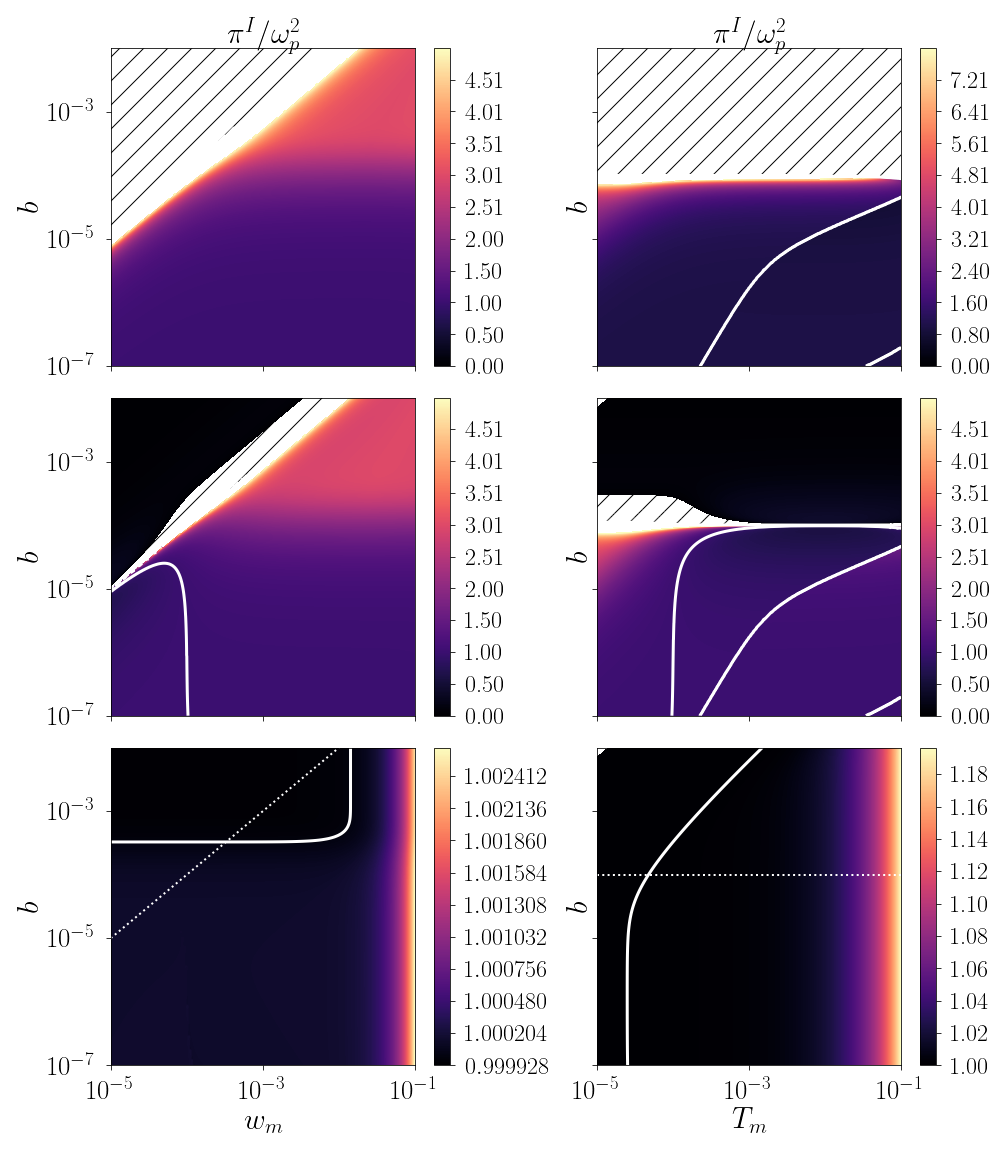}
\vspace{-0.7cm}
\caption{The contour plot of the ratio of $\pi^I/\omega_p^2$ where $\pi^I$ is the $I$th eigenvalue of the plasma mixing matrix with $I=1,2,3$. The left column corresponds to the $\omega_m$–$b$ plane with $T_m = 10^{-4}$ while the right column depicts the $T_m$–$b$ plane with $\omega_m = 10^{-4}$. All panels are computed assuming $n \equiv k/\omega = 10^{-2}$ and $\theta_B = \pi/4$. The pole at $b \sim \omega_m$ appears as a white dotted line in each panel while the white solid line indicates the contours where the anisotropic result matches the isotropic one with $\pi^I = \omega_p^2$. In the hatched regions, $\pi^I <0$ such that level-crossings with BSM particles are kinematically forbidden.\label{fig:classical_modes_contour}}
\end{figure}

In the {low-frequency} regime where
\begin{equation}
  \omega_m^2\ll b \ll 1,
\end{equation}
we have $c_-\approx c_+$, and the low-frequency form factors further reduce to
\begin{equation}
\label{pi-classical-lowf}
\mathrm{Re}[\pi_{\perp}^{\mathrm{LF}}]\approx\frac{\omega_{p}^{2}\omega_{m}^{2}}{\omega_{m}^{2}-b^{2}},\quad\mathrm{Re}[\pi_{\times}^{\mathrm{LF}}]\approx\frac{\omega_{p}^{2}\omega_{m}b}{\omega_{m}^{2}-b^{2}},\quad\mathrm{Re}[\pi_{\parallel}^{\mathrm{LF}}]\approx\omega_{p}^{2}.
\end{equation}
The pole at $\omega_m \sim b$ (or $\omega_B\sim \omega$) apparent in the low-frequency expressions of \Eq{pi-classical-lowf} corresponds to the well-known phenomenon of \textit{cyclotron resonance} \cite{stix1962theory, melrose1986instabilities, DGSwanson:2003} and is depicted in \Fig{fig:classical_modes_vs_b}. These classical, sub-critical ($b\ll 1$) low-frequency expressions correspond to the standard cold, magnetized plasma dielectric functions~\cite{stix1962theory,melrose1986instabilities}, which are often employed in the literature on photons mixing with BSM particles in astrophysical environments. However, the low-frequency expressions may differ substantially from the full result in certain parts of phase space, for instance $\pi_\perp$ in the $\omega_m \gtrsim b$ region as depicted in \Fig{fig:classical_modes_vs_b}.

\begin{figure}[t!]
\includegraphics[width=\textwidth]{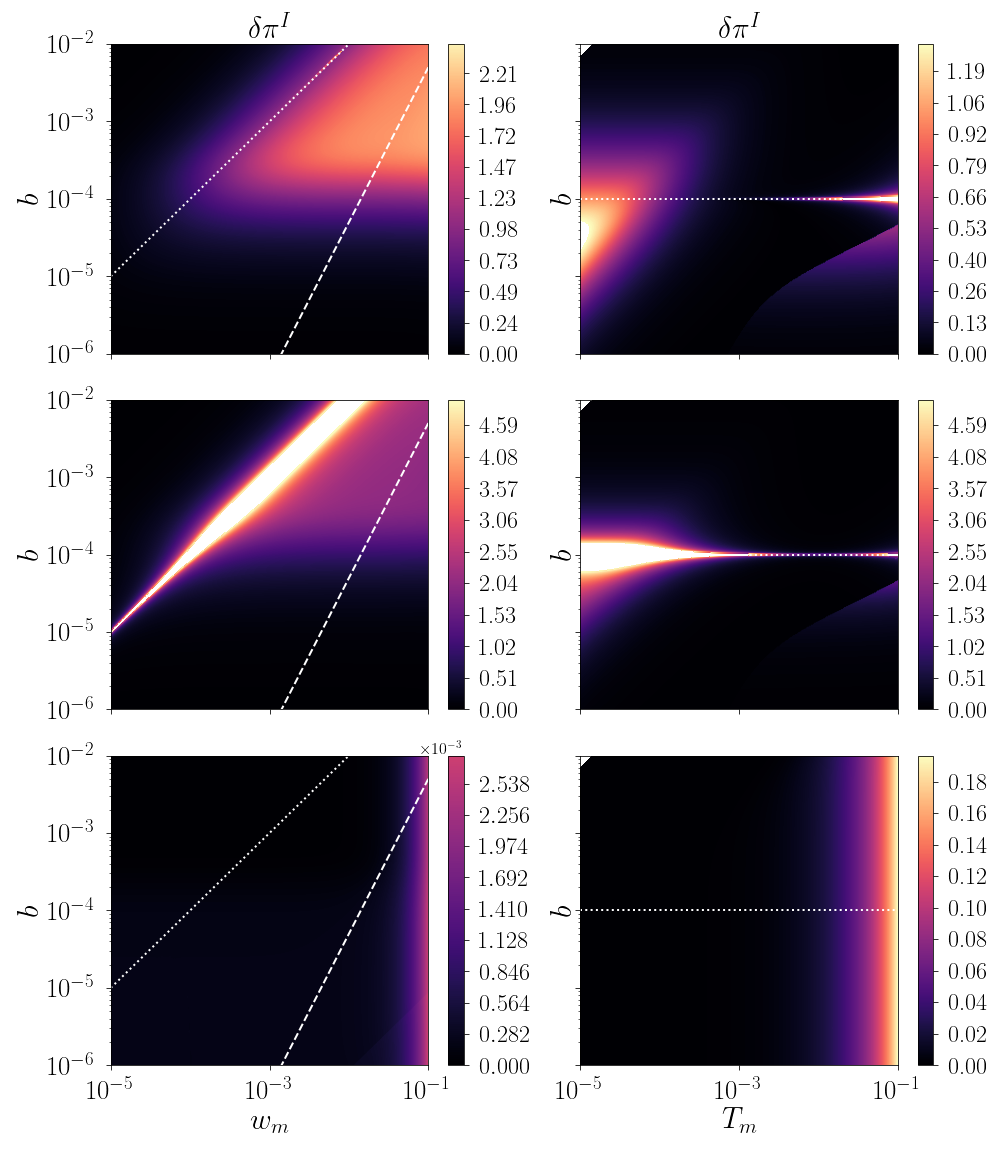}
\vspace{-0.7cm}
\caption{The relative difference $\delta\pi^I$ between the full eigenvalues of the plasma mixing matrix in the classical limit and those computed using the low-frequency approximation. The left column corresponds to the $\omega_m$–$b$ plane with $T_m = 10^{-4}$ while the right column depicts the $T_m$–$b$ plane with $\omega_m = 10^{-4}$. All panels are computed assuming $n \equiv k/\omega = 10^{-2}$ and $\theta_B = \pi/4$. The pole at $\omega_m \sim b$ appears as a white dotted line in each panel. The white dashed line at $b = \omega_m^2/2$ indicates the boundary where the low-frequency expressions should be valid, applicable only in the region to the left of this line and above. \label{fig:classical_modes_contour_delta}}
\end{figure}

To asses the accuracy of the low-frequency approximation and the impact on determining the propagating plasma eigenmodes ($\pi^I_{AA}$), in Fig.~\ref{fig:classical_modes_contour_delta} we show the fractional difference $\delta \pi^I\equiv \left|\left(\pi_{AA}^I\right)/\left(\pi_{AA}^I\right)^\mathrm{LF}-1\right|$ between the eigenvalues of the plasma mixing matrix $\pi^{IJ}_{AA}$ computed in the classical regime using \Eq{pi-classical} and the low-frequency counterpart of this expression in \Eq{pi-classical-lowf} in the $\omega_m-b$ and $T_m-b$ planes. For the two non-degenerate, mostly transverse modes, \(\pi_1\) and \(\pi_2\), we observe significant deviations at sufficiently high frequencies (\(\omega_m \gtrsim b\)) and magnetic fields larger compared to the temperature (\(b \gtrsim T_m\)). This indicates that the low-frequency expressions indeed lose accuracy in these regions of phase space. In some regions, the differences exceed $\mathcal{O}(1)$, likely due to the omission of terms proportional to \((c_- - c_+)\) in \Eq{pi-classical-lowf}. For the mostly longitudinal mode \(\pi_3\), the low-frequency expression remains reliable in most of the parameter space.

Finally, in \Fig{fig:disp_classical}, we present the dispersion relations (computed numerically using \Eq{disp_theta0}) for a representative classical plasma with $T_m = 10^{-4}$, $\omega_p/m_e =10^{-6}$ and with the photon propagation direction parallel to the magnetic field ($\theta_B = 0$). In the presence of a strong magnetic field such that $\omega_B \gtrsim \omega_p$, the two degenerate transverse modes split. The effective mass shifts from the plasma frequency $ \omega_p$ in an unmagnetized plasma to $\omega_B$ and $\omega_p^2/\omega_B$ for the transverse modes $\pi_1$ and $\pi_2$, respectively. The longitudinal mode remains largely unaffected, with $\omega \sim \omega_p$ for $0 < k < \omega_p$.

\begin{figure}[t]
\centering
\includegraphics[width=.55\textwidth]{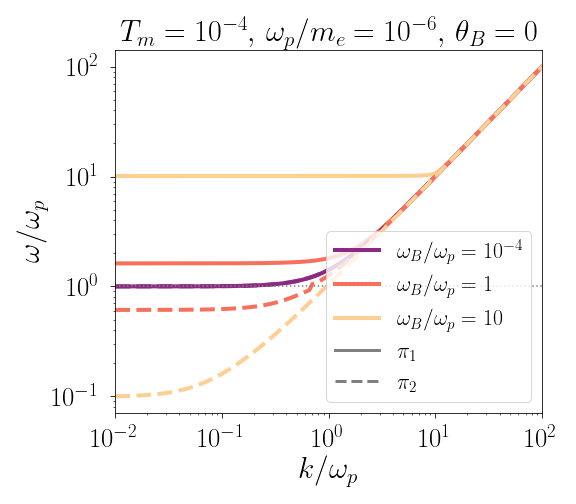}
\caption{The dispersion relations for the two transverse modes $\pi_1$ and $\pi_2$ in a classical magnetized plasma with $T_m = 10^{-4}$ and $\omega_p/m_e =10^{-6}$ for a photon propagating parallel to the magnetic field. 
\label{fig:disp_classical}}
\end{figure} 

\subsubsection{Degenerate regime}
In the degenerate limit, the chemical potential is much higher than the plasma temperature, $\xi_\mu=(\mu - m_e)/T \gg 1$. In this limit, the electron Fermi-Dirac distribution becomes a step-function and the positron distribution becomes negligible
\begin{equation}
  f_e(E) + f_{\bar{e}}(E) \approx \Theta(\mu - \sqrt{m_e^2+q_\parallel^2+q_\perp^2}).
\end{equation}
Substituting this form of the distribution function into Eqs.~\eqref{neB_0} and \eqref{neB_neq0}, we obtain the number densities for the LLL and HLL as follows
\begin{subequations}
\begin{align}
& n_e^{(n=0)}=\frac{2m_e^3 b}{4\pi^2}\sqrt{\mu_m^2 - 1} \\
& n_e^{(n>0)}=\frac{m_e^3}{3\pi^2}\left(\mu_m^2-1-2b\right)^{3/2}
\end{align}
\end{subequations}
where $\mu_m \equiv \mu/m_e$ is the non-dimensionalized chemical potential. Since $b\ll 1$, we can safely ignore the contributions of the LLL, as inferred from \Fig{fig:zeta}. The chemical potential can then be written in terms of the electron number density as
\begin{equation}
  \mu_m = \sqrt{1+\frac{(3\pi^2 n_e)^{2/3}}{m_e^2}}.
\end{equation}
Given the negligible LLL density, we can work in the quasi-isotropic limit and the phase space integrals can be written as
\begin{equation} 
\label{degenerate_approx}
\int\frac{d^{3}q}{(2\pi)^{3}}f_{e}(E)=\frac{1}{(2\pi)^{2}}\int_{0}^{q_{F}}dq_{\perp}q_{\perp}\int_{-\sqrt{q_F^2-q_{\perp}^{2}}}^{\sqrt{q_F^2-q_{\perp}^{2}}}dq_{\parallel} \\
=\frac{1}{(2\pi)^{2}}\int_{0}^{q_{F}}dq\,q^{2}\int_{0}^{\pi}d\vartheta\sin\vartheta
\end{equation}
where $q_F=\sqrt{\mu^2 - m_e^2}$ is the Fermi momentum. The integrals in \Eq{pi_quasi_iso} can then be done analytically. The results \textit{exactly matches} with the expressions in \Eq{pi_quasi_iso3} and \Eq{J_a} with $v_* = v_F$, where $v_F=q_F/E_F$ is the Fermi velocity.  

\begin{figure}[t]
\centering
\includegraphics[width=.479\textwidth]{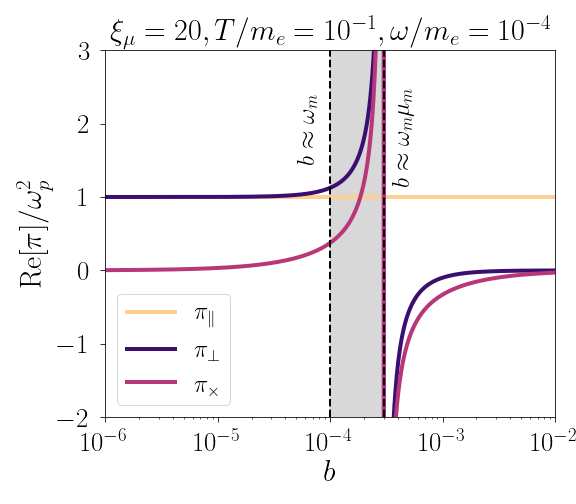}
\includegraphics[width=.49\textwidth]{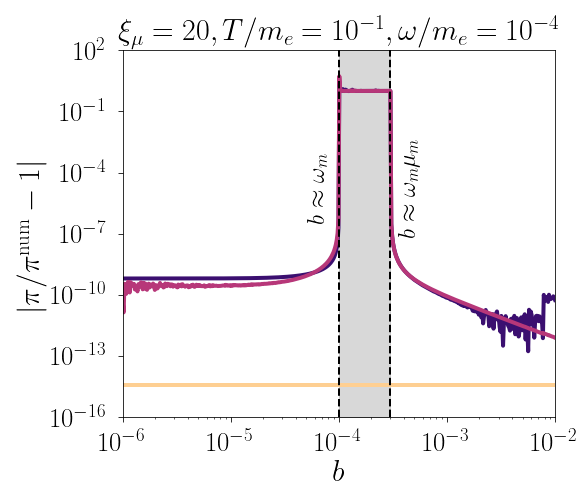}
\caption{Analytic form factors of a magnetized plasma in the degenerate regime (left) and the relative error between the analytic and numerical result (right). The grey regions are bounded by $b = \omega_m$ and $b = \omega_m\mu_m$, where the integral is highly oscillatory, leading to potentially large errors. }\label{fig:pi_degenerate}
\end{figure}

\begin{figure}[t]
\centering
\includegraphics[width=\textwidth]{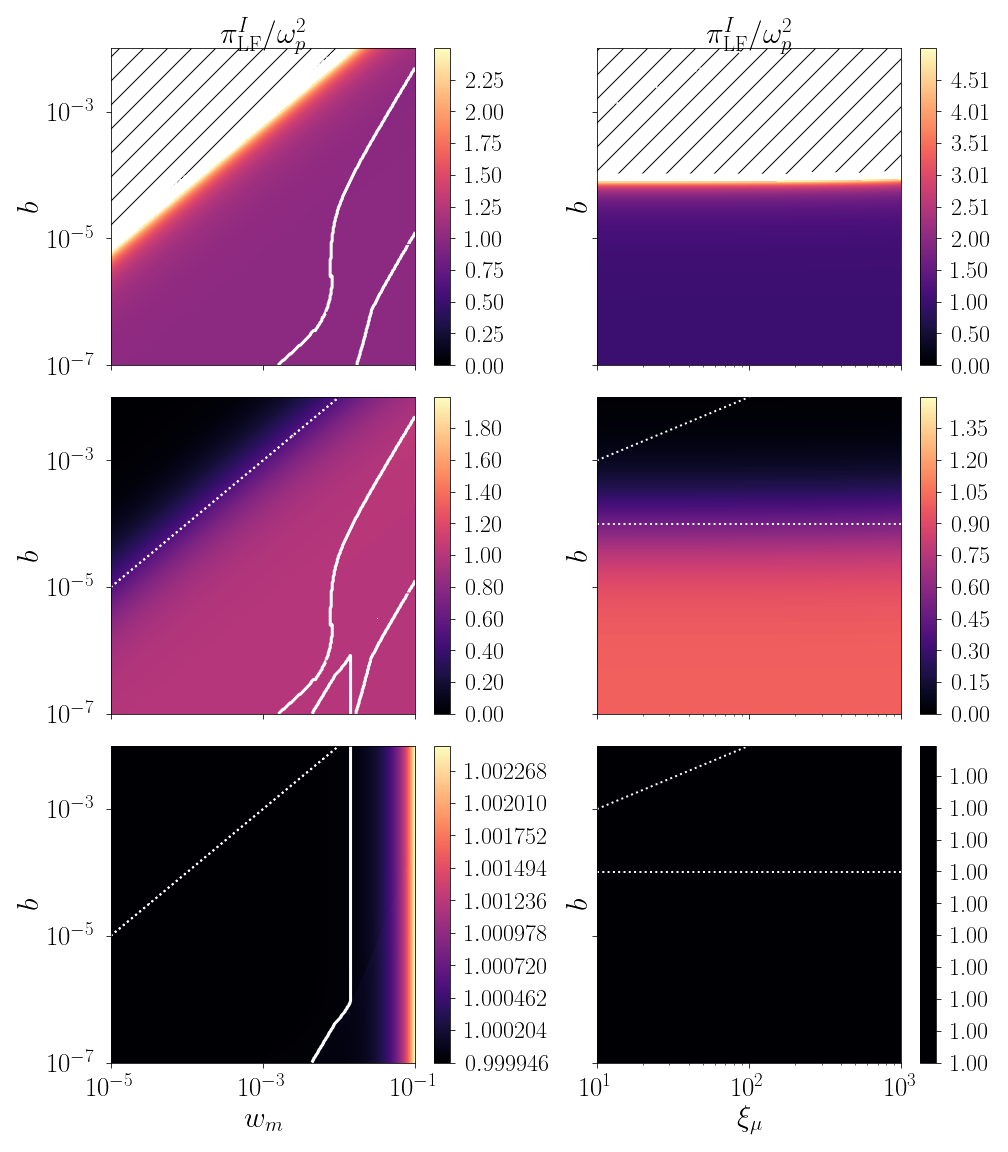}
\caption{Contour plots of the ratio $\pi^I_\mathrm{LF}/\omega_p^2$ in the degenerate regime are shown in the $\omega_m$–$b$ plane (left panels, with $\xi_\mu = 20$) and the $\xi_\mu$–$b$ plane (right panels, with $\omega_m = 10^{-4}$). All plots are generated for $n \equiv k/\omega = 10^{-3}$, $T_m = 10^{-4}$, $\theta_B = \pi/4$. The poles at $b \sim \omega_m$ and $b \sim \omega_m \mu_m$ appear as white dotted lines in each plot between which the analytic low-frequency expressions might not be accurate. Note that for the plots in the left panel $\mu_m = 1+\xi_\mu T_m \sim 1$ and hence both the white dotted lines coincide. \label{fig:degenerate_modes_contour_delta}}
\end{figure}

In \Fig{fig:pi_degenerate}, we compare our analytical estimates (for $\omega_m^2 \ll b$) from the sharp peak approximation with the result from numerical integration. The left panel of \Fig{fig:pi_degenerate} illustrates how the analytical form factors in a magnetized plasma vary with the magnetic field. The cyclotron resonance is shifted to $\omega \sim \omega_B/\mu_m$ in a degenerate plasma. For $b \lesssim \mu_m\omega_m$, the form factors approach their isotropic value, corresponding to the plasma frequency $\omega_{p}$ in degenerate plasmas, whereas the gyrotropic term vanishes. In contrast, for strong magnetic fields ($b \gtrsim \omega_m \mu_m$), the transverse form factors are suppressed, leaving only the parallel component ($\pi_\parallel$) as expected. Therefore, the isotropic results from Ref.~\cite{braaten1991calculation} should be applied with caution in degenerate, magnetized environments such as the interiors of white dwarfs. The analytic low-frequency expressions show excellent agreement with the numerical results across much of the parameter space, as demonstrated in the right panel of \Fig{fig:pi_degenerate}. However, in the region between $b \sim \omega_m $ and $b \sim \omega_m\mu_m$ (the shaded grey region in \Fig{fig:pi_degenerate}), the numerical integral oscillates rapidly, making it difficult to assess the accuracy of the analytical estimates from \Eq{pi_quasi_iso3}.

The differences between the normal modes of the magnetized degenerate plasma and the corresponding isotropic result are captured by the contour plots of the ratio \((\pi^I_{AA})_{\mathrm{LF}}/\omega_p^2\) in \Fig{fig:degenerate_modes_contour_delta}. We see significant departures for mixing matrix eigenvalues $\pi_1$ and $\pi_2$ for $b\gtrsim \omega_m$ due to the suppression of transverse modes in strong magnetic fields. The longitudinal eigenmode $\pi_3$ similar to the one derived assuming an isotropic medium, as expected from the results presented above. 

\subsubsection{Relativistic regime}
\label{sec:rel}

\begin{figure}[t]
\centering
\includegraphics[width=.55\textwidth]{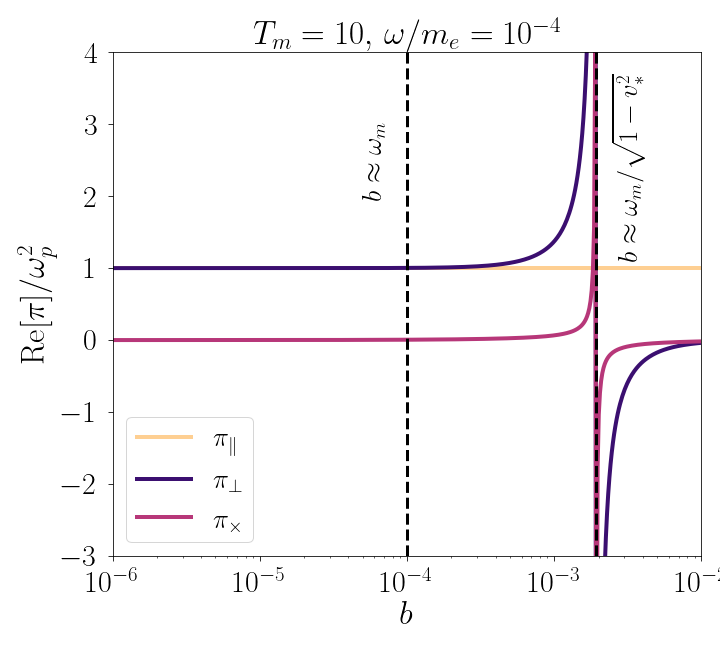}
\caption{form factors in a relativistic magnetized plasma (in the limit $\mu_m \ll 1$) as a function of the magnetic field value. The solid lines are the analytic expressions from \Eq{pi_quasi_iso3} with $v_* = \omega_1/\omega_p$ in the sub-critical field limit. The vertical dashed black lines highlight the location of the shifted pole at $b\sim \gamma_*\omega_m$ along with the location of the corresponding pole for a classical plasma at $b\sim \omega_m$.
\label{fig:rel_modes_vs_b}}
\end{figure}

In a relativistic plasma, either the temperature or chemical potential are higher than the electron rest mass and consequently we have $\frac{T}{m_e},\frac{\mu}{m_e}\gtrsim 1 \gg b$. In this regime, the occupancy of the LLL is always negligible (see \Fig{fig:zeta}) due to isotropization. Hence, the form factors in this relativistic case are exactly the same as in \Eq{pi_quasi_iso} with the Landau level energies now given by $E=\sqrt{m_e^2+q^2}$. 

In the ultra-relativistic limit ($T\gg m_e$ or $\mu \gg m_e$), we can even ignore the electron mass and the energy becomes $E\rightarrow \sqrt{q_\parallel^2+\Vec{q}^2_\perp} = |\vec{q}|$. The form factors in \Eq{pi_quasi_iso} thus reduce to much simpler 1D-integrals
\begin{subequations}\label{pi_urel}
\begin{align}
&\real[\pi_{\perp}]\approx e^{2}\int_{0}^{\infty}\frac{dq}{2\pi^{2}}\left[f_{e}(q)+f_{\bar{e}}(q)\right]\left[\omega^2 \left(C_-+C_+\right)+eB\left(C_--C_+\right)\right]\frac{2q^{3}}{3}\\
&\real[\pi_{\times}]\approx e^{2}\int_{0}^{\infty}\frac{dq}{2\pi^{2}}\left[f_{e}(q)-f_{\bar{e}}(q)\right]\left[\omega eB\left(C_-+C_+\right)-\omega\left(\frac{\omega^{2}}{2}-\frac{2q^{2}}{3}\right)\left(C_--C_+\right)\right]q^{2}\\
&\real[\pi_{\parallel}]\approx-2e^{2}\int\frac{dq}{2\pi^{2}}\left[f_{e}(q)+f_{\bar{e}}(q)\right]\frac{1}{\left(\omega^{2}-4q^{2}\right)}\frac{8q^{3}}{3}.
\end{align}
\end{subequations}

For $\omega_m^2 \ll b$, we can again use the sharp peak approximation results of \Eq{pi_quasi_iso3} with $v_* = \omega_1/\omega_p$. For illustration, in \Fig{fig:rel_modes_vs_b} we show the form factors for some typical relativistic plasmas. Notice that the cyclotron resonance has now shifted from $b \sim \omega_m$ in the classical plasma case to $b \sim \gamma_*\omega_m$ (or $\omega \sim \omega_B/\gamma_* $, where $\gamma_* = 1/\sqrt{1 - v_*^2}$). Despite this shift, the overall structure remains largely intact. This behavior is expected, as the cyclotron frequency in a relativistic plasma becomes $\omega_B/\langle\gamma\rangle$, consistent with similar findings derived from classical kinetic theory \cite{Gurevich_Beskin_Istomin_1993}. Given that $\gamma_* \gg 1$ in relativistic plasmas, the frequency at which the cyclotron resonance occurs is significantly reduced. In other words, as shown in Fig.~\ref{fig:rel_modes_vs_b}, the transverse modes can only be neglected (with their form factors approaching zero) for much higher values of the magnetic field as compared to the classical case. This implies that studies involving NS magnetospheres that ignore the transverse mode are possibly missing unaccounted-for resonances due to the relativistic nature of the plasma. Furthermore, large regions of NS magnetospheres can have sub-critical magnetic fields. Using the quasi-isotropic approximation of \Sec{sec:quasi_isotropic} with $v_*\rightarrow 1$ as appropriate for a relativistic plasma, the longitudinal form factor becomes $\pi_\parallel \sim \omega_p^2$, rather than $\omega_p^2/\langle\gamma\rangle$ as derived using the LLL approximation in Ref.~\cite{Gurevich_Beskin_Istomin_1993}, which is only valid in the super-critical limit. In plasmas where $\gamma \gg1$, this means that the resonance conditions can be very different from what has typically been assumed in the literature. In combination with the inclusion of the transverse mode, this could have profound implications for studies of axion and DP conversions in relativistic plasmas within neutron star magnetospheres.

These relativistic form factors could also prove invaluable for studying the early Universe, especially in the presence of cosmic magnetic fields \cite{Kernan:1995bz,Grasso:2000wj}, as well as relativistic plasmas in supernovae, or ultra-degenerate conditions in high-mass white dwarfs near the Chandrasekhar limit.

\section{Discussion and Conclusions}
\label{sec:conclusions}
In this article, we have demonstrated that the presence of magnetic fields induces substantial qualitative changes in the dispersive behaviour of photons in plasmas, which in turn affects BSM particle production in magnetized environments. The influence of magnetic fields in astrophysical plasmas is often not treated accurately or overlooked entirely. For instance, by omitting the transverse modes, many current studies involving axions in NS magnetospheres and white dwarfs may be missing resonances. These subtle issues highlight the need for a more comprehensive treatment in future research.

In this work, we have additionally provided detailed derivations of the photon self-energy tensor $\Po$ in a magnetized medium in the LWL using methods of FTFT. Per the recipe presented in Section~\ref{sec:recipe}, we can then project $\Po$ into mode space to derive the mixing matrix $\pi^{IJ}$ in the T/L basis (with the Hermitian part computed explicitly in \Eq{pimixB}), find the eigenvectors of this matrix to determine the plasma normal modes, and compute the emission rates of DPs and axions.

We have further taken various simplifying limits of the plasma properties (classical, degenerate, etc.), which will allow for fast evaluation of $\Po$ and the corresponding normal modes. This is significant because in any realistic astrophysical system, the ambient plasma will vary spatially, leading to normal modes whose polarizations and dispersion relations are rotated and shifted as particles propagate through. This qualitatively different picture may have implications for the phenomenology of BSM particles propagating in magnetized astrophysical environments, where there could be more opportunities for emission and level-crossing to occur.

\acknowledgments

It is a pleasure to thank Asher Berlin, Simon Caron-Huot, Aditya Chugh, Saniya Heeba, Yoni Kahn, Ani Prabhu, and Hugo Sch\'erer for useful conversations and correspondence pertaining to this work. NB was supported in part by a doctoral research scholarship from the Fonds de recherche du Qu\'ebec – Nature et technologies and by the Canada First Research Excellence Fund through the Arthur B. McDonald Canadian Astroparticle Physics Research Institute. NB and KS acknowledge support from a Natural Sciences and Engineering Research Council of Canada Subatomic Physics Discovery Grant and from the Canada Research Chairs program. KS thanks the Kavli Institute for the Physics and Mathematics of the Universe (IPMU) and the Kavli Institute for Theoretical Physics (KITP) (supported by grant NSF PHY-2309135) for their hospitality during the completion of this work. This analysis made use of \texttt{Numpy} \cite{harris2020array}, \texttt{Scipy} \cite{virtanen2020scipy}, \texttt{Matplotlib} \cite{hunter2007matplotlib}, and \texttt{Mathematica}~\cite{wolfram1991mathematica}.

\appendix
\section{Axion and DP self-energies from the thermal photon propagator}\label{App:thermal}
\begin{figure}[t]
\centering
\includegraphics[width=.47\textwidth]{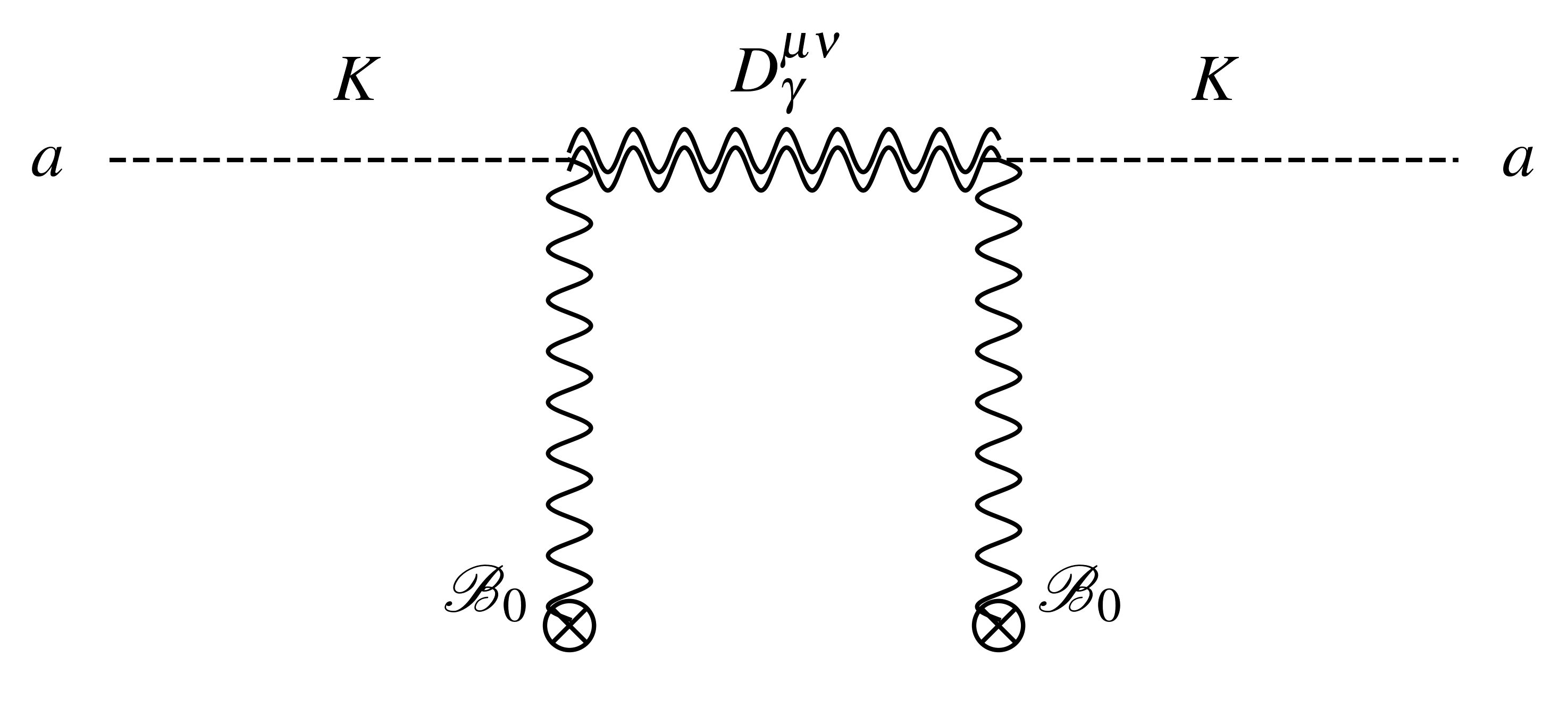}
\qquad
\includegraphics[width=.47\textwidth]{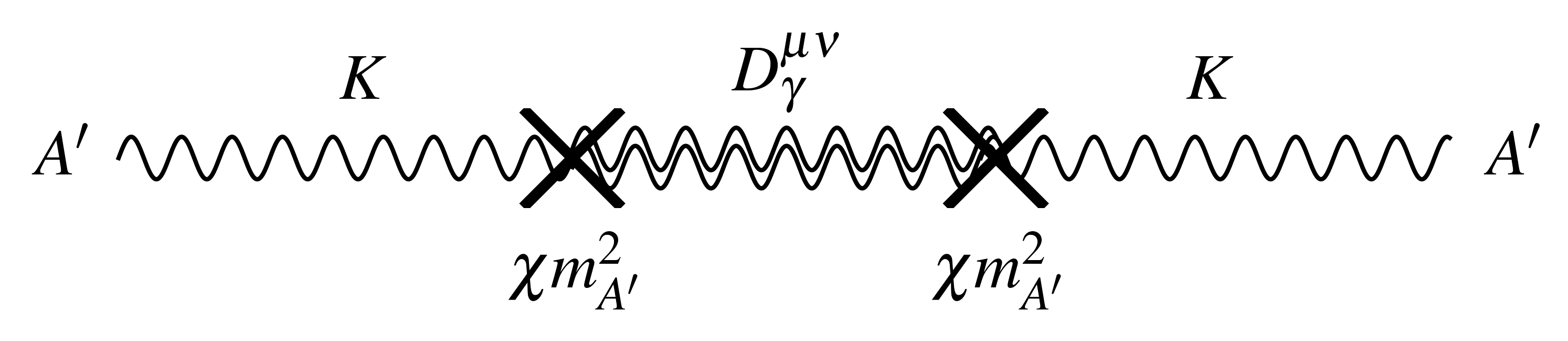}
\caption{Self-energy diagrams for axions $a$ (left panel) and DPs $\Ap$ (right panel) with the in-medium photon propagator $D^{\mu\nu}_\gamma$ represented by the wavy double lines. \label{fig:BSM_self_energy}}
\end{figure}

To leading order in the couplings, the contribution to the self-energies of axions and DPs in a medium are depicted in \Fig{fig:BSM_self_energy}. Via their coupling to photons, both self-energies can be written in terms of the full photon propagator $\mathcal{D}^{\mu\nu}_\gamma$ of the interacting theory as
\begin{align}
\label{self-energy-ax}
&\Pi_a (K) = m_a^2 + g_{a\gamma}^2 \fourvec{K}^\mu \fourvec{K}^\nu \tilde{F}^{0}_{\mu\alpha} \tilde{F}^{0}_{\nu\beta} D_\gamma^{\alpha\beta} \\
\label{self-energy-DP}
&\Po_{\Ap}(K) = \mAp^2 \delta^{\mu\nu} + \chi^2\mAp^4 D^{\mu\nu}_\gamma .
\end{align}
In FTFT, the inverse of the full photon propagator in a medium is given by \cite{Das:1997gg, Bellac:2011kqa, Kapusta:2006pm}
\begin{equation}
\left(D_{\gamma}\right)_{\mu\nu}^{-1}=\left(D_{\gamma}^{0}\right)_{\mu\nu}^{-1}-\Pi_{\mu\nu}
\end{equation}
where $D_{\gamma}^{0}$ is the vacuum photon propagator and can be written in the covariant gauge as~\cite{Kapusta:2006pm}
\begin{equation}
\left(D_{\gamma}^{0}\right)_{\mu\nu}^{-1}=K^{2}g_{\mu\nu}-\left(\frac{\xi_{g}-1}{\xi_{g}}\right)K_{\mu}K_{\nu}
\end{equation}
for gauge parameter $\xi_g$. In its eigenbasis, the photon self-energy is $\Po = \sum_{I=1}^3 \pi^I_{AA}\tilde{P}^{\mu\nu}_I$ where $\pi^I_{AA}$ are the eigenvalues and the eigen-projection tensors are expressed in terms of the eigenvectors $\eps^\mu_I$ (as described in \Sec{sec:2.2}) as $\tilde{P}_I^{\mu\nu}=\eps^\mu_I(\eps^\nu_I)^*$. The full photon propagator thus becomes
\begin{equation}
\label{photon-prop}
D^{\mu\nu}_\gamma(\fourvec{K})=-\sum_{I}\frac{\tilde{P}_{I}^{\mu\nu}}{\fourvec{K}^{2}-\pi_{AA}^{I}}+\xi_{g}\frac{\fourvec{K}^{\mu}\fourvec{K}^{\nu}}{\fourvec{K}^{4}} .
\end{equation}
Putting the above expression of $\mathcal{D}^{\mu\nu}_\gamma$ in the self-energy expressions of \Eq{self-energy-ax} and \Eq{self-energy-DP} yields the same result from \Eq{X-selfenergy}.

\section{Dispersive and absorptive parts of the self-energy} \label{App:dispersive}
In \Sec{NMS}, we demonstrated that the dispersion and damping equations of Eqs.~\eqref{A-dispersion1} and \eqref{X-dispersion1} are determined by the real and imaginary parts of the eigenvalues, $\pi^I_{AA}$, of the plasma mixing matrix $\pi^{IJ}_{AA}=-(\eps_\mu^{I})^* \Pi^{\mu\nu}_R \epsilon^{J}_\nu$. Given that the eigenvalues of a Hermitian (anti-Hermitian) matrix are purely real (imaginary), we obtain $\real\,\pi^I_{AA}$ ($\imag\,\pi^I_{AA}$) from the Hermitian (anti-Hermitian) part of $\pimix$, denoted as $\herm[\pimix]$ ($\Aherm[\pimix]$).

The Hermitian part of the mixing matrix originates from the Hermitian part of the self-energy,
\begin{equation}
\begin{split}\herm\left[\pimix\right] & =\frac{1}{2}\left[\pimix+\left(\pi_{AA}^{JI}\right)^{*}\right]\\
 & =\frac{1}{2}\left[-(\epsilon_{\mu}^{I})^*\Pi_{R}^{\mu\nu}\epsilon_{\nu}^{J}-(\epsilon_{\mu}^{I})^*\left(\Pi_{R}^{\nu\mu}\right)^{*}\epsilon_{\nu}^{J}{}\right]\\
 & =-\left(\epsilon_{\mu}^{I}\right)^*\herm\left[\Pi_{R}^{\mu\nu}\right]
 \epsilon_{\nu}^{J}
\end{split}
\end{equation}
and similarly for the anti-Hermitian part. Hence, it is \textit{not} the real (imaginary) part but the \emph{Hermitian} (\emph{anti-Hermitian}) part of the self-energy that describes the dispersion (damping). This distinction is only strictly moot when the plasma mixing matrix is expressed in its eigenbasis. 

Now, consider the self-energy to be decomposed into symmetric and anti-symmetric parts (the latter of which generally occurs in anisotropic plasmas, for instance the terms proportional to $P_\times = i\eps^{\mu\nu\alpha\beta}u_\alpha b_\beta$ projection tensor in a magnetized plasma), with projection tensors $S^{\mu\nu}$ (symmetric) and $iA^{\mu\nu}$ (anti-symmetric) such that
\begin{equation}
\Po=\pi_{S}S^{\mu\nu}+i\pi_{A}A^{\mu\nu}
\end{equation}
where $\pi_{S,A}$ are the corresponding complex-valued form factors. The factor of $i$ is to ensure that the antisymmetric part is Hermitian i.e. $\left(iA^{\nu\mu}\right)^*=iA^{\mu\nu}$. Then the dispersive and absorptive part of the self-energy are respectively given by
\begin{equation}\label{disp&abs}
\begin{aligned}\herm\left[\Po\right] & =\real[\pi_{S}]S^{\mu\nu}+i\real[\pi_{A}]A^{\mu\nu}\\
\Aherm\left[\Po\right] & =\imag[\pi_{S}]S^{\mu\nu}+i\imag[\pi_{A}]A^{\mu\nu}.
\end{aligned}
\end{equation}
Therefore, for the dispersive component, the symmetric (anti-symmetric) part corresponds to purely real (imaginary) form factors, while the absorptive component shows the opposite relationship.

\section{Vacuum contributions}\label{App:vac}
\subsection{Pure vacuum contribution}
For completeness and to establish our notation, we briefly outline the standard calculation of $\left(\Pi^V_{(11)}\right)^{\mu\nu}$ below.

The pure vacuum contributions to the photon self energy can be obtained by inserting only the vacuum part $S^0_V$ of the free electron propagator from \Eq{S_V} in \Eq{self1-noB}:
\begin{equation}\label{Pi_V}
\left(\Pi_{(11)}^{V}\right)^{\mu\nu}(K)=ie^{2}\int\frac{d^{4}Q}{(2\pi)^{4}}\frac{T^{\mu\nu}(Q,K)}{\left(\left(Q+K\right)^{2}-m^{2}+i\epsilon\right)\left(Q^{2}-m^{2}+i\epsilon\right)}
\end{equation}
where $T^{\mu\nu}$ is defined in \Eq{Tmunu} and is evaluated to be
\begin{equation}
T^{\mu\nu}(Q,K)=4\left[(K+Q)^{\mu}Q^{\nu}+Q^{\mu}(K+Q)^{\nu}-g^{\mu\nu}\left(Q\cdot\left(K+Q\right)-m^{2}\right)\right].
\end{equation}
Implementing the standard Feynman parameterization to combine the denominators in \Eq{Pi_V}, we obtain the following expression
\begin{equation}
\left(\Pi_{(11)}^{V}\right)^{\mu\nu}(K)=ie^{2}\int_{0}^{1}dx\int\frac{d^{d}Q}{(2\pi)^{d}}\Lambda^{2-d/2}\left.\frac{T^{\mu\nu}(Q,K)}{\left(\left(Q+xK\right)^{2}-\Delta_-\right)^{2}}\right|_{d\rightarrow4}
\end{equation}
where $\Delta_-(x,K)\equiv m^2 -x(1-x)K^2$ is a scalar function only dependent on the photon momentum $K$.
Here, the spacetime dimension has been changed from 4 to $d$ for the purpose of dimensional regularization, and appropriate factors of the scale parameter $\Lambda$ have been introduced to ensure the correct overall dimension.
Changing the momentum integral variable $Q\rightarrow (Q-xK)$, $T^{\mu\nu}$ becomes
\begin{equation}
T^{\mu\nu}(Q,K)=4\left[2Q^{\mu}Q^{\nu}-g^{\mu\nu}Q^{2}-S^{\mu\nu}\right]+\text{linear terms}
\end{equation}
where $S^{\mu\nu}$ is tensor function quadratic in $K$ but independent of $Q$ and is given by
\begin{equation}
S^{\mu\nu}(x,K)\equiv2x(1-x)K^{\mu}K^{\nu}-g^{\mu\nu}\Delta_{+}(x,K)
\end{equation}
with $\Delta_+(x,K) \equiv m^2 +x(1-x)K^2$.
Using the following $d$-dimensional momentum integrals 
\begin{subequations}\label{d_dim_int}
\begin{align}
\int\frac{d^{d}Q}{(2\pi)^{d}}\frac{1}{\left(Q^{2}-\Delta_{-}\right)^{n}} & =\frac{i(-1)^{n}}{\left(4\pi\right)^{d/2}}\frac{\Gamma\left(n-d/2\right)}{\Gamma(n)}\left(\frac{1}{\Delta_{-}}\right)^{n-d/2}\\
\int\frac{d^{d}Q}{(2\pi)^{d}}\frac{Q^{2}}{\left(Q^{2}-\Delta_{-}\right)^{n}} & =\frac{i(-1)^{n-1}}{\left(4\pi\right)^{d/2}}\left(\frac{d}{2}\right)\frac{\Gamma\left(n-1-d/2\right)}{\Gamma(n)}\left(\frac{1}{\Delta_{-}}\right)^{n-1-d/2}\\
\int\frac{d^{d}Q}{(2\pi)^{d}}\frac{Q^{\mu}Q^{\nu}}{\left(Q^{2}-\Delta_{-}\right)^{n}} & \frac{i(-1)^{n-1}}{\left(4\pi\right)^{d/2}}\left(\frac{g^{\mu\nu}}{2}\right)\frac{\Gamma\left(n-1-d/2\right)}{\Gamma(n)}\left(\frac{1}{\Delta_{-}}\right)^{n-1-d/2}
\end{align}
\end{subequations}
we can perform the integral in \Eq{Pi_V} to obtain
\begin{equation}
\left(\Pi_{(11)}^{V}\right)^{\mu\nu}(K)=\frac{e^{2}}{4\pi^{2}}\int_{0}^{1}dx\left(\frac{4\pi\Lambda}{\Delta_{-}}\right)^{\varepsilon}\left[g^{\mu\nu}\left(\varepsilon-1\right)\Gamma\left(\varepsilon-1\right)\Delta_{-}-S^{\mu\nu}\Gamma\left(\varepsilon\right)\right]\left.\right|_{\varepsilon\rightarrow 0}
\end{equation}
where we have defined $\varepsilon=2-d/2$. Taking the $\varepsilon \rightarrow 0$ limit to recover $4$-dimensional spacetime, we can write the vacuum polarization tensor as $\left(\Pi_{(11)}^{V}\right)^{\mu\nu} = \left(g^{\mu\nu}K^2 - K^\mu K^\nu \right)\Pi^V$, with the form factor $\Pi^V$ given by 
\begin{equation}
\Pi^{V}(K^2)=-\frac{e^{2}}{4\pi^{2}}\int_{0}^{1}dx\,2x(1-x)\left(\frac{1}{\varepsilon}-\gamma_{E}+\log\frac{4\pi\Lambda}{\Delta_{-}}\right)
\end{equation}
where $\gamma_E$ is the Euler-Mascheroni constant and the $1/\varepsilon$ pole arises due to the well-known logarithmic divergence in the momentum integral. Since only the $Q^2$-dependence of $\Pi^V$ is observable, it can be regularized by defining
\begin{equation}\label{reg_Pi_V}
\hat{\Pi}^V(K^2)\equiv \Pi^V(K^2) - \Pi^V(0)=-\frac{e^{2}}{4\pi^{2}}\int_{0}^{1}dx\,2x(1-x)\log\frac{m^{2}}{\Delta_{-}}
\end{equation}
such that $\hat{\Pi}^V(0)\rightarrow 0$ as needed for photons to be massless in vacuum.

\subsection{Magnetized vacuum contribution}
In analogy to the pure vacuum case, the photon self energy in a magnetized vacuum can be obtained by inserting the modified free vacuum electron propagator $S^0_{BV}$ from \Eq{S_BV} in \Eq{self1-noB}:
\begin{equation}\label{Pi_BV}
\left(\Pi_{(11)}^{BV}\right)^{\mu\nu}(K)=ie^2\sum_{n,l}\int\frac{d^{2}q_\parallel}{(2\pi)^{2}}\int\frac{d^{2}q_\perp}{(2\pi)^{2}}\frac{T^{\mu\nu}_{n,l}(Q,K)}{\left(\left(q_\parallel+k_\parallel\right)^{2}-m^{2}+i\epsilon\right)\left(q_\parallel^{2}-m^{2}+i\epsilon\right)}
\end{equation}
where now the modified trace term $T^{\mu\nu}_{n,l}$ is given by \Eq{Tmunu-nl1}. Taking the $k_\perp \rightarrow 0$ as the first step of LWL, the transverse integrals in the trace term can be performed to obtain \Eq{int_tmununl}. After substituting this result in \Eq{Pi_BV} and applying the Feynman parameterization, we obtain
\begin{equation}
\left(\Pi_{(11)}^{BV}\right)^{\mu\nu}_{k_\perp \rightarrow 0}=i\frac{e^{3}B}{4\pi}\sum_{n,l}(-1)^{n+l}\int_{0}^{1}dx\int\frac{d^{d}q_{\parallel}}{(2\pi)^{d}}\Lambda^{1-d/2}\left.\frac{\tilde{T}_{n,l}^{\mu\nu}}{\left(\left(q_{\parallel}+xk_{\parallel}\right)^{2}-\Delta^{n,l}_-\right)^{2}}\right|_{d\rightarrow2}
\end{equation}
where we have defined
\begin{equation}
\Delta^{n,l}_-(x,k_\parallel) \equiv \Delta_-(x,k_\parallel) + 2eB\{l-x(l-n)\}.  
\end{equation}
Performing the $d$-dimensional momentum integrals using \Eq{d_dim_int} and defining the dimensional regularization parameter to be $\varepsilon = 1-d/2$, we obtain
\begin{multline}\label{Pi_BV2}
\left(\Pi_{(11)}^{BV}\right)^{\mu\nu}=-\frac{e^{2}}{8\pi^{2}}\int_{0}^{1}dx\sum_{n,l}(-1)^{n+l}2eB\left(\frac{4\pi\Lambda}{\Delta_{-}^{n,l}}\right)^{\varepsilon}\left[4neB\frac{\Gamma(1+\varepsilon)}{\Delta_{-}^{n,l}}g_{\parallel}^{\mu\nu}\delta_{l-1}^{n-1}\right.\\
\left.-\left\{ \varepsilon\Gamma(\varepsilon)g_{\parallel}^{\mu\nu}+\frac{\Gamma(1+\varepsilon)}{\Delta_{-}^{n,l}}S_{\parallel}^{\mu\nu}\right\}\left(\delta_{l}^{n}+\delta_{l-1}^{n-1}\right)-\left\{ \Gamma(1+\varepsilon)\frac{\Delta_{+}}{\Delta_{-}^{n,l}}+\Gamma(\varepsilon)(1-\varepsilon)\right\} g_{\perp}^{\mu\nu}\left(\delta_{l-1}^{n}+\delta_{l}^{n-1}\right)\right]_{\varepsilon\rightarrow 0}
\end{multline}
where $S_\parallel^{\mu\nu}(x,k_\parallel)\equiv 2x(1-x)k_\parallel^{\mu}k_\parallel^{\nu}-g_\parallel^{\mu\nu}\Delta_{+}(x,k_\parallel)$ and
we have discarded the anti-symmetric terms since the charge conjugation symmetry forbids any gyrotropic contribution sourced by these terms in vacuum. Moreover, taking the LLL limit, \Eq{Pi_BV2} exactly reduces to \ref{pi_para_BV} corroborating the findings of \cite{Hattori:2022uzp}.

In the case of a magnetized vacuum, a key new step involves performing the Landau level summations in \Eq{Pi_BV2}. The summation over $l$ can be handled using the Kronecker $\delta$s, after which the summation over $n$ can be expressed in terms of the Hurwitz zeta functions, which are defined as
\begin{equation}
\zeta(s, a) = \sum_{n=0}^\infty \frac{1}{(n + a)^s}.
\end{equation}
By defining $z_B \equiv \frac{\Delta_-}{2eB}$, we finally obtain
\begin{multline}
\left(\Pi_{(11)}^{BV}\right)^{\mu\nu}=-\frac{e^{2}}{8\pi^{2}}\int_{0}^{1}dx\left(\frac{4\pi\Lambda}{2eB}\right)^{\varepsilon}\left[4eB\,\Gamma(1+\varepsilon)g_{\parallel}^{\mu\nu}\left\{ \zeta(\varepsilon,z_{B})-z_{B}\zeta(1+\varepsilon)\right\} \right.\\
\left.-2eB\,\varepsilon\Gamma(\varepsilon)g_{\parallel}^{\mu\nu}\left\{ 2\zeta(\varepsilon,z_{B})-z^{-\varepsilon}\right\} -\Gamma(1+\varepsilon)S_{\parallel}^{\mu\nu}\left\{ 2\zeta(1+\varepsilon)-z^{-(1+\varepsilon)}\right\} \right.\\
\left.+\left\{ \Delta_{+}\Gamma(1+\varepsilon)+2eB(1-\varepsilon)\Gamma(\varepsilon)\right\} g_{\perp}^{\mu\nu}\left\{ \zeta(1+\epsilon,z+x)+\zeta(1+\varepsilon,z-x)-\frac{1}{(z-x)^{1+\varepsilon}}\right\} \right].
\end{multline}
\begin{figure}[t!]
\centering
\includegraphics[width=.47\textwidth]{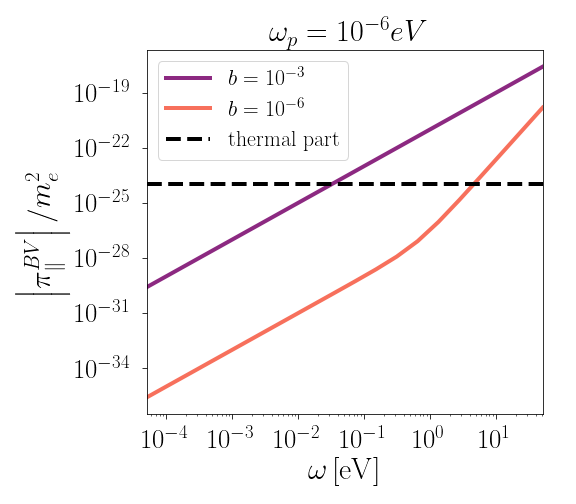}
\quad
\includegraphics[width=.47\textwidth]{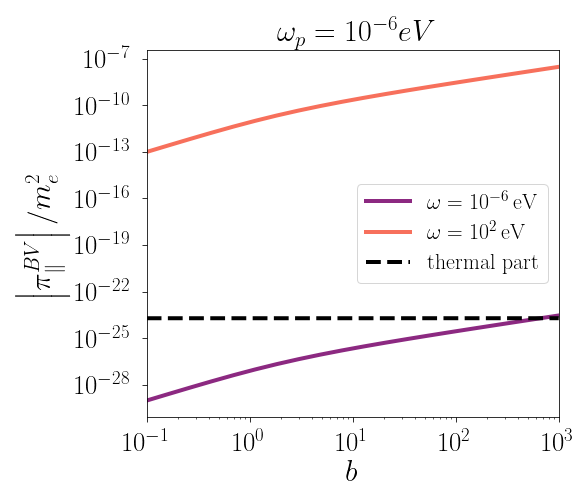}
\vspace{-0.4cm}
\caption{{The longitudinal form factor in a magnetized vacuum as a function of the frequency $\omega$ (left) and magnetic field $b$ (right). The black-dashed line indicates the thermal contribution of a classical plasma to the longitudinal part of the self-energy which is approximately equal to the plasma frequency $\omega_p$ which we fix to a value of $\omega_p = 10^{-6}\, \mathrm{eV}$. }\label{fig:ratio_piBV_wp}}
\end{figure}
As with the pure-vacuum case, taking the $\varepsilon \rightarrow 0$ limit we obtain
\begin{multline}\label{Pi_BV_eps0}
\left(\Pi_{(11)}^{BV}\right)^{\mu\nu}=-\frac{e^{2}}{8\pi^{2}}\int_{0}^{1}dx\left[\tilde{P}_{\parallel}^{\mu\nu}4k_{\parallel}^{2}x(1-x)\left(\frac{1}{\epsilon}-\gamma_{E}+\log\left(\frac{4\pi z_B\Lambda}{\Delta_{-}}\right)-\frac{1}{2z}-\psi^{(0)}(z_B)\right)\right.\\
\left. +g_{\perp}^{\mu\nu}\left(4k_{\parallel}^{2}x(1-x)\left\{ \frac{1}{\epsilon}+\log\left(\frac{4\pi z_B\Lambda}{\Delta_{-}}\right)\right\} +\Delta_{+}\left\{ \frac{1}{x-z_B}-\left(H_{n}(z_B-x-1)+H_{n}(z_B+x-1)\right)\right\} \right.\right.\\
\left.\left.+\frac{\Delta_{-}}{z_B}\left\{ 2(1+\gamma_{E})z_B+\log(z_B-x)+\zeta^{(1,0)}(0,z_B-x)+\zeta^{(1,0)}(0,z_B+x)\right\} \right)\right]
\end{multline}
where we see the appearance of the logarithmically divergent $1/\eps$ term from pure-vacuum QED. A word of caution, all the logarithmic functions here has to be interpreted as complex logarithm. The presence of magnetic fields does not introduce any new divergent behaviour. We can then regularize the self-energy tensor by following the procedure in Ref.~\cite{melrose1976vacuum,Bakshi:1976vd}, where we subtract the zero-field limit ($b\rightarrow 0\, \text{or}\, z_B\rightarrow \infty$) of \Eq{Pi_BV_eps0} from \Eq{Pi_BV_eps0} itself and then add the regularized pure vacuum polarization tensor, $\left(\hat{\Pi}_{(11)}^{V}\right)^{\mu\nu}=\left(k_{\parallel}^{2}g^{\mu\nu}-k_{\parallel}^{\mu}k_{\parallel}^{\nu}\right)\hat{\Pi}^V(k_\parallel^2)$ (in the $k_\perp = 0$ limit) with $\hat{\Pi}^V$ from \Eq{reg_Pi_V}, \textit{i.e.}
\begin{equation}
\left(\hat{\Pi}_{(11)}^{BV}\right)^{\mu\nu}=\left(\Pi_{(11)}^{BV}\right)^{\mu\nu} - \left(\Pi_{(11)}^{BV}\right)^{\mu\nu}_{b\rightarrow 0}+\left(\hat{\Pi}_{(11)}^{V}\right)^{\mu\nu}
\end{equation}
which has the correct zero-field vacuum limit, as expected. We finally obtain the regularized polarization tensor in magnetized vacuum as
\begin{multline}\label{Pi_BV_reg}
\left(\hat{\Pi}_{(11)}^{BV}\right)^{\mu\nu}=-\frac{e^{2}}{8\pi^{2}}\int_{0}^{1}dx\left[\tilde{P}_{\parallel}^{\mu\nu}4k_{\parallel}^{2}x(1-x)\left(\log\left(\frac{z_{B}m^{2}}{\Delta_{-}}\right)-\frac{1}{2z_{B}}-\psi^{(0)}(z_{B})\right)\right.\\
\left.+g_{\perp}^{\mu\nu}\left(4k_{\parallel}^{2}x(1-x)\left\{ \log\left(\frac{z_{B}m^{2}}{\Delta_{-}}\right)-\gamma_{E}\right\} +\Delta_{+}\left\{ \frac{1}{x-z_{B}}-\left(H_{n}(z_{B}-x-1)+H_{n}(z_{B}+x-1)\right)\right\} \right.\right.\\
\left.\left.+\frac{\Delta_{-}}{z_{B}}\left\{ 2(1+\gamma_{E})z_{B}+\log(z_{B}-x)+\zeta^{(1,0)}(0,z_{B}-x)+\zeta^{(1,0)}(0,z_{B}+x)\right\} \right)\right].
\end{multline}

{In \Fig{fig:ratio_piBV_wp}, we present the longitudinal form factor associated with $\tilde{P}^{\mu\nu}$, keeping the plasma frequency fixed at $\omega_p = 10^{-6}\, \mathrm{eV}$ by adjusting the temperature and magnetic field. As shown, the vacuum contributions dominate at high energies ($\omega \gg \omega_p$), while for low energies ($\omega \sim \omega_p$), they become negligible. This validates their omission in previous studies on BSM particle conversion in neutron star magnetospheres and white dwarfs \cite{Hook:2018iia,Hardy:2022ufh}. It is worth noting that the strength of the magnetic field—whether super-critical or subcritical—does not determine the relevance of vacuum contributions. However, stronger magnetic fields cause the vacuum terms to start dominating at lower frequencies. Hence, caution should always be exercised regarding the specific parameter space under consideration.}

\bibliographystyle{unsrt}
\bibliography{bib}
\end{document}